\documentclass{emulateapj}

\shorttitle{Using C I fine structure to probe physical conditions in \dlas}
\shortauthors{Jorgenson et al.}

\def\aap{A \& A}
\def\aj{AJ}
\def\apj{ApJ}
\def\apjl{ApJL}
\def\apjs{ApJS}

\def\araa{ARAA}
\def\mnras{MNRAS}

\def\lya{Ly$\alpha$ }

\def\etal{et al. }
\def\kms{km~s$^{-1}$ }

\def\s-1{s$^{-1}$}
\def\Hz-1{Hz$^{-1}$}

\def\sci#1{{\; \times \; 10^{#1}}}

\def\l1l2{\lambda_2 \; {\rm and} \; \lambda_1}

\def\intl{\int\limits}

\def\eht1{{\rm \hat e_1}}

\def\ltk{\left [ \,}
\def\ltp{\left ( \,}
\def\ltb{\left \{ \,}
\def\rtk{\, \right  ] }
\def\rtp{\, \right  ) }
\def\rtb{\, \right \} }

\def\d3x{d^3 x}

\def\cmma{\;\;\; ,}
\def\perd{\;\;\; .}

\def\cm#1{\, {\rm cm^{#1}}}

\usepackage{graphicx}
\usepackage{epstopdf}
\usepackage{graphicx}
\usepackage{lscape} 

\begin{document}

\def\intl{\int\limits}
\def\nstat{$\approx $}
\def\perd{\;\;\; .}
\def\cmma{\;\;\; ,}
\def\ltk{\left [ \,}
\def\ltp{\left ( \,}
\def\ltb{\left \{ \,}
\def\rtk{\, \right  ] }
\def\rtp{\, \right  ) }
\def\rtb{\, \right \} }
\def\jnu{$J_{\nu}$}
\def\jnuphot{$J_{\nu}^{phot}$}
\def\jnuciistar{$J_{\nu}^{\rm CII^{*}}$}
\def\junit{ergs cm$^{-2}$ s$^{-1}$ Hz$^{-1}$ sr$^{-1}$}
\def\jnutot{$J_{\nu}$$^{total}$}
\def\jnubkd{$J_{\nu}$$^{Bkd}$}
\def\jnuloc{$J_{\nu}$$^{local}$}
\def\jnulw{$J_{\nu}$$^{LW}$}
\def\jnutotciistr{$J_{\nu }$$^{total, C\,II^*}$}
\def\jnulocciistr{$J_{\nu } $$^{local, C\,II^*}$}
\def\jnutotci{$J_{\nu }$$^{total, C\, I}$}
\def\jnulocci{$J_{\nu }$$^{local, C\, I}$}
\def\jnutothtwo{$J_{\nu }$$^{total, H_2}$}
\def\jnulochtwo{$J_{\nu }$$^{local, H_2}$}
\newcommand{\snrlim}{SNR$_{lim}$}
\newcommand{\nhi}{$N_{\rm HI}$}
\newcommand{\mnhi}{N_{\rm HI}}
\newcommand{\flls}{f_{\rm HI}^{\rm LLS}}
\newcommand{\fdla}{f_{\rm HI}^{\rm DLA}}
\newcommand{\llls}{$\ell_{\rm LLS}$}
\newcommand{\ldla}{\ell_{\rm DLA}}
\newcommand{\fnhi}{$f_{\rm HI}(N,X)$}
\newcommand{\mfnhi}{f_{\rm HI}(N,X)}
\newcommand{\Nth}{2 \sci{20} \cm{-2}}
\newcommand{\taux}{$d\tau/dX$}
\newcommand{\gz}{$g(z)$}
\newcommand{\nz}{$n(z)$}
\newcommand{\nx}{$n(X)$}
\newcommand{\omg}{$\Omega_g$}
\newcommand{\ostr}{$\Omega_*$}
\newcommand{\momg}{\Omega_g}
\newcommand{\olls}{$\Omega_g^{\rm LLS}$}
\newcommand{\odla}{$\Omega_g^{\rm DLA}$}
\newcommand{\oneut}{$\Omega_g^{\rm Neut}$}	
\newcommand{\ohi}{$\Omega_g^{\rm HI}$}
\newcommand{\olwz}{$\Omega_g^{\rm 21cm}$}
\newcommand{\ndla}{71}
\newcommand{\cmk}{cm$^{-3}$ K }
\newcommand{\ci}{C\,I}
\newcommand{\cistr}{C\,I$^{*}$}
\newcommand{\mcistr}{C\,I^{*}}
\newcommand{\cistrstr}{C\,I$^{**}$}
\newcommand{\mcistrstr}{C\,I^{**}}
\newcommand{\citot}{(C\,I)$_{tot}$}
\newcommand{\mcitot}{(C\,I)_{tot}}
\newcommand{\cli}{Cl\,I}
\newcommand{\clii}{Cl\,II}
\newcommand{\cii}{C\,II}
\newcommand{\ciistr}{C\,II$^*$}
\newcommand{\dla}{DLA}
\newcommand{\dlas}{DLAs}
\newcommand{\htwo}{H$_{\rm 2}$}
\newcommand{\he}{He\, I}
\newcommand{\sii}{Si\,II}
\newcommand{\siistr}{Si\,II$^{*}$}
\newcommand{\hi}{H\, I}
\newcommand{\ctwo}{C\,II}
\def\lc{${\ell}_{c}$}
\def\lcunit{ergs s$^{-1}$ H$^{-1}$}

\title{Understanding Physical Conditions in High Redshift Galaxies through C I Fine Structure Lines: Data and Methodology\altaffilmark{1}}

\author{Regina A. Jorgenson\altaffilmark{2, 3}, Arthur M. Wolfe\altaffilmark{3}, J. Xavier Prochaska\altaffilmark{4}}

\altaffiltext{1}{Some of the data presented herein were obtained at the W.M. Keck Observatory, which is operated as a scientific partnership among the California Institute of Technology, the University of California and the National Aeronautics and Space Administration. The Observatory was made possible by the generous financial support of the W.M. Keck Foundation.}

\altaffiltext{2}{Institute of Astronomy, University of Cambridge, Madingley Road, Cambridge, CB3 0HA, UK; raj@ast.cam.ac.uk}

\altaffiltext{3}{Department of Physics, and 
Center for Astrophysics and Space Sciences, 
University of California, San Diego, 
9500 Gilman Dr., La Jolla; CA 92093-0424}

\altaffiltext{4}{Department of Astronomy and Astrophysics, 
UCO/Lick Observatory;
University of California, 1156 High Street, Santa Cruz, CA  95064}

\begin{abstract}
We probe the physical conditions in high redshift galaxies, specifically, the Damped Lyman-alpha Systems  (\dlas ) using neutral carbon (\ci ) fine structure lines and molecular hydrogen (\htwo ).  We report five new detections of \ci\ and analyze the \ci\ in an additional 2 \dlas\ with previously published data. 
We also present one new detection of \htwo\ in a \dla .  
 We present a new method of analysis that simultaneously constrains \emph{both} the volume density and the temperature of the gas, as opposed to previous studies that 
a priori assumed a gas temperature.  We use only the column density of \ci\ measured in the fine structure states and the assumption of ionization equilibrium in order to constrain the physical conditions in the gas.  We present a sample of 11 \ci\ velocity components in 6 \dlas\ and compare their properties to those derived by the global \ciistr\ technique.  The resulting median values for this sample are: $<$n(\hi )$>$ = 69 cm$^{-3}$, $<$T$>$ = 50 K, and $<$log(P/k)$>$ = 3.86 \cmk , with standard deviations,  $\sigma$$_{n(\hi )}$ = 134 cm$^{-3}$, $\sigma$$_T$ = 52 K, and $\sigma$$_{log(P/k)}$ = 3.68 \cmk.  
This can be compared with the integrated median values for the same \dlas : $<$n(\hi )$>$ = 2.8 cm$^{-3}$, $<$T$>$ = 139 K, and $<$log(P/k)$>$ = 2.57 \cmk , with standard deviations $\sigma$$_{n(\hi )}$ = 3.0 cm$^{-3}$, $\sigma$$_T$ = 43 K, and $\sigma$$_{log(P/k)}$ = 0.22 \cmk .  Interestingly, the pressures measured in these high redshift \ci -clouds are similar to those found in the Milky Way.  
We conclude that the \ci\ gas is tracing a higher-density, higher-pressure region, possibly indicative of post-shock gas or a photodissociation region on the edge of a molecular cloud.  We speculate that these clouds may be direct probes of the precursor sites of star formation in normal galaxies at high redshift.   
\end{abstract}


\section{Introduction}

The high redshift neutral gas layers known as the Damped Lyman-$\alpha$ Systems (\dlas ) are simultaneously well-understood and mysterious.  On the one hand, the large SDSS survey has identified nearly 1,000 \dlas\ and produced a statistically significant description of the \hi\ column density distribution function, the line density, and the contribution to the neutral gas mass density of the Universe, out to redshifts of $z$ $\sim$ 4
\citep{phw05,pw09}.  
On the other hand, the precise nature of \dlas\ is still not well understood and basic physical properties such as the volume densities, temperatures, physical sizes, and masses are difficult to constrain due to the nature of their detection as absorption imprints on background quasar spectra.  

As the dominant source of neutral gas in the Universe between $z$=[0, 5], the \dlas\ represent a key link in the history of galaxy formation, as they likely provide the source of neutral gas to fuel star formation at high-$z$ (\cite{wolfe03b}, hereafter WGP03).  
The purpose of this paper is to constrain the physical conditions in \dlas\ by taking advantage of a relatively simple three-level system, the neutral carbon (\ci ) fine structure states.  

The ground electronic state of \ci , 2s$^2$ 2p$^2$ $^3$P$_J$, is split into three fine structure levels denoted as $^3$P$_0$, $^3$P$_1$ and $^3$P$_2$.  Following convention, we will refer to these fine structure states as \ci , \cistr , and \cistrstr .  The excited fine structure states are only 23.6 K and 62.4 K above the ground state, which makes them sensitive probes of conditions in cold gas.  The advantage of using the fine structure states of \ci\ is that rather than determining only a line of sight column density, the relative excitation of the \ci\ fine structure states allows for a determination of local physical conditions such as volume density, temperature and pressure.

For almost two decades, the utility of \ci\ fine structure transitions in probing cold gas has been recognized and applied to the high redshift Universe.  
The \ci\ fine structure levels are populated by collisional excitation and de-excitation, radiative decay, UV pumping and direct excitation by the Cosmic Microwave Background (CMB) radiation.  Because of this sensitivity to the CMB, CI fine structure lines in \dlas\ were first used to determine the temperature of the CMB at high redshifts as a test of Big Bang Cosmology (i.e. \cite{songalia94}).  However, because the ionization potential (IP) of neutral carbon is below a Rydberg (IP$_{\ci }$ = 11.3 eV $<$ IP$_{\hi }$ = 13.6 eV), singly ionized carbon, \ctwo , is the dominant state of carbon in the ISM, and \ci\ is not commonly found in \dlas.  
Additionally, because the \ci\ fine structure transitions are generally weak and suffer significant blending, it is only recently with modern high-resolution echelle spectrographs that detailed measurements of the lines have been possible.  

\ci\ has been detected in several high-$z$ \dlas , often as a byproduct of a search for molecular hydrogen (\htwo ), i.e. ~\cite{srianand2005, noterdaeme07b}.  Because they are photo-ionized and photo-dissocated, respectively, by photons of similar energy, the two species tend to be found together.  Previous analyses of \ci\ fine structure states in high-$z$ \dlas\ have generally assumed a gas temperature in order to calculate the gas volume density and have generally found that for reasonable assumptions of the gas temperature, \ci\ detections require relatively high densities.
In this paper, we introduce a technique in which we make no a priori assumptions about the gas temperature.

Inspired by the work of \cite{jenkins79} and \cite{jenkins01} who used \ci\ fine structure absorption to probe pressures in the local Milky Way ISM, we implement their technique on high redshift \dlas .    
Most recently, \cite{jenkins07} find that most of the \ci\ in the Milky Way is in gas at pressures between $3 <$ log (P/k) $<$4 cm$^{-3}$ K, with the distribution centered at P/k = 2700 cm$^{-3}$ K. We will show that, interestingly, our results for the high redshift \dlas\ are similar to those of \cite{jenkins07} for the Milky Way ISM.  
In addition, by invoking the sensible assumption of ionization equilibrium, we can use the population of the \ci\ fine structure levels to constrain the total radiation field in some cases.  This in turn acts as a check on the radiation field as determined by the `\ciistr\ technique' (WPG03; WGP03) -- a way of measuring the radiation field due to star formation in a \dla\ by equating the cooling rate measured via the CII* fine structure transition with the heating rate.

This is the first in a series of two papers focused on deriving the physical conditions in \dlas\ via the analysis of \ci\ fine structure lines.  In this first paper, we report the detection of \ci\ in five high redshift \dlas , and analyze an additional 2 \dlas\ with previously published \ci\ data.  We derive limits on density, temperature, and pressure as well as the radiation field contained in the \ci -bearing gas of each \dla .  In the systems for which we have coverage of \htwo , we compare the \ci\ fine structure results to those derived by an analysis of the \htwo\ rotational J level populations.  In the second paper of this series we will extend our analysis to the \ci -bearing \dlas\ already published in the literature, analyze the broader physical implications of the \ci\ data and finally, propose a physical picture of \ci -bearing \dlas .
    
This paper is organized as follows:
We discuss our data and data analysis techniques in \S~\ref{sec:data}.  We present our procedure for analyzing the \ci\ data through the specific example of \dla\ 0812$+$32 at z$_{abs}$ = 2.626 in \S~\ref{sec:technique}.  We then summarize the results for each of the \ci -bearing \dlas\ in \S~\ref{sec:results}.  \S~\ref{sec:discussion} contains discussion of the \dlas\ presented here and we conclude in \S~\ref{sec:conclusion}.  Throughout this paper we make the standard assumption that the ratio of column density of element X is equal to the ratio of volume density, $\frac{N(X^a)}{N(X^b)} = \frac{n(X^a)}{n(X^b)}$, i.e. we are analyzing average conditions along the line of sight. 

\section{Data and Methodology}~\label{sec:data}

The \ci\ sample presented here represent serendipitous discoveries made during the course of an on-going campaign to obtain high resolution spectra of \dla\ targets for detailed study.  
Data for this paper was taken primarily with the HIRES spectrograph (Vogt et al. 1994) on the Keck I telescope, with a typical decker that resulted in an instrumental resolution FWHM = 6.25 \kms .  Details of specific observations are given in Table~\ref{tab:observations}.   Data were reduced and continuum fit using the standard XIDL \footnote{http://www.ucolick.org/$\sim$xavier/IDL/} packages.  Because of the complex blending of many close fine structure transitions that typically exhibit several velocity components, we were not able to successfully apply the apparent optical depth technique ~\citep{savage91} for a measurement of the column densities.  We also did not have the high signal to noise ratio and resolution of the ~\cite{jenkins01} study of interstellar \ci , in which they developed a modified AODM technique \citep{jenkins01}.  Instead we used the VPFIT package, version 9.5 \footnote{http://www.ast.cam.ac.uk/$\sim$rfc/vpfit.html} to measure the \ci\ fine structure column densities, N(\ci ), N(\cistr ), and N(\cistrstr ), redshifts, and Doppler parameters (b), where b=$ \sqrt{2}$$\sigma$, and $\sigma$ is the velocity dispersion in \kms .  Wavelengths and f-values were taken from ~\cite{morton03} \footnote{Note, however, that there is some confusion over the correct f-values for the \ci\ transitions given a more recent set determined by E. B. Jenkins (2006, private communication).  Since we cannot determine which set of f-values is more correct, in this paper we use the ~\cite{morton03} values to be consistent with what has previously been used in the literature.}.
  
In general, we simultaneously fit as many \ci\ multiplets as possible. In all cases, we included all multiplets that fell redward of the \lya\ forest, which generally included the strongest multiplets at 1656\AA\ and 1560\AA\ , and usually the multiplet at 1328\AA .  In cases where a multiplet fell within the Lyman $\alpha$ forest, yet did not contain any obvious blending with a forest line, we included that multiplet in our analysis.  We rejected sections of spectra that contained obvious blending.  

To determine upper limits for non-detected fine structure states, generally for N(\cistrstr ), we used VPFIT, which allows for an estimation of the 1$\sigma$ upper limit by inputting a "reasonable guess" of the column density and re-running the fit to give linear errors.  The error is then taken as the 1$\sigma$ upper limit.  For a conservative reasonable guess we used the measured column density of the line with the smallest measured column density, usually N(\cistr ), since it is almost always the case that N(\cistrstr ) $<$ N(\cistr ).  This results in a conservative upper limit to the N(\cistrstr ) value.  

We present one new detection of \htwo\ in \dla\ 2340$-$00, and we analyze the \htwo\ in \dla\ 0812$+$32.  
While it is likely that the other systems contain measurable \htwo\ as well, we lack
the spectral coverage to confirm this speculation. 

All other information about each \dla , including metallicity, dust-to-gas ratio, and log N(C II) 
was determined using the Apparent Optical Depth Method, AODM ~\citep{savage91}, unless otherwise stated (i.e. in the cases where we performed a full component by component analysis).  To measure the log$\frac{C II}{C I}$ we must employ the conventional method of measuring N(\cii ) by proxy using N(Si II), because the \ctwo\ $\lambda$1334 transition is saturated in every \dla\ presented in this paper.  As in \cite{wolfe04} we let [C/H] \footnote{The abundance ratio with respect to solar is defined as [X/Y] = log(X/Y) - log(X/Y)$_{\odot}$} = [Si/H] $+$ [Fe/Si]$_{int}$ where [Fe/Si]$_{int}$ = $-$0.2 for a minimal depletion model or [Fe/Si]$_{int}$ = 0.0 for a maximal depletion model.  While \cite{wolfe04} estimate the error in the measurement of $\frac{\ctwo }{\ci }$ to be 0.1 dex, we use the more conservative estimate of 0.2 dex.

Details of the observations are given in Table~\ref{tab:observations}, while the details of the \ci\ measurements are summarized in Table~\ref{tab:cidata}.  In the following we provide a brief summary of the data analysis of each \dla\ in our sample:

$\bullet {\bf \dla\ 0812+32, z_{abs} =2.62633:}$   \dla\ 0812$+$32 at z$_{abs}$ =2.62633 has been studied extensively (see ~\cite{pro2003}), and boasts one of the highest known \dla\ metallicities and \hi\ column densities.   ~\cite{pro2003} used the many available transitions to show that the relative
elemental abundance pattern is similar to that of the Milky Way.  ~\cite{jorg09} presented the first direct evidence of cold (T$_{thermal}$ $\leq$ 78 K, (115 K, 2$\sigma$)) gas at high redshift using a curve of growth analysis of a sub-resolution, narrow velocity \ci\ component in this \dla , and recently, ~\cite{tumlinson2010} presented the detection of HD.
 
\begin{figure}
\plotone{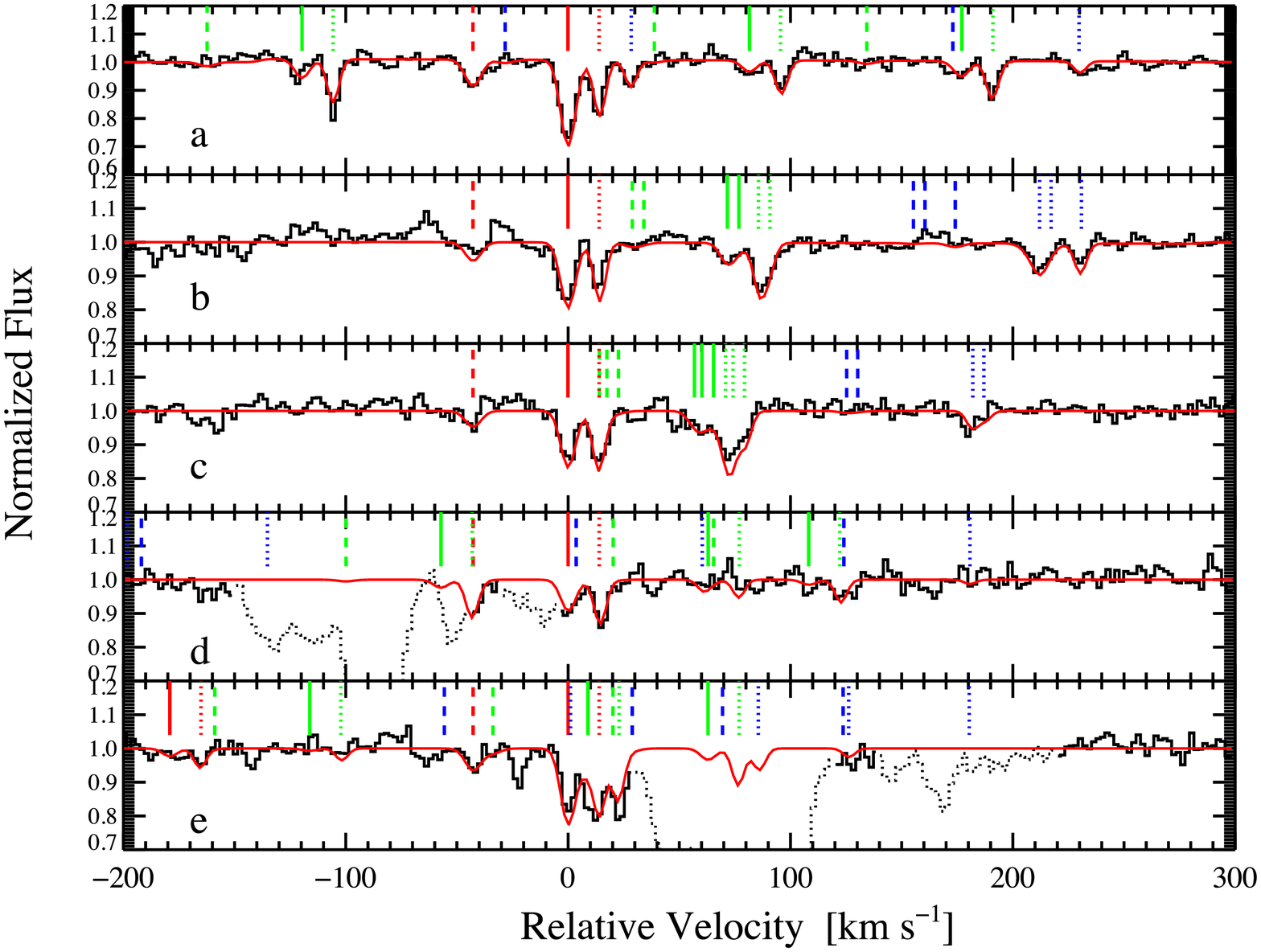}
\caption{ \dla\ 0812$+$32 \ci\ velocity structure. Spectral regions covering the five \ci\ multiplets used in the analysis of \dla\ 0812$+$32: The multiplets are a) 1656\AA , b) 1560\AA , c) 1328\AA , d) 1280\AA , and e) 1277\AA .  Black is the data, red is our fit.  The different fine structure transitions are color coded: \ci\ = red, \cistr\ = green, \cistrstr\ = blue, while the three velocity components are denoted by different linestyles: component 1 at $v\sim -43$\kms , or z$_{abs}$ =  2.625808 is dashed, component 2 at v = 0\kms , or z$_{abs}$ = 2.6263247 is solid, while component 3 at $v\sim +14$\kms , or z$_{abs}$ = 2.626491 is dotted.  Interloper lines are denoted by dotted black lines.
}
\label{fig:j0812_spec}
\end{figure}

 In this work we simultaneously fit data from three Keck HIRES runs with details given in Table~\ref{tab:observations}.  Eight multiplets were used to fit the \ci\ fine structure lines; $\lambda$1656, $\lambda$1560, $\lambda$1328, $\lambda$1280, $\lambda$1279, $\lambda$1277, $\lambda$1276, and $\lambda$1270.  The $\lambda$1276 $-$ $\lambda$1280 multiplets were blended with a C IV doublet at z$_{abs}$ = 1.992.  
Figure~\ref{fig:j0812_spec} contains the spectral data of \dla\ 0812$+$32, in black, overlaid with our fit, in red, and the \ci\ fine structure lines marked.  Each velocity component is denoted by a different linestyle (solid, dashed, dotted, etc.) while each fine structure state is denoted by a different color (\ci = red, \cistr = green, \cistrstr = blue).  It is apparent from this figure that 1) in some cases the \ci\ fine structure transitions fall so close to each other that they are blended, and 2) the multi-component velocity structure found in most \dla s further complicates the analysis.  
As denoted in Figure~\ref{fig:j0812_spec}, three \ci\ velocity components are required to fit this \dla .  Component 1 at z$_{abs}$=2.625808, or $v \sim -43$ \kms , with Doppler parameter of b = 3.25\kms , is the weakest component and contains the largest errors.   Component 2 at z$_{abs}$=2.6263247 with b = 2.57\kms , is located at $v = 0$ \kms , and has an upper limit on N(\cistrstr ).  Component 3 at $v \sim +14$ \kms , or z$_{abs}$=2.626491, is the narrow component reported in ~\cite{jorg09} with b = 0.33 \kms .

\begin{figure}
\plotone{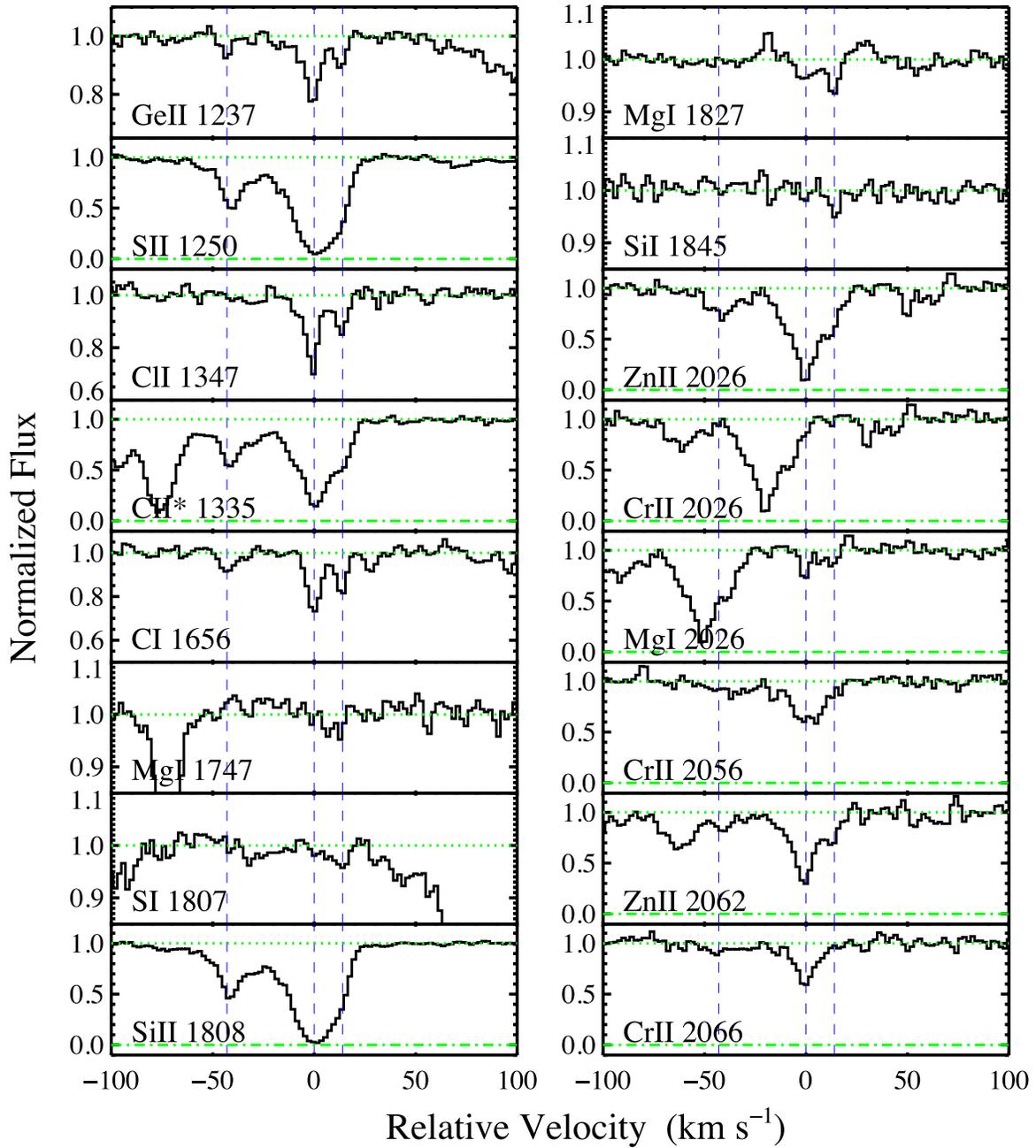}
\caption{ \dla\ 0812$+$32 low ions.  Blue vertical line at v = 0 \kms\ marks component 2 at z$_{abs}$ = 2.6263247, while the narrow Doppler parameter, cold component 3, is located at $+$14 \kms .
}
\label{fig:J0812_spec_lowions}
\end{figure}

While the velocity structure of \ci\ in \dla\ 0812$+$32 is well constrained by the fitting of many multiplets, our confidence in the fit is increased by the similar velocity structure seen in neutral chlorine, \cli\ $\lambda$1347.  Because of their similar ionization potentials, both below a Rydberg, \ci\ and \cli\ $\lambda$1347 are often observed to have the same velocity structure ~\citep{jura74a}.  This is seen in Figure~\ref{fig:J0812_spec_lowions} along with several other sub-Rydberg and low-ion transitions that trace the velocity structure of \ci , namely, Ge II $\lambda$1237, Mg I $\lambda$1827, Si I $\lambda$ 1845 and Zn II $\lambda$ 2062.

This \dla\ also contains relatively strong \htwo , with a total log N(\htwo ) = 19.90 cm$^{-2}$, giving a molecular fraction of f = 0.067, where f is defined as f = $\frac{2 N(H_{2})}{N(H I) + 2 N(H_{2})}$.  The \htwo\ velocity components are consistent with those of \ci .

\begin{figure}
\plotone{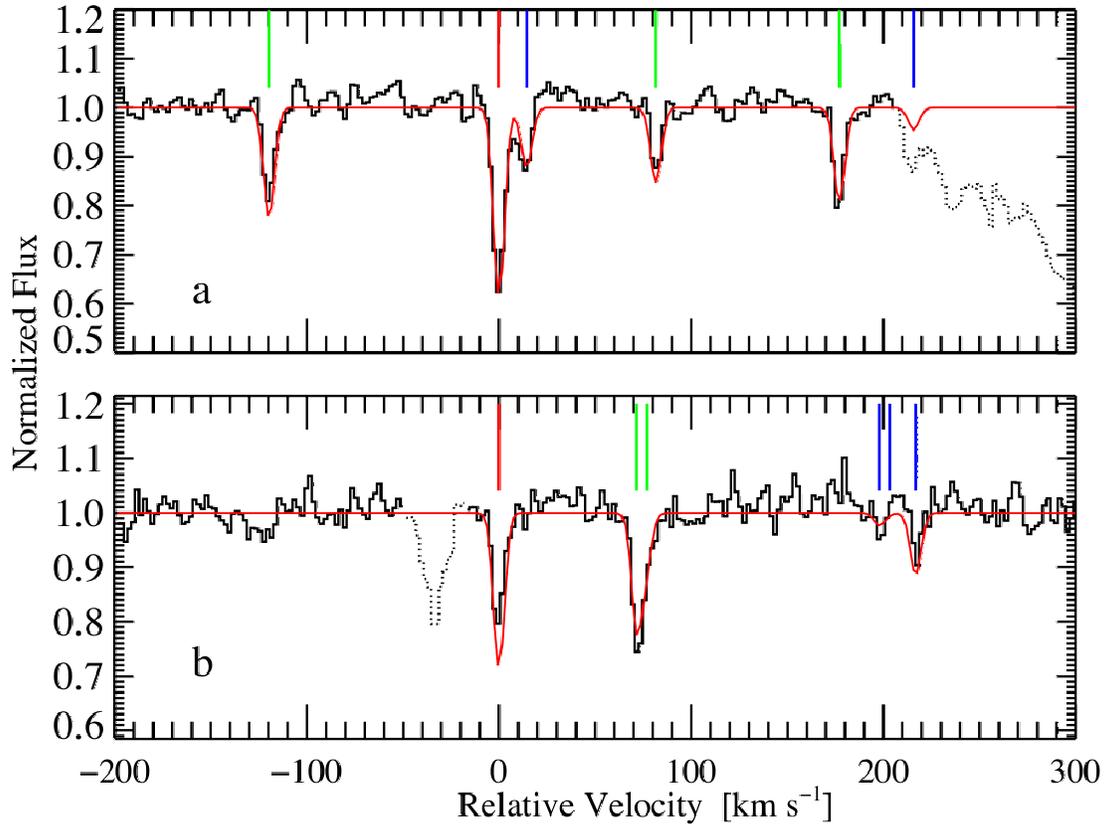}
\caption{ \dla\ 0812$+$32 z$_{abs}$=2.06 \ci\ velocity structure.  Notation is as in Figure 1.  Only the 1656$\AA$ and 1560$\AA$ multiplets were used in the fit because of the serious blending with the Lyman $\alpha$ forest in lower wavelength multiplets.  The single \ci\ component is located at $v$ = 0 \kms , with z$_{abs}$= 2.066780.
}
\label{fig:j0812_z206_spec}
\end{figure}

\begin{figure}
\plotone{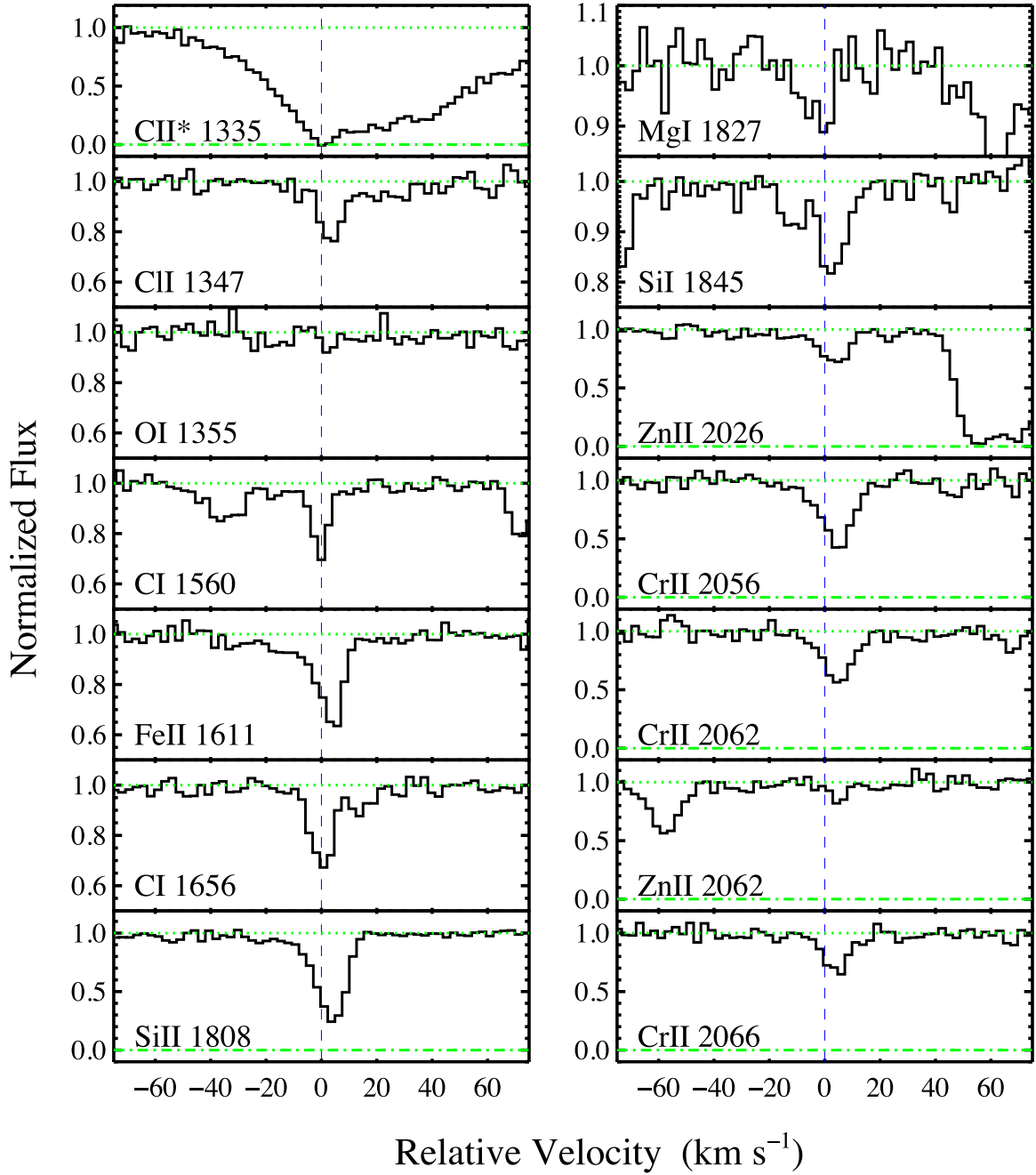}
\caption{ \dla\ 0812$+$32 z$_{abs}$=2.066780.  Note the velocity offset between \ci\ and the other low ions.  Also note the blending of \ciistr .
}
\label{fig:J0812_lowz_spec_lowions}
\end{figure}

$\bullet {\bf \dla\ 0812+32, z_{abs} =2.066780:}$  Data for this \dla\ is the same as that for \dla\ 0812$+3$2, z$_{abs}$ = 2.62633 discussed above.  
There is one \ci\ velocity component in this \dla .  To fit \ci , we used only the $\lambda$1656 and $\lambda$1560 multiplets because of heavy blending with the Lyman $\alpha $ forest blueward of restframe $\sim$$\lambda$ 1467 \AA , see Figure~\ref{fig:j0812_z206_spec}.   This \dla\ also contains \cli , however it is interesting to note that the centroid of the \cli\ profile is displaced by $\approx$$+$5\kms\ with respect to $v$ = 0 \kms\ at the \ci\ centroid of z$_{abs}$= 2.066780.  
While this is unexpected, it is also true of all of the low-ions, see Figure~\ref{fig:J0812_lowz_spec_lowions}.  However, Si I and Mg I appear to align more closely with \ci . 

$\bullet  {\bf \dla 1331+17: }$  \cite{wolfe79} discovered 21 cm absorption towards QSO1331$+$17 at z$_{abs}$=1.77642, a similar redshift at which a \dla\ had previously been discovered at optical wavelengths ~\citep{carswell75}.  They deduced a spin temperature of T$_s$ $\approx$ 770 $-$ 980 K by combining the 21 cm line equivalent width with the N(\hi ) obtained from Lyman $\alpha$ absorption. 

One of the first attempts to measure the CMB temperature at high redshift using \ci\ was made by \cite{meyer86} using \dla\ 1331$+$17.  They measured \ci\ and put an upper limit on the ratio n(\cistr )/n(\ci ) that allowed them put an upper limit on the temperature of the CMB, T$_{CMB} <$16 K at $z_{abs} = 1.776$.  Later, \cite{songalia94} used the Keck telescope to make more precise measurements.  They succeeded in measuring the \cistr\ transition and derived a CMB temperature of T = 7.4 K that agreed well with the theoretical prediction.  More recently, \cite{cui05} discovered an unusually high level of molecular hydrogen (\htwo ) in \dla\ 1331$+$17 using Hubble Space Telescope.  They detect a molecular fraction of 5.6$\% \pm$0.7$\%$ in a component at z$_{abs}$=1.776553.  They derive an excitation temperature of the \htwo - bearing component of T$_{ex}$ = 152 K.  Recently ~\cite{carswell09} discovered a narrow velocity component of \ci\ at z$_{abs}$ = 1.776525, that 
requires gas with T$_{thermal}$ $\leq$ 218 K (1$\sigma$). 

 To maximize the UV coverage for \dla\ 1331$+$17, we used a Keck HIRES spectrum with instrumental resolution FWHM = 6.25 \kms in combination with a bluer UVES spectrum of resolution FWHM = 7.0 \kms kindly provided by R. F. Carswell.  This provided coverage of the maximum number of \ci\ multiplets, down to the restframe \ci\ $\lambda$1277 multiplet at observed wavelength $\lambda _{obs}$ $\approx$ 3546 \AA .  We included in the fit the following multiplets:   $\lambda$1656, $\lambda$1560, $\lambda$1328, $\lambda$1280, and $\lambda$1277.

\begin{figure}
\plotone{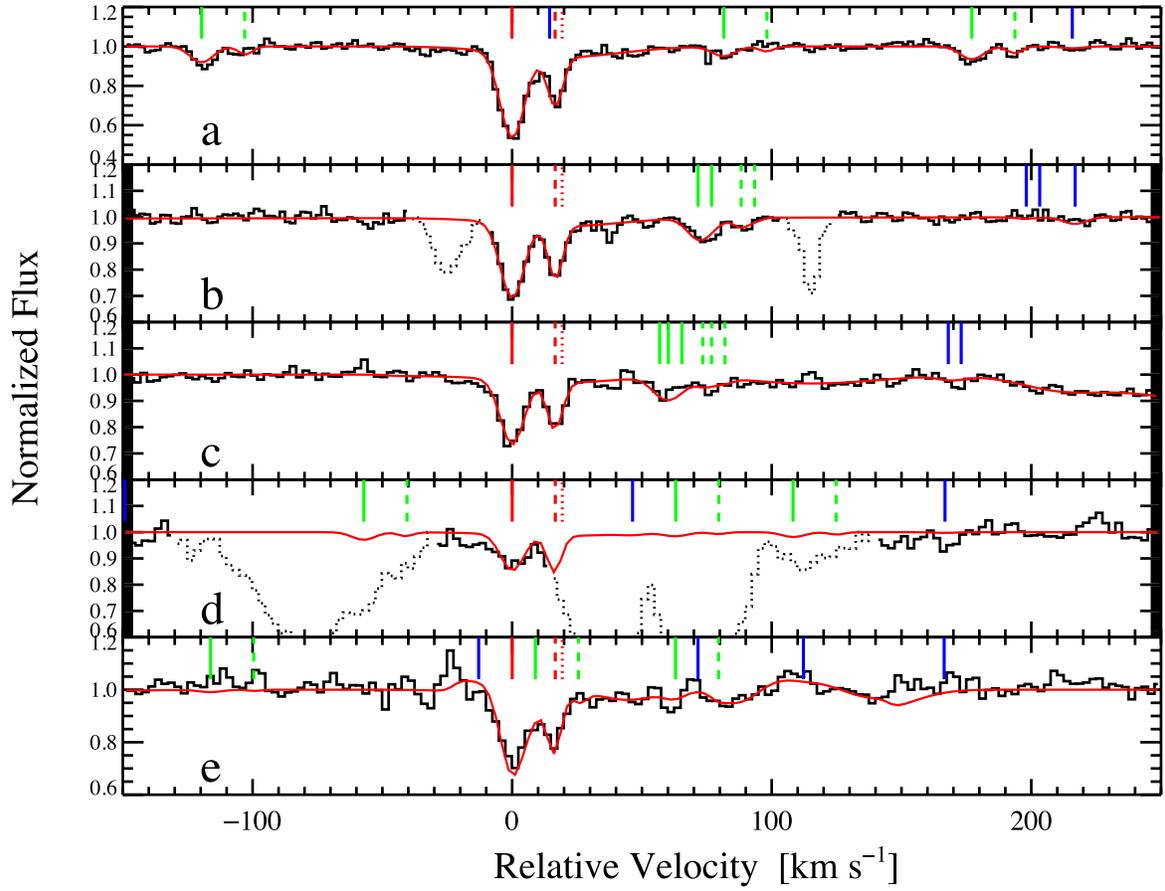}
\caption{Spectral regions covering the five \ci\ multiplets used in the analysis of \dla\ 1331$+$17.  Notation is as in Figure 1.  This system requires 3 \ci\ components: component 1 at $v=0$\kms , or z$_{abs}$ = 1.77637 is solid,   component 2 at $v\sim$17\kms , or z$_{abs}$ = 1.77652 is dashed, while component 3 at $v\sim$20 \kms , or z$_{abs}$ = 1.77659 is dotted.  Interloper lines are denoted by dotted black lines.
}
\label{fig:q1331spec}
\end{figure}

\begin{figure}
\plotone{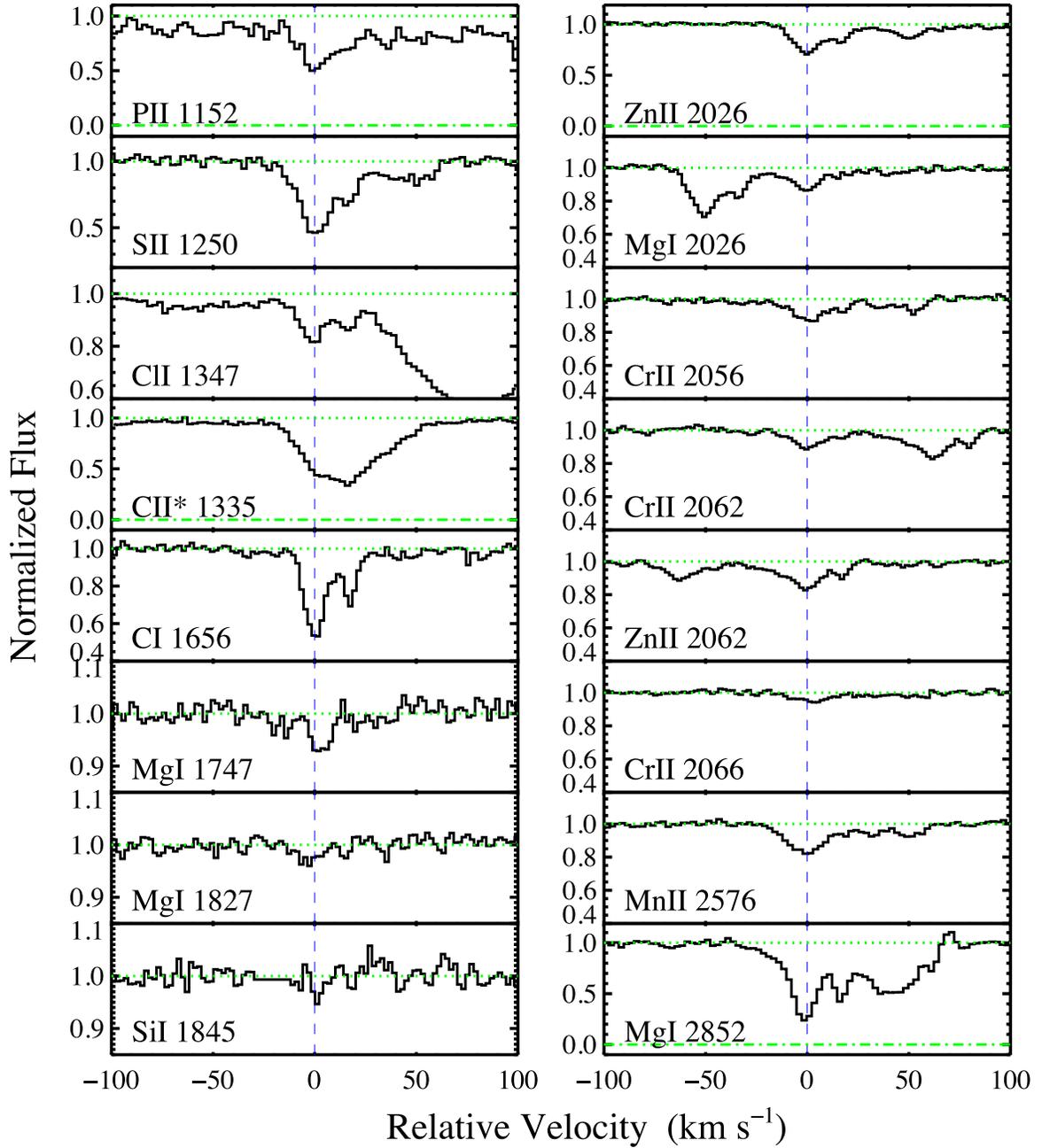}
\caption{Resonance lines and low ions that trace the \ci\ velocity structure of \dla\ 1331$+$17.  $v = 0$ \kms located at component 1 at z$_{abs}$ = 1.77637.  Note that the \ciistr\ transitions is likely blended with a forest line due to its different velocity profile.
}
\label{fig:1331_spec_other_ions}
\end{figure}

 As shown in Figure~\ref{fig:q1331spec}, \dla\ 1331$+$17 requires three \ci\ velocity components.  Component 1, at $v\sim$0 \kms , is the strongest component in terms of column density (see Table ~\ref{tab:cidata}) with b = 5.08 $\pm$ 0.24 \kms .  Component 2, at $v\sim$17 \kms , has a narrow velocity structure (b = 0.55 $\pm$ 0.13 \kms ), and is discussed in detail in ~\cite{carswell09}.  The third component, at $v\sim$20 \kms , with b = 24.45 $\pm$ 6.2 \kms , does not exhibit measurable \cistr\ or \cistrstr\ absorption.  Lending confidence to the fit is the presence of \cli , shown in Figure~\ref{fig:1331_spec_other_ions} along with several other sub-Rydberg and low-ion transitions that trace the velocity structure of \ci , namely, P II $\lambda$1152, S II $\lambda$1250, Mn II $\lambda$2576, and Mg I $\lambda$2852.

We can compare our results to another recent measurement of \ci\ absorption in this system.  \cite{dessauges04} used VLT and Keck data to measure \ci\ in two components, log N(\ci ) = 13.12 $\pm$ 0.02 cm$^{-2}$ at $z_{abs}$ = 1.776365 and log N(\ci ) = 12.72 $\pm$ 0.02 cm$^{-2}$ at $z_{abs}$ = 1.776523.  While in the latter case, what we call component 2, their measurements agree with ours to within 1$\sigma$ errors, the component 1 is in disagreement, with small errors, by 0.12 dex.  However this difference is reasonable considering continuum placement uncertainties and the fact that they did not include fitting of the \ci\ fine structure states which involve considerable blending with the resonance line.

$\bullet {\bf \dla\ 1755+578:}$  \dla\ 1755$+$578, at z$_{abs}$ = 1.9692, contains 8 \ci\ components, making it one of the more complex \ci\ systems.  It is interesting in its own right because we have discovered the presence of \siistr\ absorption in this \dla .  While \siistr\ has been observed in GRB-\dlas\ pumped by the UV radiation field of the GRB afterglow \citep{sava04,pcb06},  
this is the first known case of \siistr\ observed in a high redshift QSO-\dla .  Because \siistr\ is generally thought to arise in warm gas -- its excitation energy is 413K -- this system is not only unique but also extremely interesting since it appears that several velocity components contain \emph{both} \ci\ and \siistr\ absorption, indicating that this is one of the rare \dlas\ exhibiting both CNM \emph{and} WNM gas at high-$z$. 
We discuss this object in greater detail in a future paper.  

$\bullet {\bf \dla\ 2100-06:}$  While this \dla , at z$_{abs}$ = 3.09237, exhibits measurable \ci , our data are not of sufficiently high quality to obtain good measurements of the \ci\ fine structure states.  This is in part because the \ci\ lines are not strong, and in part because the strongest multiplet, $\lambda$1656, falls close to an order gap and as a result is contained in a lower signal to noise region.   

\begin{figure}
\plotone{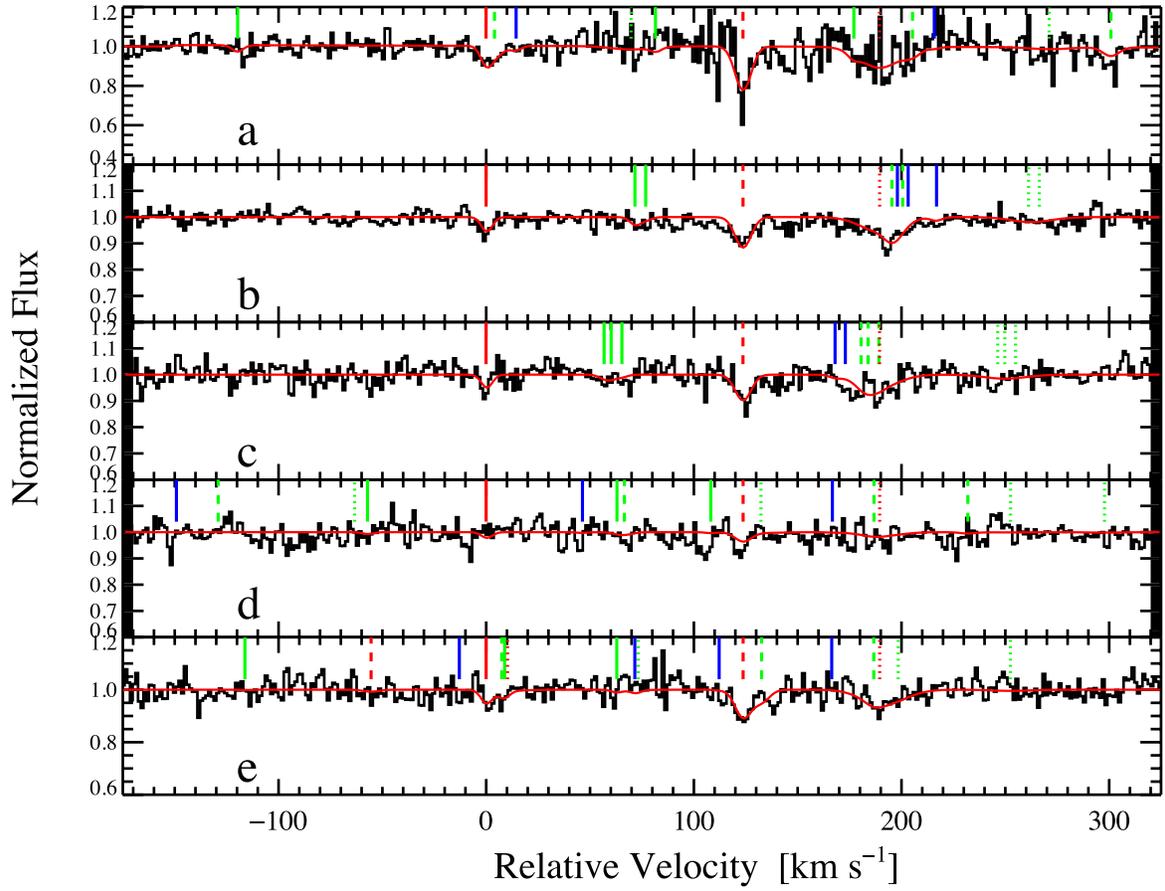}
\caption{Spectra of \dla\ J2100$-$06 covering the 1656$\AA$, 1560$\AA$, 1328$\AA$, 1280$\AA$, and 1277$\AA$ multiplets.  Notation as in Figure 1.  The best fit requires three \ci\ velocity components.  The lower S/N in 1656$\AA$ multiplet (panel a) is due to its proximity to the order gap.  
}
\label{fig:J2100_spec}
\end{figure}

All \ci\ multiplets $\lambda$1277 and redwards are included in the fit.  The fit requires three \ci\ velocity components, only one of which has measurable N(\cistrstr ), albeit with large enough errors that it might best be considered as
an upper limit.  We present spectral coverage 
of the multiplets used in Figure~\ref{fig:J2100_spec}.

\begin{figure}
\plotone{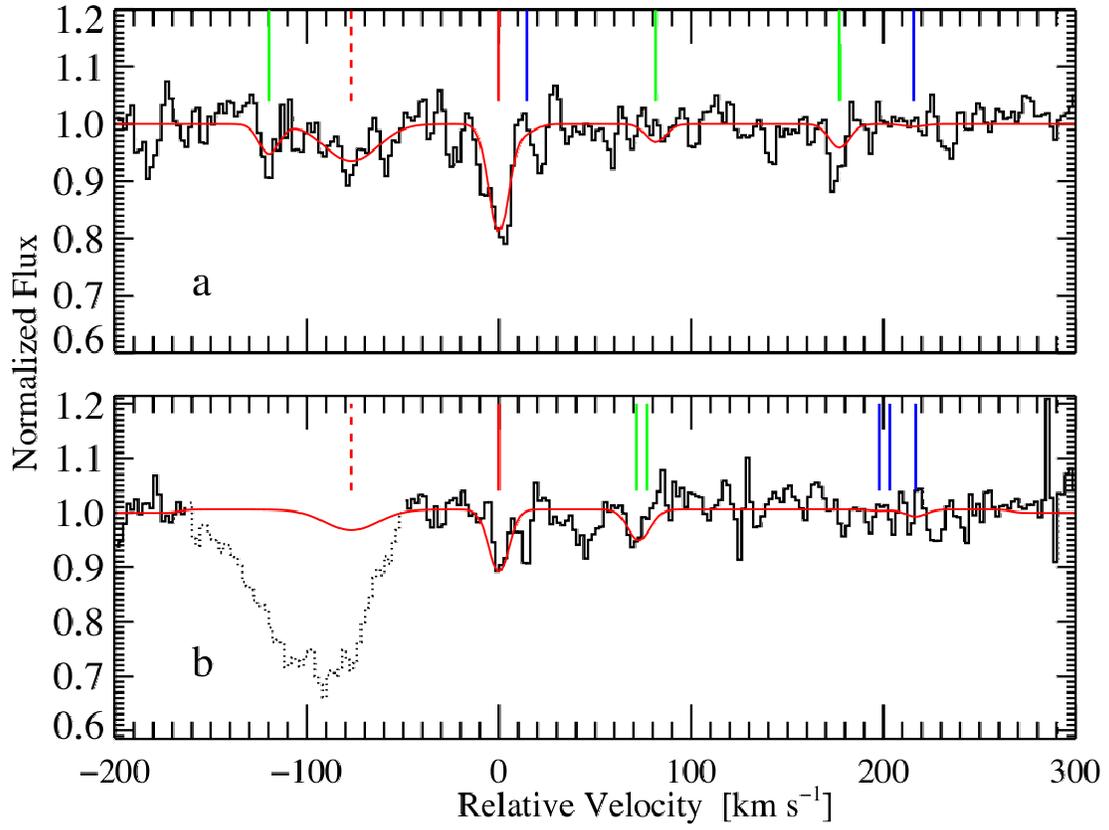}
\caption{Spectra of \dla\ 2231$-$00 covering the 1656$\AA$ and 1560$\AA$ multiplets.  Notation as in Figure 1.  The best fit requires two \ci\ velocity components:  component 1 at $v$ = $-$77 \kms , or z$_{abs}$ = 2.06534, with no measurable fine structure lines, and component 2 at $v$ = 0 \kms , or z$_{abs}$ = 2.066122, with detected \ci\ fine structure lines (however, N(\cistrstr ) is technically an upper limit).  
}
\label{fig:q2231_spec}
\end{figure}

\begin{figure}
\plotone{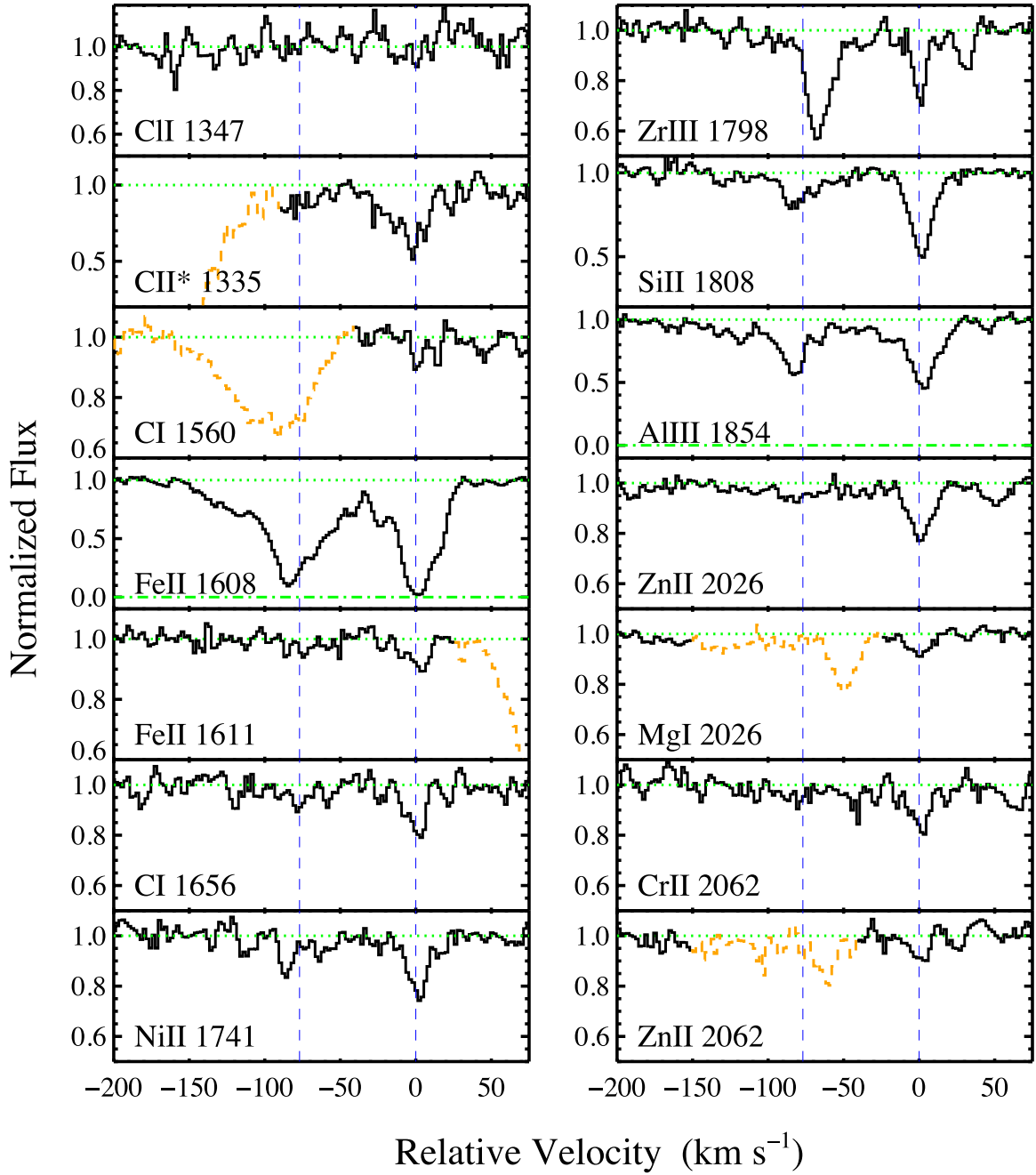}
\caption{Spectra of \dla\ 2231$-$00 low ions.  $v = 0$ \kms is located on component 2 at z$_{abs}$ = 2.066122 and marked by a vertical dashed blue line.  Component 1 is located at v$\sim$$-$77 \kms .
}
\label{fig:J2231_spec_lowions}
\end{figure}

$\bullet {\bf \dla\ 2231-00:}$  \dla\ 2231$-$00 at $z_{abs} =$ 2.066 was analyzed by ~\cite{pro99a}, who reported the detection of titanium, and
~\cite{pro99b} analyzed a Lyman-limit system at z$_{abs}$=2.652 in this sightline.  With an absorption redshift of z$_{abs}$ = 2.066 in the line of sight to a quasar with emission redshift of z$_{em}$ = 3.02, both the $\lambda$1328 and the $\lambda$1560 \ci\ multiplets of this \dla\ fall in the Lyman-$\alpha$ forest.  While most of the $\lambda$1560 multiplet is unblended, the $\lambda$1328 multiplet contains a blend with a large Lyman-alpha forest line.  Our analysis includes only the $\lambda$1656 and $\lambda$1560 multiplets, which are shown in Figure~\ref{fig:q2231_spec}.  The best fit required two \ci\ velocity components:  component 1 at $v$ = $-$77 \kms , or z$_{abs}$ = 2.06534, with no measurable fine structure lines, and component 2 at $v$ = 0 \kms , or z$_{abs}$ = 2.066122, with detected \ci\ fine structure lines.  These two \ci\ components trace the two strongest components of the other low ions, as seen in Figure~\ref{fig:J2231_spec_lowions}.  While \cli\ is located in a region of low S/N, the detection of Mg I $\lambda$ 2026 lends confidence to the \ci\ detection.  We note that while the large error of the N(\cistrstr) measurement of component 2 indicates that it should probably be considered an upper limit, for the purposes of the present analysis we will treat the value as a detection.  

\begin{figure}
\epsscale{0.8}
\plotone{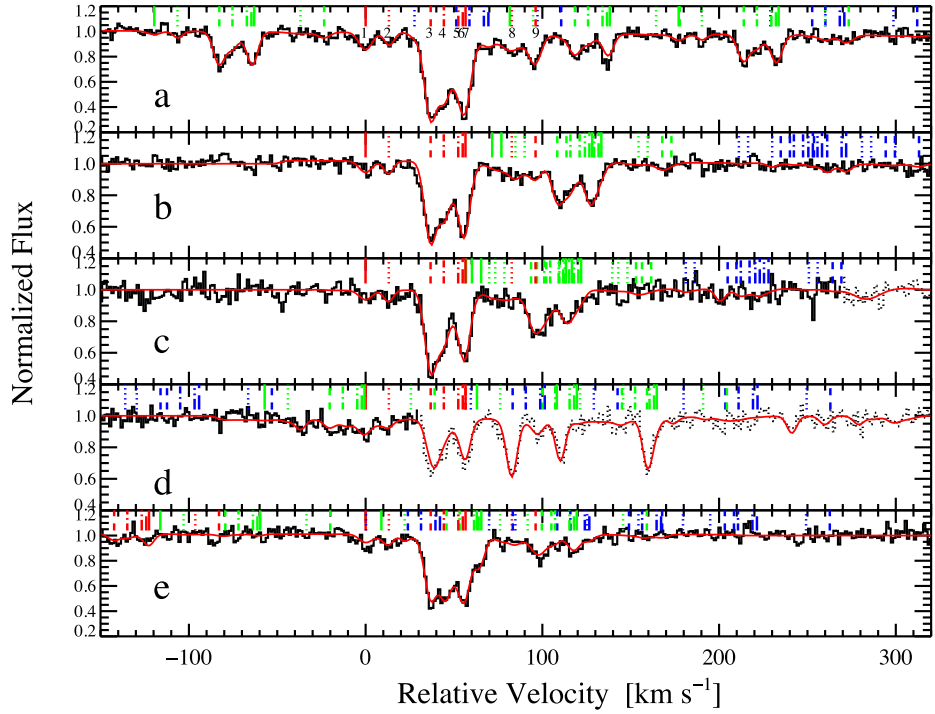}
\caption{J2340 \ci\ velocity profiles over five multiplets, a) 1656$\AA$, b) 1560$\AA$, c) 1328$\AA$, d) 1280$\AA$, e) 1277$\AA$.  Notation is as in Figure 1.  The components are labeled in (a), as 1 $-$ 9 from lowest to highest relative redshift (relative to the arbitrarily chosen v = 0 \kms at z$_{abs}$ = 2.054151) and located at the following velocities: $v \sim$ 0 \kms, 13 \kms , 37 \kms , 44 \kms , 52 \kms , 55 \kms , 57 \kms , 83 \kms , and 96 \kms .  
}
\label{fig:j2340_spec}
\end{figure}

$\bullet {\bf \dla\ 2340-00, z_{abs} = 2.054924:}$  The complex multi-component nature of the \ci\ in this z$_{abs}$ = 2.054924 \dla\ made it difficult to fit despite the fact that seven \ci\ multiplets ($\lambda$1270, $\lambda$1276, $\lambda$1277, $\lambda$1279, $\lambda$1328, $\lambda$1560 and $\lambda$1656) have been included in the fit.  The $\lambda$1279 multiplet was only partially used due to blending with interlopers.  Nine \ci\ components were required in the fit, see Figure ~\ref{fig:j2340_spec}.  The components are labeled  1 $-$ 9 and located at the following velocities relative to an arbitrarily chosen z$_{abs}$ = 2.054151 for v = 0 \kms : $v \sim$ 13 \kms , 37 \kms , 44 \kms , 52 \kms , 55 \kms , 57 \kms , 83 \kms , and 96 \kms .  

\dla\ 2340$-$00 also contains a relatively high total column density of molecular hydrogen, logN(\htwo ) = 18.20 cm$^{-2}$, giving an \htwo\ fraction f = 0.014, making it one of the few relatively \htwo - rich \dlas . 

$\bullet {\bf ABSL \ 2340-00, z_{abs} = 1.36:}$  The sightline towards J2340$-$00 contains a second set of \ci\ absorption lines that require 4 \ci\ velocity components to fit. Only the $\lambda$1656 and $\lambda$1560 \ci\ multiplets were available to constrain the fit because of the low redshift.   Because of blending, in the case of $\lambda$1656 multiplet with the \ci\ $\lambda$1279/$\lambda$1280 multiplet of the high-$z$ absorber, and in the case of $\lambda$1560 multiplet with the Lyman $\alpha$ forest, the \ci\ lines were fit by tying them to their associated Mg I transitions.  The results are given in Table~\ref{tab:cidata}.


\section{Applying the \ci\ fine structure technique}~\label{sec:technique}

In this section we detail the processes by which we determined the constraints on the physical conditions in each \dla .  

\subsection{The Steady State Equation}
We developed an in-house code based on POPRATIO \citep{silva01} to calculate the theoretical \ci\ fine structure level populations by making the standard assumption of steady state
.  
Following \cite{silva01} we considered spontaneous radiative decay, direct excitation by the CMB, UV pumping, and collisional excitation and de-excitation.  As in POPRATIO, the rate equations leading to steady state populations of state i is given by

\begin{equation}
\begin{array}{rcl}
\sum_j n_j (A_{ji} + B_{ji} u_{ji} + \Gamma_{ji} + \sum_k n^k q^k_{ji}) = \\
n_i \sum_j (A_{ij} + B_{ij} u_{ij} + \Gamma_{ij} + \sum_k n^k q^k_{ij})
\end{array}
\label{eqn:steadystate}
\end{equation}

\noindent where A$_{ij}$ are the spontaneous decay transition probabilities, B$_{ij}$ are the Einstein coefficients, u$_{ij}$ is the spectral energy density of the radiation field, $\Gamma_{ij}$ is the indirect excitation rate due to fluorescence and is defined by \cite{silva01}.
The quantity $n^k$ is the volume density of the collision partner $k$, where $k$ = (H$^0$, n$_e$, n$_p$) and q$^k_{ij}$ = $< \sigma v>$, is the collision rate coefficient.  The reverse rates are calculated using the assumption of detailed balance.  All coefficients were taken to be the same as those used in POPRATIO, with the exception of collisions with neutral hydrogen, for which we used the more recent rate coefficients calculated using the analytical formula by \cite{Abrahamsson07} extended to the temperature range of T = 10,000 K.   For the sake of simplicity, we did not consider excitation by collisions with either molecular hydrogen (\htwo )  or helium (\he ).  In the case of the former, the paucity of \htwo\ found in \dlas , at typical fractions of less than $\approx 10^{-5}$ renders the effect of collisions with \htwo\ so small as to be negligible.  
However, even in \dlas\ in which the \htwo\ fraction is relatively large (for \dla\ 1331$+$17, the molecular fraction was determined by  \cite{cui05} to be f = 0.056 or 5.6$\%\pm0.7\%$), \htwo\ does not have a large effect on the collisional excitation of \ci .  \footnote{I.e. the rate of excitation due to collisions with neutral hydrogen at T = 100 K is n$^{\hi}$ q$_{01}^{\hi}$ = 2.976 $\times$10$^{-10}$ n(\hi ) s$^{-1}$, where we use the conventional notation of representing \ci , \cistr , \cistrstr\ by the indices 0, 1, 2 respectively.    If we take the fraction of \htwo\ to be $5.6\%$ of \hi , the rate for collisions with \htwo\ is n$^{H2}$ q$_{01}^{H2}$ = 6.7 $\times$10$^{-11}$ (0.056) n(\hi ) = 3.75 $\times$10$^{-12}$ n(\hi ) s$^{-1}$, which is roughly two orders of magnitude smaller than q$_{01}^{\hi}$.}    
In the latter case of \he , collision rates are significantly lower than those of other partners.  Additionally, the density of \he\ compared with that of \hi\ is typically n(\he ) = 0.0975 n(\hi ) \citep{anders89}, making collisions with \he\ relatively unimportant. 

Direct excitation by the CMB is calculated assuming the standard cosmology and a CMB temperature of $T = T_{0} (1+ z)$ where $T_{0}$ = 2.725 K \citep{mather99}.  At high redshift, the CMB radiation generally has the strongest effect on the \ci\ fine structure level populations because of the small temperature difference between the ground and first fine structure states.  However, depending on the physical circumstances, other mechanisms such as UV pumping or collisions can dominate.

We included UV pumping due to a radiation field consisting of two components that we will call \jnubkd\ , and \jnuloc\ , and let \jnutot\ = \jnuloc\ $+$ \jnubkd .  \jnubkd\  is the background radiation due to the integrated contribution from high $z$ galaxies and quasars, known as the Haardt-Madau background (\cite{haardt96}; and more recently using CUBA \footnote{CUBA (Haardt \& Madau, 2003) is available at: http://pitto.mib.infn.it/$\sim$haardt/cosmology.html}).  In all cases, the minimum value of the total radiation field is determined by \jnubkd\ .   In each case the value of \jnubkd\  is calculated based on the redshift of the \dla , and these values are summarized in Table~\ref{tab:ciistarsolutions}.  For an explanation of these values, see Figure 1 of \cite{wolfe04}.  Because each \ci\ -bearing \dla\ also contains strong \ciistr\ $\lambda$1335.7 absorption, we used the \ciistr\ technique (see WPG03 and \cite{wolfe04}) to estimate the local radiation field due to star formation, \jnulocciistr\ , and included this contribution in the UV pumping.  The \jnulocciistr\ is calculated at $\lambda$ = 1500\AA , or 8.27 eV, and in \S~\ref{sec:direct} we explain the estimated error on this value, $\sim$$\pm$50\% .  Note that in general, \jnulocciistr\ = \jnuloc . We introduce the notation \jnulocciistr\ to specify how the local radiation field is measured, i.e. in this case it is the local star formation rate per unit area measured via the \ciistr\ technique.  

\begin{figure}
\plotone{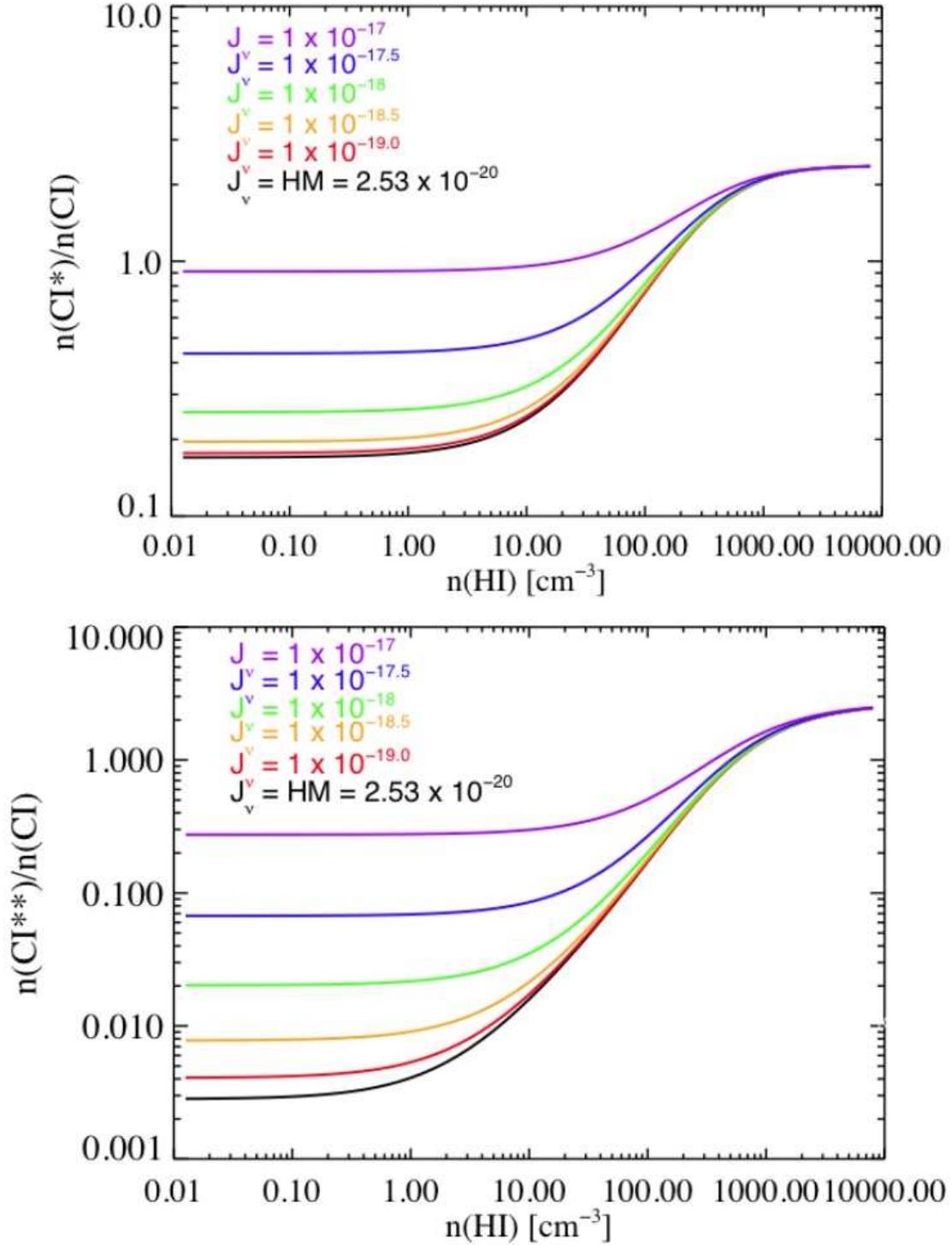}
\caption{Excitation of the fine structure level \cistr\  (top) and \cistrstr\ (bottom) caused by increasing the strength of the radiation field.  
We plot the ratio $\frac{n(C I*)}{n(C I) }$ and $\frac{n(C I**)}{n(C I) }$ and have included spontaneous radiative decay, excitation by the CMB at z$_{abs}$=2, and collisions with neutral hydrogen at a temperature of T=100K.  HM is the value of the Haardt-Madau background at $z=2$ and is therefore a minimum total radiation field.  Note that the radiation field must be $\sim$$\geq$ 1 $\times$ 10$^{-18.5}$ \junit\  in order to cause significant effects to the level populations of the \ci\ fine structure states at low densities (i.e. when collisions are not the dominant mechanism).
}
\label{fig:rad_fields}
\end{figure}

To quantitatively determine the effects of the radiation field on the \ci\ fine structure level populations, we plot the $\frac{n(\mcistr )}{n(\ci )} $  and $\frac{n(\mcistrstr )}{n(\ci )} $ versus neutral hydrogen density for excitation by a wide range of radiation fields in Figure~\ref{fig:rad_fields}.  
In this example we have considered collisions with neutral hydrogen at T = 100 K, spontaneous radiative decay, and the excitation by the CMB at z = 2, in addition to a radiation field of varying strengths as denoted in Figure~\ref{fig:rad_fields}.  At low density, excitation by the CMB is dominant, unless the input UV radiation field is strong, in which case, the UV dominates.  At higher densities, i.e. n(\hi ) $>$ 10 cm$^{-3}$, collisional excitation becomes important and finally at n(\hi ) $\gtrsim$1000 cm$^{-3}$ the levels thermalize.  In other words, at low densities the CMB sets the floor of $\frac{n(\mcistr )}{n(\ci )} $.  Only when the radiation field exceeds a total strength of approximately \jnutot\ $\gtrsim$ 10$^{-18.5}$ \junit , does $\frac{n(\mcistr )}{n(\ci )} $ begin to significantly exceed that caused by the CMB alone.   Since \jnubkd\ rarely exceeds 10$^{-19}$ \junit , the Haardt-Madau background alone at a redshift of z $\sim$ 2, has essentially no effect on the \ci\ fine structure excitation.  
 
\subsection{Steady State Solution}

The steady state densities in each of the \ci\ fine structure states are found by solving three homogeneous equations with three unknowns.  To solve the three homogeneous equations, we solve for the ratio of each excited state relative to the ground state.  We denote all terms involving CMB excitation, UV pumping and collisions by the shorthand, R$_{ij}$ = B$_{ij}$ u$_{ij}$ + $\Gamma_{ij}$ + $\sum_k n^k q^k_{ij}$ which is summed over the k different collision partners, where all terms were defined in the previous subsection.  Reverse reaction rates are calculated through the principle of detailed balance.  Following ~\cite{jenkins01}(see their equations 10 and 11), we find:

\begin{equation}
\begin{array}{lr}
\frac{n(\mcistr )}{n(\ci )} = \\
\frac{(R_{0,1})(A_{2,1} + A_{2,0} + R_{2,1} + R_{2,0}) + (R_{0,2})(A_{2,1} + R_{2,1})}{(R_{1,2})(A_{2,0} + R_{2,0}) + (A_{1,0} + R_{1,0})(A_{2,1} + A_{2,0} + R_{2,1} + R_{2,0})}
\end{array}
\label{eqn:cistrratio}
\end{equation}

and

\begin{equation}
\begin{array}{lr}
\frac{n(\mcistrstr )}{n(\ci )} = \\
\frac{(R_{0,2})(A_{2,0} + R_{1,0} + R_{1,2})+(R_{0,1})(R_{1,2})}{(R_{1,2})(A_{2,0} + R_{2,0}) + (A_{1,0} + R_{1,0})(A_{2,1} + A_{2,0} + R_{2,1} + R_{2,0})}
\end{array}
\label{eqn:cistrstrratio}
\end{equation}

\noindent where the states \ci , \cistr , \cistrstr\ are denoted by the indices 0, 1, 2 respectively.  The resulting theoretical solutions are functions of the density of neutral hydrogen, n(\hi ), and the temperature. 
Following ~\cite{jenkins79} we define, 

\begin{equation}
f1 \equiv \frac{n(\mcistr )}{n\mcitot\ }  =  \frac{\frac{n(\mcistr )}{n(\ci )}}{1.0 + \frac{n(\mcistr )}{n(\ci )} +  \frac{n(\mcistrstr )}{n(\ci )}} 
\label{eqn:...}
\end{equation}

and

\begin{equation}
 f2 \equiv  \frac{n(\mcistrstr )}{n\mcitot\ }  =  \frac{\frac{n(\mcistrstr )}{n(\ci )}}{1.0 + \frac{n(\mcistr )}{n(\ci )} +  \frac{n(\mcistrstr )}{n(\ci )}} 
 \label{eqn:...}
 \end{equation}
 
\noindent where $n\mcitot\  = n(\ci ) + n(\mcistr ) +  n(\mcistrstr\ $).  We give the values of (f1, f2) for each component of each \dla\ in Table~\ref{tab:cidata}.  

\begin{figure}
\epsscale{0.8}
\plotone{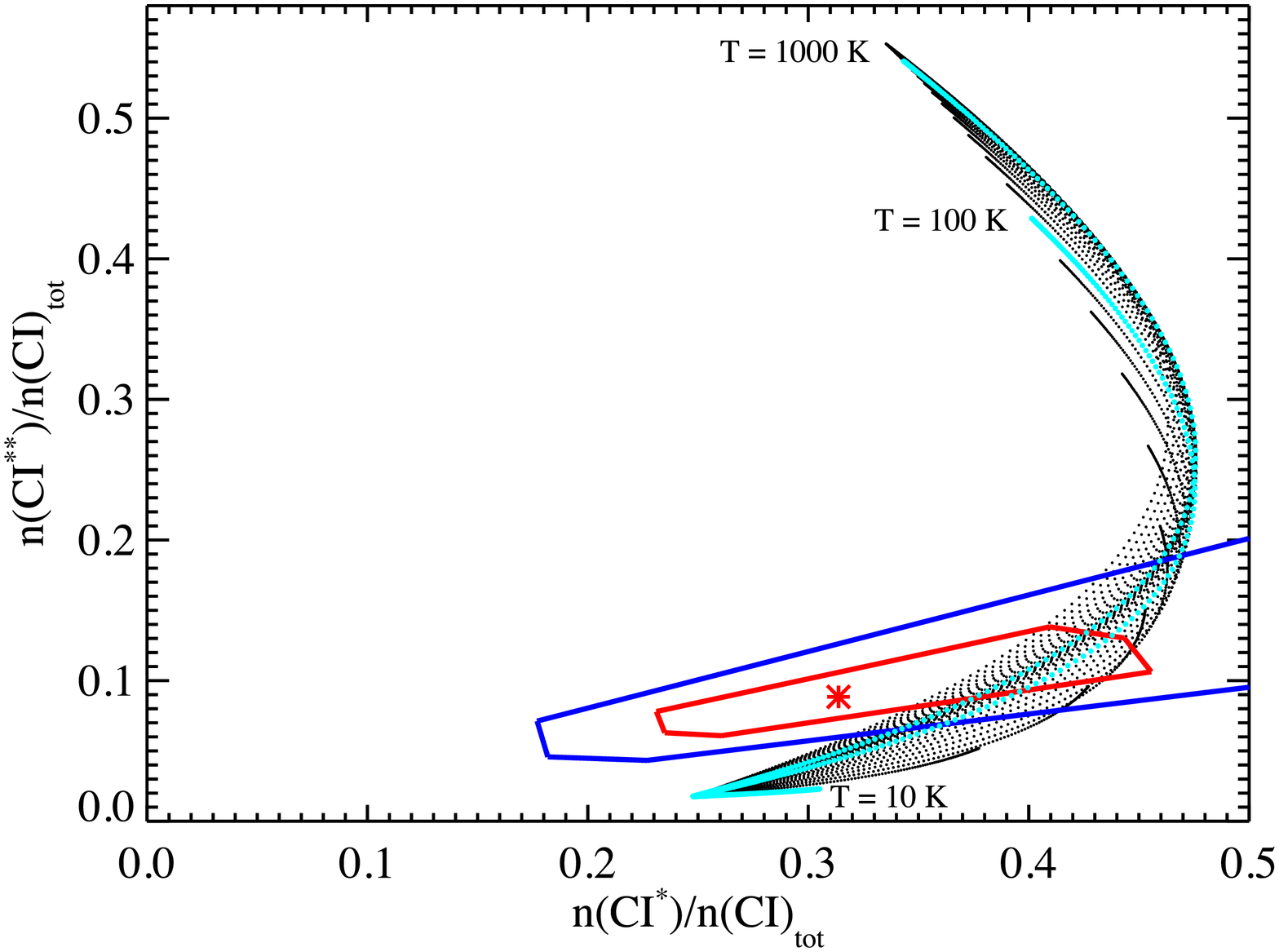}
\caption{f1 verus f2 for component 3 of \dla\ 0812$+$32, where f1 = n(\cistr )/n(\ci )$_{tot}$, f2 = n(\cistrstr )/n(\ci )$_{tot}$, and n(\ci )$_{tot}$ = n(\ci ) + n(\cistr ) + n(\cistrstr ).  The data point is marked by a red asterisk and the 1$\sigma$ error polygon is marked in red, while the 2 $\sigma$ error polygon is blue.  Theoretical tracks are indicated by black points and run from T = 10 - 10,000 K and n(\hi ) = 10$^{-3.5} - 10^{4.1}$ cm$^{-3}$.  For clarity we have highlighted the tracks corresponding to T = 10, 100 and 1000 K in cyan.  
}
\label{fig:q1331_f1vsf2}
\label{fig:J0812_f1vsf2}
\end{figure}

\begin{figure}
\plotone{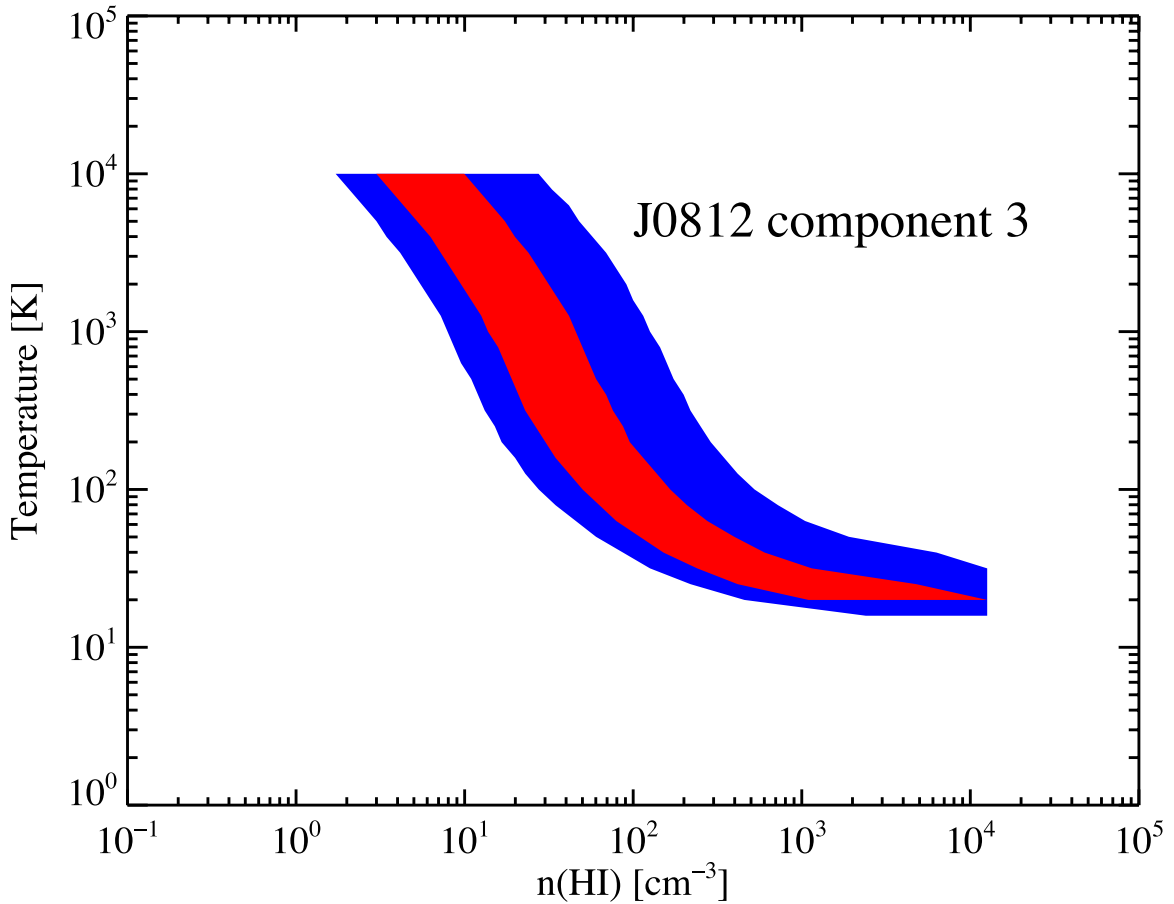}
\caption{n(\hi ) versus temperature for the allowed solutions of \dla\ 0812$+$32 component 3 from the \ion{C}{1} analysis.  1$\sigma$ results are in red while 2$\sigma$ is in blue.  
}
\label{fig:J0812_nvst} 
\end{figure}

In Figure~\ref{fig:J0812_f1vsf2} we plot the theoretical solutions in the (f1, f2) plane for the example case of component 3 of \dla\ 0812$+$32.  In this case, \jnutotciistr\ as derived from the \ciistr\ technique, is \jnutotciistr\ = 7.4 $\times$10$^{-19}$ \junit.  The solutions follow tracks, one for each temperature from T = 10 K $-$ 10,000 K increasing in steps of 0.1 dex.  
Along each track, n(\hi ) increases from $10^{-3.5}$ cm$^{-3}$ to $10^{4.1}$ cm$^{-3}$ in steps of 0.02 dex, each density being represented by an individual point.  We chose the ranges of temperatures and densities to ensure that we had broad coverage of the entire plane of possible solutions.  While the \ci\ data may allow higher T values, these would imply a primarily ionized gas which is highly unlikely for this material.  The data point at (f1, f2) $\approx$ (0.31, 0.09) is determined by our curve of growth analysis (in the case of the resonance line \ci ) and VPFIT fits to the data for component 3 (see Table~\ref{tab:cidata}).  1$\sigma$ (2$\sigma$) errors are determined by calculating the values of f1 and f2 using the upper and lower allowed values as determined by the 1$\sigma$ (2$\sigma$) error bars for each column density measurement.  
The resulting red (blue) polygon defines a region of space in the f1 versus f2 plane that contains the acceptable 1$\sigma$(2$\sigma$) solutions.  
  In Figure~\ref{fig:J0812_nvst} we plot the predicted theoretical solutions that fall within the 1$\sigma$ (in red) and 2$\sigma$ (in blue) error polygons of the data, on a graph of density versus temperature.  These are the allowed density and temperature combinations derived from the \ci\ data under the assumption of a single phase absorber.

\subsection{Ionization Equilibrium}

To further constrain the \ci\ solutions we introduce the assumption of ionization equilibrium and utilize the measured $\frac{\ctwo }{\ci } $ ratio, which we assume is equal to $\frac{n(\ctwo )}{n(\ci )} $ = $\frac{N(\ctwo )}{N(\ci )} $.  Ionization equilibrium is a reasonable assumption because the densities in these clouds are generally relatively high, thereby ensuring that the recombination times are shorter than the typical dynamical timescales.   In basic form, ionization equilibrium can be written: n$_e$ n(X$^{+}$) $\alpha$ = n(X) $\Gamma$, where $\alpha$ is the case A recombination coefficient of element X$^{+}$ to X and $\Gamma$ represents the ionization rate.  From this equation we get, $\frac{n(\ctwo )}{n(\ci )} $ = $\frac{\Gamma}{n_e \alpha}$, i.e. the ratio $\frac{\ctwo }{\ci } $ is a function of $\Gamma$ (where $\Gamma$ is proportional to the radiation intensity), n$_e$, the electron density, and $\alpha$, which is a function of temperature.  Therefore, for a given radiation field and temperature, we determine $\frac{\ctwo }{\ci }$ for a range of possible electron densities, n$_e$  (Details of the ionization equilibrium are given in WPG03, and we discuss the sensitivity to the radiation field in section~\ref{sec:ion_constrain}).  We then use the measured $\frac{\ctwo }{\ci } $ of each \dla\ to constrain our allowed solutions.  While we measure N(\ci ) directly, we must employ the conventional method of measuring N(\cii ) by proxy using N(Si II), because the
available \ctwo\ $\lambda$1334 transition is saturated in all the \dlas\ considered here.  We generally measure the Si II or S II and Fe II using the standard AODM, which is well suited for cases such as these in which we have more than one transition of an ion.  As in \cite{wolfe04} we let [C/H] = [Si/H] $+$ [Fe/Si]$_{int}$ where the intrinsic (nucleosynthetic) ratio [Fe/Si]$_{int}$ = $-$0.2 for a minimal depletion model or [Fe/Si]$_{int}$ = 0.0 for a maximal depletion model.  We follow ~\cite{murphy04} and adopt the minimal depletion model in this work, and analyze the implications of the minimal versus maximal depletion model in section~\ref{sec:model}.  While \cite{wolfe04} estimate the error in the measurement of $\frac{\ctwo }{\ci }$ to be 0.1 dex, we use the more conservative estimate of 0.2 dex.  We also note that, assuming there is no hidden saturation of metals, this is the maximum N(\cii ) that could be associated with the \ci\ gas, and we discuss the implications of this assumption further in section~\ref{sec:ciiovci}.

\begin{figure}
\epsscale{0.8}
\plotone{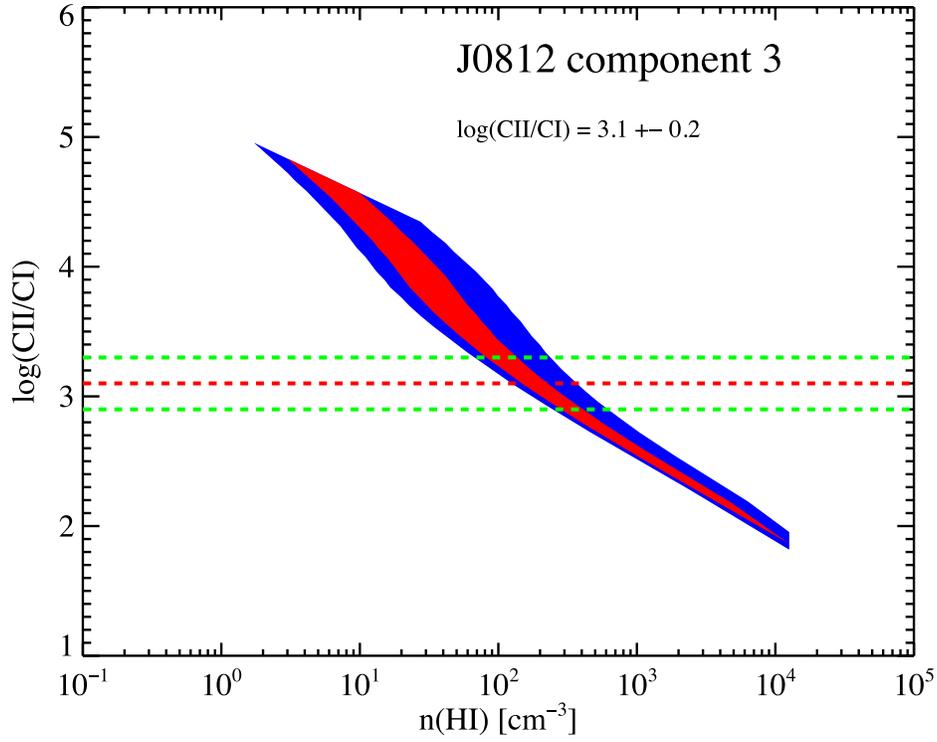}
\caption{n(\hi ) versus  $\frac{n(\ctwo )}{n(\ci )} $ for the allowed solutions of \dla\ 0812$+$32 component 3.  The measured $\frac{n(\ctwo )}{n(\ci )} $ is indicated by the red dashed line with green dashed lines indicating the errors of $\pm$0.2 dex.  1$\sigma$ results are in red while 2$\sigma$ is in blue.    
}
\label{fig:J0812_c2ovc1} 
\end{figure}

\begin{figure}
\epsscale{0.8}
\plotone{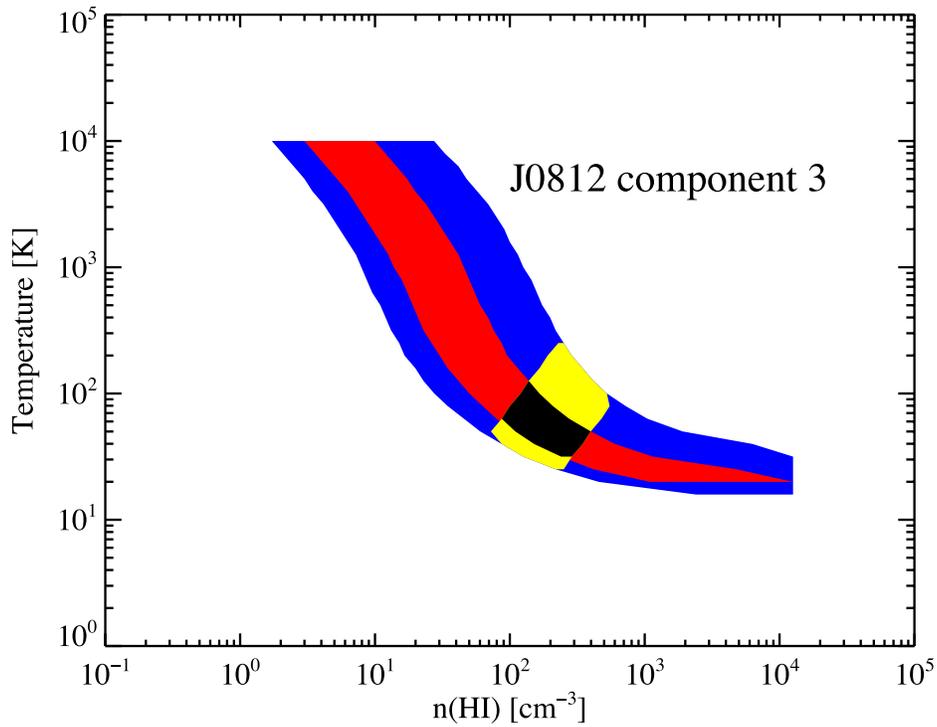}
\caption{n(\hi ) versus temperature for the allowed solutions of \dla\ 0812$+$32 component 3.  The final solutions, as constrained by $\frac{\ctwo }{\ci }$, are shown in black (1$\sigma$) and yellow (2$\sigma$).
}
\label{fig:J0812_nhivst_good}
\end{figure}

To demonstrate the constraints imposed by ionization equilibrium, in Figure~\ref{fig:J0812_c2ovc1} we plot the \ci\ solutions in terms of n(\hi ) versus log($\frac{\ctwo }{\ci }$) for the example case of component 3 in \dla\ 0812$+$32.  The measured log($\frac{\ctwo }{\ci }$) = 3.10 is indicated by the red dashed line with a range of $\pm$0.2 dex indicated by green dashed lines.  It is obvious that the region of allowed \ci\ fine structure solutions is further constrained by invoking ionization equilibrium.  For clarification, in Figure ~\ref{fig:J0812_nhivst_good} we re-plot the density versus temperature diagram, however now we denote the final solutions, those constrained by the $\frac{\ctwo }{\ci }$ ratio, in black (1$\sigma$) and yellow (2$\sigma$).
As a result, invoking ionization equilibrium results in even tighter constraints on the densities and temperatures of the \ci\ -bearing cloud without making any assumptions about the gas temperature.  In this example case we constrain the density to be 72 $\lesssim$ n(\hi ) $\lesssim$ 549 cm$^{-3}$ while the temperature is constrained to be 25 K $\lesssim$ T $\lesssim$ 251 K, and the pressure 3.54 $\lesssim$ log (P/k) $\lesssim$ 4.80 cm$^{-3}$ K.  $\chi $$^2$ minimization finds the best-fitting solution:  n(\hi ) = 100 cm$^{-3}$, T = 79 K, and log (P/k) = 3.90 cm$^{-3}$ K.

\subsubsection{Using Ionization Equilibrium to Constrain the Radiation Field}\label{sec:ion_constrain}

Until now we have assumed the UV radiation field due to star formation, \jnuloc\ , is determined by the \ciistr\ technique.  
We will now relax this constraint and allow the input local radiation field to vary, repeat the above analysis for each case, and examine the results as a function of radiation field.  We show that for some \ci\ -bearing clouds we can place upper and lower limits on the allowed radiation field utilizing only the \ci\ fine structure data and the assumption of ionization equilibrium.  

We use a grid of \jnuloc\  values from \jnuloc\  $\approx$ 10$^{-21}$ $-$ $10^{-16}$ \junit .  To differentiate these radiation fields from those predicted by the \ciistr\ technique, we will call them \jnulocci .  For each \jnulocci\  we first add the \jnubkd\  to obtain a \jnutotci\ , and then rerun the above analysis, calculating the f1 versus f2 tracks for each \jnutotci , followed by the ionization equilibrium analysis.   
  
\begin{figure}
\epsscale{1.1}
\plotone{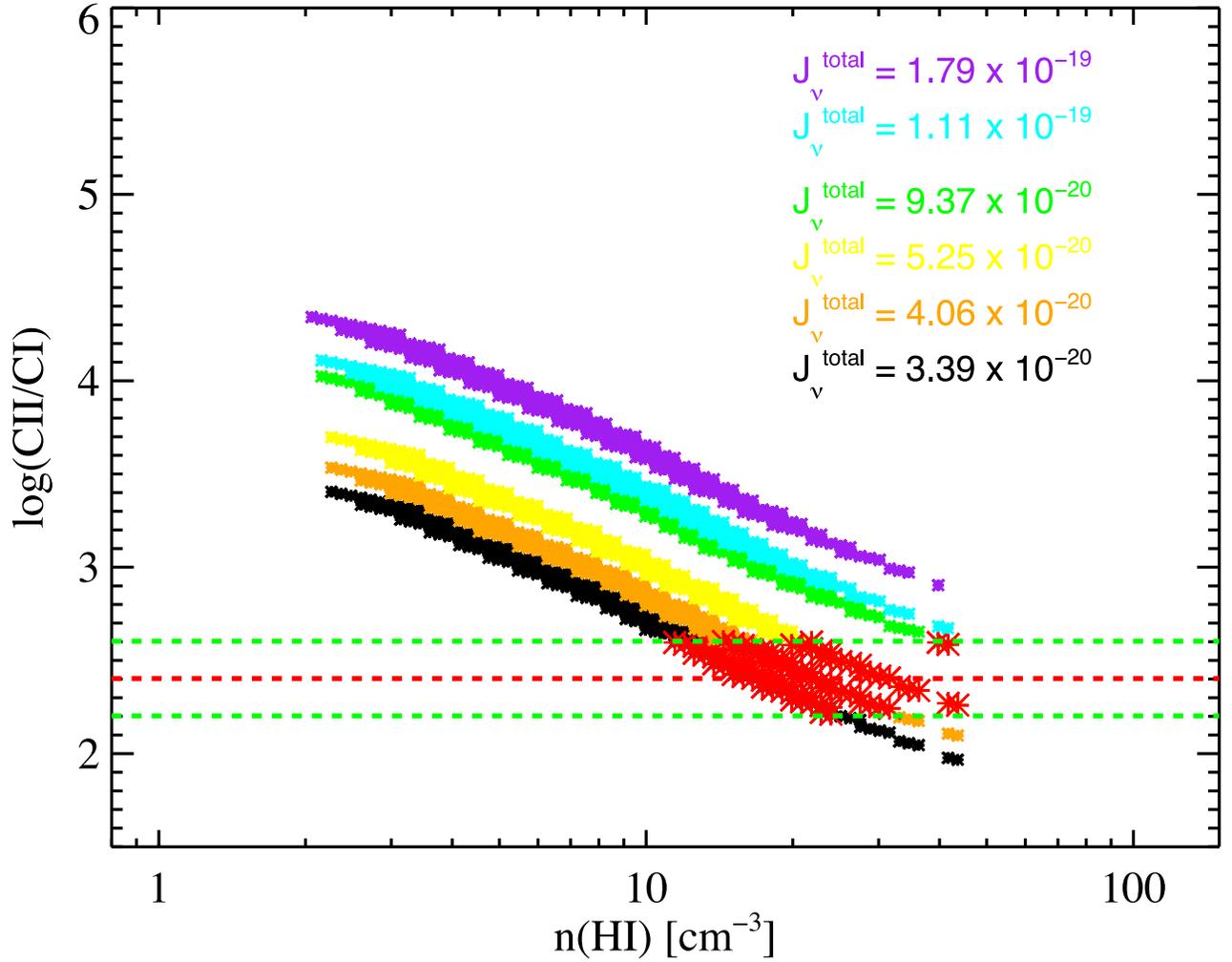}
\caption{Example of a case, shown here for DLA 1331$+$17, component 1, where it is possible to constrain the radiation field using only the \ci\ fine structure states and the assumption of ionization equilibrium.  Log($\frac{\ctwo\ }{\ci\ }$) versus n(\hi ) for a selection of \jnu\ values.  Solutions falling within the allowed range of log($\frac{\ctwo\ }{\ci\ }$) are marked in red.  It is seen Ïthat for \jnutot\ $\geq$$\sim$1$\times$10$^{-19}$ \junit , there are no acceptable solutions.  Note: This is for the 1$\sigma$ \ci\ solutions and the value of the Haardt-Madau background for this object is 2.53 $\times$10$^{-20}$ \junit .  It is also apparent that the range of allowed densities is constrained to be 10 $\leq$ n $\leq$ 50 cm$^{-3}$.
}
\label{fig:select_jnus}
\end{figure}

 In Figure~\ref{fig:select_jnus} we plot the 1$\sigma$ \ci\ results for the example case of \dla\ 1331$+$17 on a graph of $\frac{\ctwo }{\ci }$ versus n(\hi ) where for clarity we show only a sub-sample of the entire \jnutotci\  grid results.  
Solutions corresponding to each \jnutotci\  are represented by different colors.  Again, we use the $\frac{\ctwo }{\ci }$ data to constrain the allowed solutions, and it is apparent that for \jnutotci\  $\gtrsim$ 1 $\times$10$^{-19}$\junit , there are no acceptable solutions.  This limit on \jnutotci\  is actually quite strict considering that the Haardt-Madau background is only $\sim$1 order of magnitude lower than this value.  
 
 This technique places an upper limit on the local radiation field, and hence the star formation rate, using only \ci\ fine structure absorption and the assumption of ionization equilibrium, without invoking any assumptions about star formation from the \ciistr\ technique.  We note in this example, of \dla\ 1331$+$17, component 3, that the strength of the radiation field as determined by the \ciistr\ technique (\jnutotciistr\ = 3.3 $\times$10$^{-19}$ \junit\ ) is just slightly higher than the 1$\sigma$ upper limit derived independently from the \ci\ data.  However, in the case of DLA 1331$+$17, the \ciistr\ transition is likely blended with a Lyman$-\alpha$ forest line, forcing us to estimate the true N(\ciistr ) by assuming it traces the velocity structure of other low ions and fitting it together with the forest line as explained in section ~\ref{sec:1331}.

A byproduct of this analysis places limits on the density, temperature and pressure of the cloud.  In this example case of \dla\ 1331$+$17, at the 1$\sigma$ level the density is limited to the range, 
11 $\lesssim$ n(\hi ) $\lesssim$ 44 cm$^{-3}$ while the temperature is constrained to be 79 $\lesssim$ T $\lesssim$ 794 K, and the pressure 3.50 $\lesssim$ log (P/k) $\lesssim$ 4.04 cm$^{-3}$ K.  Again, we stress that these limits are derived independently of the results of the \ciistr\ technique, using \emph{only} the \ci\ fine structure data and the assumption of ionization equilibrium for a range of possible radiation fields.  

\section{Results}~\label{sec:results}

A summary of the details of each \dla\ are presented in the following tables:  Table~\ref{tab:cidata}: a summary of the \ci\ data, Table~\ref{tab:otherdata}: a summary of general \dla\ information, Table~\ref{tab:ciistarsolutions}: the resultant radiation fields derived from the \ciistr\ technique, and in Table~\ref{tab:cisolutions}: the final \ci\ fine structure solutions giving allowed ranges of densities, temperatures and pressures.  Finally, in Table~\ref{tab:cisolutions_fullgrid}, we give the results of lifting the \ciistr\ constraint on the local radiation field.  

We now briefly describe the results for each \ci\ -bearing \dla .  The casual reader may wish to skip directly to the more general discussion of the results in \S~\ref{sec:discussion}.


\subsection{FJ0812$+$32, z$_{abs}$ =2.62633}

This \dla\ contains three distinct \ci\ velocity components whose \ci\ fine structure levels we will analyse individually.  In measuring the metallicity, N(C II) and N(\ciistr ) we will take two approaches; The first, discussed here, and the second, the individual component analysis, discussed in section ~\ref{sec:0812indiv}.
Heavy blending of line profiles and the uncertainty/degeneracy of line profile fitting techniques have led to the customary use of the AODM technique to measure the amount of metals over the entire, blended \dla\ profile.  Additionally, this avoids any question about the distribution of the N(\hi ), which by definition is damped and therefore kinematically unknowable.  Because we are measuring the metals over the entire (blended) \dla , we refer to these measurements as 'global', and we apply them to each \ci\ component, which we consider reasonable given that the low-ions such as S II and Si II track the velocity components of \ci , see Figure ~\ref{fig:J0812_spec_lowions}.

We measure log N(\ciistr ) = 14.30$\pm$0.01 cm$^{-2}$ which along with the \ciistr\ technique determines that the radiation due to stars \jnulocciistr\  = 7.17$\times$10$^{-19}$\junit .  Adding in the contribution from the background, \jnubkd\  = 2.45 $\times$ 10$^{-20}$\junit , gives \jnutotciistr\  $\sim$7.4$\times$10$^{-19}$\junit .  The excitation rates due to the local star formation alone are $\Gamma$$_{01}$  = 3.17$\times$10$^{-9}$s$^{-1}$, $\Gamma$$_{02}$ = 2.40 $\times$10$^{-9}$s$^{-1}$, and $\Gamma$$_{12}$ = 3.12 $\times$10$^{-9}$s$^{-1}$.  These can be compared with those of the Haardt-Madau background at the redshift of this \dla\ that produces excitation rates $\Gamma$$_{01}$ = 1.09 $\times$10$^{-10}$s$^{-1}$, $\Gamma$$_{02}$ = 8.23$\times$10$^{-11}$s$^{-1}$, $\Gamma$$_{12}$ = 1.07$\times$10$^{-10}$s$^{-1}$.  We used the total radiation field and the assumption of ionization equilibrium to determine solutions for the three \ci\ components that are labeled 0812$+$32$_{global}$ in Table~\ref{tab:cisolutions}.  We are unable to perform an analysis of component 2 given that the upper limit of N(\cistrstr ) $\leq$ 12.39 cm$^{-2}$ is relatively large, resulting in a (f1, f2) combination with no limitation on n and T.  

We can use the constraints on the volume density to estimate the physical size of the cloud.  Using the neutral hydrogen column density logN(\hi ) = 21.35 cm$^{-2}$ and, for example, the best-fit volume density for component 3, n(\hi )$\approx$ 100 cm$^{-3}$, we can estimate the size of the \ci\ -bearing cloud to be $\ell $ = N(\hi )/n(\hi ) $\approx$ 2.24 $\times 10^{19}$ cm, or $\approx$ 7 pc.  This estimation is technically an upper limit to the size of the cloud that assumes that all of the \hi\ is associated with the \ci\ component.  

When we relax the constraint of \jnutotciistr\ as derived from N(\ciistr ), as discussed in the previous section, at the 2 $\sigma$ level we constrain the density to n $\geq$ 0.1 cm$^{-3}$  and n $\geq$ 7 cm$^{-3}$ for components 1 and 3 respectively.  In both cases we place a not-so-strict upper limit on the allowed radiation field of \jnutot\ $\leq$ 773 $\times$10$^{-19}$ \ \junit .  

\subsubsection{\dla\ 0812$+$32: Individual Component Analysis}~\label{sec:0812indiv}

While the initial modeling was completed using the radiation field and metallicity as measured over the entire profile, it is clear upon careful inspection of the spectrum (see Figure ~\ref{fig:J0812_spec_lowions} and Figure 7 of ~\cite{jorg09}), that the depletion, and hence, dust-to-gas ratio, is not constant over the three components.  This motivates our attempt to analyze each velocity component individually.  Specifically, component 3, the narrow component, contains obvious Zn II, \ciistr\ and \ci , while it exhibits no evidence for Cr II.  A natural explanation is that Cr II is heavily depleted onto dust grains in component 3. However, as discussed in ~\cite{jorg09}, blending precludes a straightforward measurement of the equivalent width, and profile fitting using VPFIT, produces a Zn II column density in component 3 that is unrealistically high and ruled out by an upper limit on N(O I) assuming solar relative abundances.  We explain this by assuming the presence of another weak, broad component (for details see ~\cite{jorg09}), and model the log N(Zn II) in component 3.   

Because Zn II and Cr II are not saturated and appear to trace the velocity structure of \ci\ quite well, we use these ions to determine the metal distribution, the dust-to-gas ratio, and the N(C II) in each component separately.  To do this we first tie the redshifts and b values of each ion together and then use VPFIT to determine the column densities in each component.  It is expected that Zn II and Cr II be tied together because they are metal-line transitions arising in singly ionized species with similar ionization potentials, and thus are likely to show the same physical/velocity structure.  However, they will not necessarily exactly trace the \ci , given that \ci\ is affected by the incident \jnutot\ and n$_e$, and if there is a gradient in \jnutot\ or n$_e$, this could cause a difference between the structure of \ci\ and the other low ions.  We give the results of this fitting in Table~\ref{tab:J0812_dust2gas_comps}.  In light of the work of ~\cite{jorg09} we include an additional broad weak component, in order to achieve a realistic N(Zn II) in component 3.  

Due to the nature of damped lines, we cannot use Lyman-$\alpha$ to determine the distribution of the neutral hydrogen among these components.  Therefore, we assume that the N(\hi ) traces the low-ion metals, in this case Zn II, and that the neutral hydrogen is distributed proportionally to the metals.  This results in each component having the same metallicity, [Zn/H] = $-$0.58.  We note, that in theory, this metallicity should be the same as that determined for the global case, [Zn/H] = $-$0.81.  However, the individual component analysis reveals the presence of the narrow component 3, for which we have estimated the N(Zn II)  based upon an upper limit on N(O I).  While the exact metallicity of this component remains a mystery without higher resolution data, it is apparent that component 3 lacks significant Cr II, indicating a high level of depletion, and consequently a higher dust-to-gas ratio than components 1 and 2.  Following WPG03 (their equation 7) we define the dust-to-gas ratio relative to the Milky Way as follows,

\begin{equation}
 \kappa = 10^{[Zn/H]_{int}} (10^{[Cr/Zn]_{int}} - 10^{[Cr/Zn]_{gas}}), 
 \end{equation}
 
 \noindent where "gas" is the abundance ratio in the gas phase and "int" is the intrinsic nucleosynthetic abundance.  We calculate the dust-to-gas ratio in component 3 to be $\sim$17\% of the Milky Way (log$\kappa$ = $-$0.78).  For comparison, ~\cite{pro2003} measure $\kappa$ $\sim$ 6\% 
  (log$\kappa$=$-$1.24) over the entire profile of this \dla , and the typical dust-to-gas ratio in \dlas\ is $\sim$1/30 solar. %
We summarize information about each component in Table~\ref{tab:J0812_dust2gas_comps}.

As we might expect, given that the column densities of all low ions with the exception of Cr II, but including \ciistr ,  are much higher in component 3 (due to the fixed small Doppler parameter), the \jnulocciistr\ of component 3 deduced from the \ciistr\ technique is more than 1 order of magnitude larger than that determined for components 1 and 2. 
The results of the \ci\ models for each component are presented in 
Table~\ref{tab:cisolutions} where they are denoted by the subscript 'div'.  
Note that the radiation field for component 3 predicted by the \ciistr\ technique is quite large, at 260 $\times$ 10$^{-19}$ \junit .  The solutions require low temperatures (T $\leq$ 32 K) and densities in excess of 10$^3$ cm$^{-3}$, which imply an upper limit on the cloud size of $\ell $ $\leq$ 0.1 pc.  This small size may be in conflict with the evidence against partial covering provided by the \htwo , see section ~\ref{sec:0812h2}.  It is possible that we have over-estimated the amount of N(\ciistr ) associated with the narrow component 3, and that more of the N(\ciistr ) is associated with the broader component 4, but we would require a higher resolution spectrum to confirm this scenario.  

Relaxing the constraint of \jnuloc\ determined by the \ciistr\ technique, gives
no limits on temperature, and a large range of allowed densities for component 1 ( n(\hi ) = 0.002 - 4166 cm$^{-3}$), with a lower limit of n(\hi ) $\geq$ 6 cm$^{-3}$ for component 3.  Constraints on the allowed radiation fields are consistent with that determined by the \ciistr\ technique.

\subsubsection{\dla\ 0812$+$32 Molecular Hydrogen}~\label{sec:0812h2}

\begin{figure}
\plotone{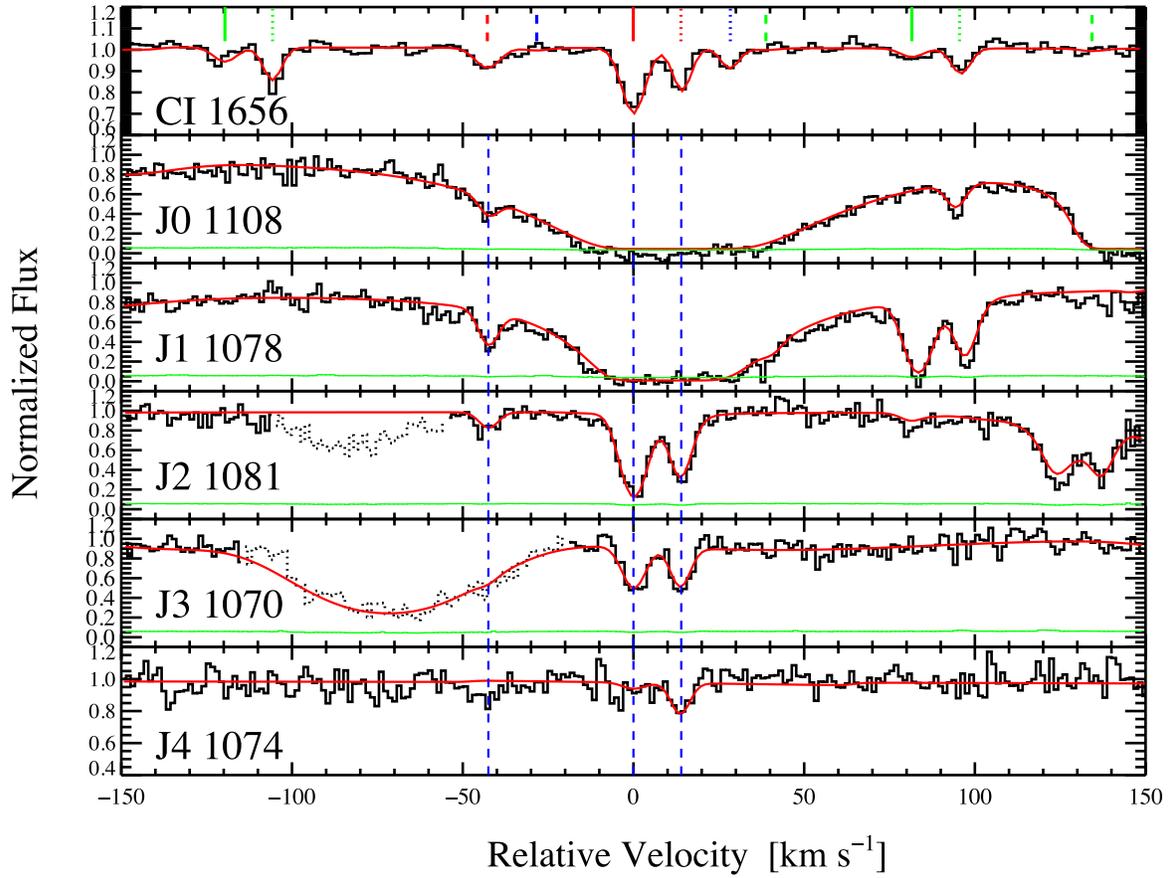}
\caption{Examples of the \htwo\ absorption in \dla\ 0812$+$32.  We show the relatively unblended sections here  (blends are denoted by dotted lines) over several rotational J levels.  Blue vertical dashed lines mark the three \htwo\ components, from left to right, components 1, 2, 3.  The close association with the three \ci\ components is clear.      
}
\label{fig:J0812_h2_vel}
\end{figure}

This \dla\ also shows evidence of relatively strong molecular hydrogen (\htwo ), see Figure ~\ref{fig:J0812_h2_vel}.  As shown in ~\cite{jorg09}, there is no evidence for partial coverage, as the \htwo\ lines are black at line center.  We used VPFIT to fit the \htwo , which required three components whose redshifts roughly agree with those of the three \ci\ components.  While the Doppler parameters do not agree with those of the \ci -- they are roughly similar to or smaller, whereas we would expect them to be larger by a factor of (6)$^\frac{1}{2}$ -- 
we will refer to these \htwo\ components as components 1, 2, and 3, and assume that they are co-spatial with the \ci\ components.  This is not the first report of such differences in the $b$-values between
\htwo\ and \ci\ components in DLAs 
\citep{noterdaeme07a,noterdaeme07b}.  Because of the close match in redshift space and the likelihood that \ci\ and \htwo\ are co-spatial, we fixed the well-determined Doppler parameter of component 3 to that of the \ci\  ($b^{CI}$ = 0.33 \kms ), scaled by the relative atomic weights as done in ~\cite{jorg09}\footnote{$b^{H_2 }$ = (6)$^{1/2}$ $b^{CI}$ = 0.81 \kms . }.  
We list the parameters of the \htwo\ fits in Table~\ref{tab:J0812}. 

It is interesting to note that like the \ci , 
the majority of the \htwo\ resides in component 3, the narrow velocity component.  The fraction of \htwo\  in component 3, f $\geq$ 0.06, or $\geq$ 6\%, with the upper limit representing the uncertainty in N(\hi ) distribution, is among the highest found in \dlas .  While we have tied the Doppler parameter to that of the \ci , we note that because of heavy saturation of the J = 0 and J = 1 lines in this component, log N is determined by the damping wings of the profile and is therefore insensitive to the choice of b.  We have verified this by artificially fixing the Doppler parameter to a range of values from b = 0.2 - 6 \kms as well as by raising the continuum by 5\%.  None of these tests change the resultant N(\htwo , J=0) and N(\htwo , J=1) by more than $\sim$0.03 dex.  While the J = 0 and J = 1 lines are completely insensitive to the Doppler parameter, the higher level J states are sensitive to the choice of Doppler parameter.  Therefore, we also report a model in which we have allowed the Doppler parameter of component 3 to be determined by the J = 3 state and we report the results, called model 2, in Table~\ref{tab:J0812}.   

\begin{figure}
\epsscale{0.6}
\plotone{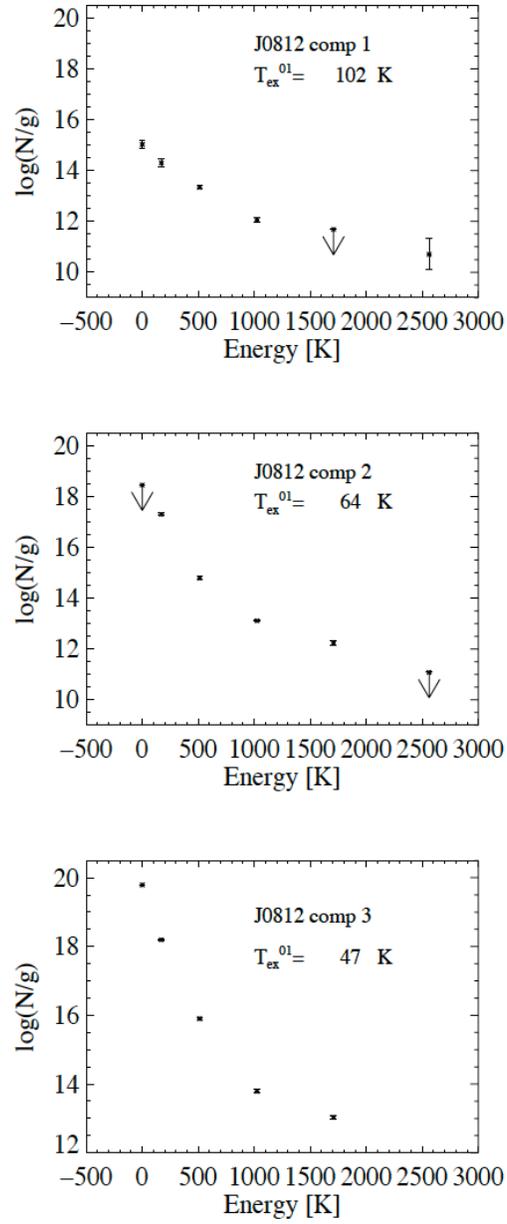}
\caption{\htwo\ excitation diagrams for the components of \dla\ 0812$+$32.    
}
\label{fig:J0812_h2}
\end{figure}

We can estimate the kinetic temperature of the clouds using the column densities of \htwo\ in the J=0 and J=1 rotational states and assuming the states are thermalized according to the Boltzmann distribution (see equation 8 in ~\cite{levshakov02}).  The excitation temperature, T$_{ex}$, is defined by, 

\begin{equation}
\frac{N(J)}{N(0)} = \frac{g(J)}{g(0)} \exp [ ^-{\frac{B_v J (J + 1)}{T_{ex}}} ]
\end{equation} 

\noindent where B$_v$ = 85.36 K for the vibrational ground state and g(J) is the degeneracy of level J.  In Figure~\ref{fig:J0812_h2} we show the standard \htwo\ excitation diagram, log(N/g) versus energy, for the J=0 to J=5 levels for each component.  As explained in Appendix 1 (on molecular hydrogen), the kinetic temperature of the gas is proportional to the negative inverse of the slope determined by the J=0 and J=1 levels, assuming that the levels are thermalized (a good assumption given the densities of these clouds), i.e. T$_{ex}$ equals the kinetic temperature.  For components 1, 2, and 3 we determine the following temperatures, T$_{ex}^{01}$ = 102 K,  T$_{ex}^{01}$ = 73 K, T$_{ex}^{01}$ = 47 K, respectively.  These are consistent with the temperatures derived from the \ci\ data, which leads us to believe there is good correspondence between the \ci\ and \htwo\ data 
and that the gas probed here is a CNM.  Also, the T$_{ex}^{01}$ = 47 K derived for the narrow component 3 is consistent with the upper limit on the thermal temperature of T$_{thermal}$ $\leq\ $ 78 K required by the curve of growth determined Doppler parameter. 

We can also use the \htwo\ data to determine the ambient/incident radiation (UV flux) field on the cloud, \jnutothtwo , using the J=0 and J=4 states (see Appendix 1 for details).  The level populations above J = 1 are unlikely to be thermalized since their larger Einstein A coefficients implies that these states are depopulated by spontaneous emission more rapidly than by collisional de-excitation; this rules out the condition of detailed balance required to establish thermal equilibrium.  Instead, these states are likely to be populated by UV pumping, which is what we shall assume.  Following ~\cite{hira05} we call this radiation field $J$$_{\nu}$$^{LW}$, as determined by absorption in the Lyman-Werner \htwo\ bands.  Since the \htwo\ measurement should reflect the total incident radiation field,  \jnuloc\ $+$ \jnubkd = $J$$_{\nu}$$^{LW}$ = \jnutothtwo .  
Results for the photoabsorption rate, $\beta$, and for the shielding terms, both self-shielding by \htwo\ and shielding due to dust, are given in Table~\ref{tab:J0812_dust2gas_comps}.  Note that $\beta _0$ is derived from the J = 4 population, while $\beta _1$ is derived from the J = 5 population.  Using the J = 4 population, we obtained the following radiation fields for components 1, 2, and 3 respectively: \jnutothtwo\ = $<$ 3.7 $\times$10$^{-20}$, 1.7 $\times$10$^{-20}$, 3.6 $\times$10$^{-20}$ \junit .  Note, that for component 1, N(\htwo, J = 4) is an upper limit, and hence, this is technically an upper limit on the radiation field. Also notice that these radiation fields are only slightly larger, or smaller in the case of component 2, than \jnubkd .  This places a rather strict limit on additional radiation from local star formation.  

Interestingly, the radiation field as calculated from the \htwo\ 
is rather independent of the choice of global versus individual component parameters.  In the above we gave the results of the individual component analysis, where we assumed that the N(\hi ) tracks the metals, in this case the N(Zn II), and used the individual component metallicity, N(\hi ) and dust-to-gas ratio to calculate the radiation field.  If we instead use the 'global' model, the N(\hi ) increases to the global value, here log N(\hi ) = 21.35 cm$^{-2}$, while $\kappa$ decreases in each component to the global value, here $\kappa$ = 0.07.  These two effects essentially cancel each other out in the calculation of shielding (see equation ~\ref{eqn:shield2} in Appendix A)
and we obtain very similar radiation fields in each case (i.e. for component 1, \jnutothtwo $_{global}$ $<$ 4.2 $\times$10$^{-20}$ \junit, compared with \jnutothtwo $_{indiv}$ $<$  3.7 $\times$10$^{-20}$ \junit  ).

We do not understand 
why the radiation field derived from the \htwo\ data is $\approx$ 20 times smaller than that derived by the \ciistr\ technique, \jnutot\ = 7.4 $\times$10$^{-19}$ \junit .  Additionally,  the radiation field predicted by the \htwo\ levels for component 2 is low enough that it is excluded by the \ci\ data and the Haardt-Madau background (however, we note that the difference is not large and it could be within the errors -- i.e. for component 2, if we consider the errors on the J = 0 and J = 4 levels we derive \jnutothtwo\ = 3.1 $\times$10$^{-20}$ \junit , consistent with the Haardt-Madau background, but still $\approx$ 20 times smaller than the \ciistr\ prediction).  This conflict could potentially be resolved if either 1) we are underestimating the population in the N(\htwo , J=4) level -- which could be caused by hidden saturation of narrow components as was demonstrated for the narrow component in \dla\ 1331$+$17 ~\citep{carswell09}, where the N(\htwo , J = 4) is likely underestimated 
because of the presence of the small Doppler component which contains the majority of the molecular gas (see further discussion in section~\ref{sec:1331} and ~\cite{carswell09}), or 2) the \ciistr\ prediction is overestimating the radiation field.  In the latter case, the radiation field is dependent upon the assumption of the equilibrium pressure existing at the geometric mean (see Section ~\ref{sec:model} for details).  If instead the equilibrium pressure is located at P$_{max}$ or P$_{min}$, the resultant radiation field can change by up to a factor of 10 (for example, see Table ~\ref{tab:compare}).  In the former case, we estimate that if the radiation field as predicted by the \ciistr\ technique is accurate and the \htwo\ feels this entire radiation field, we require the amount of \htwo\ in the J=4 level to increase by a factor of 20, to N(\htwo , J=4) $\sim$ 15.3 cm$^{-2}$.  This large amount of \htwo\ is ruled out by the data unless it exists in a very narrow component with b $\lesssim$ 0.2 \kms . 

We also derive densities from the \htwo\ data (see Appendix 1), assuming the temperature is equal to the kinetic temperature derived from the J = 0 and J =1 states.  The resultant densities, n(\hi ) = 21, 11, and 37 $\cm{-3}$ for components 1, 2, and 3 respectively, 
are nearly consistent with the \ci\ results (where for components 1 and 3 we derived n(\hi ) = 23 - 417 $\cm{-3}$ and 72 - 549 $\cm{-3}$, respectively).  Again, since the radiation fields were not consistent and they are involved in this calculation, we may expect some of the discrepancy.  We note however, that the densities are consistent with the lower limits derived from the \ci\ data alone (n(\hi ) $\geq$0.1 and $\geq$7 cm$^{-3}$ respectively).  

\begin{figure}
\plotone{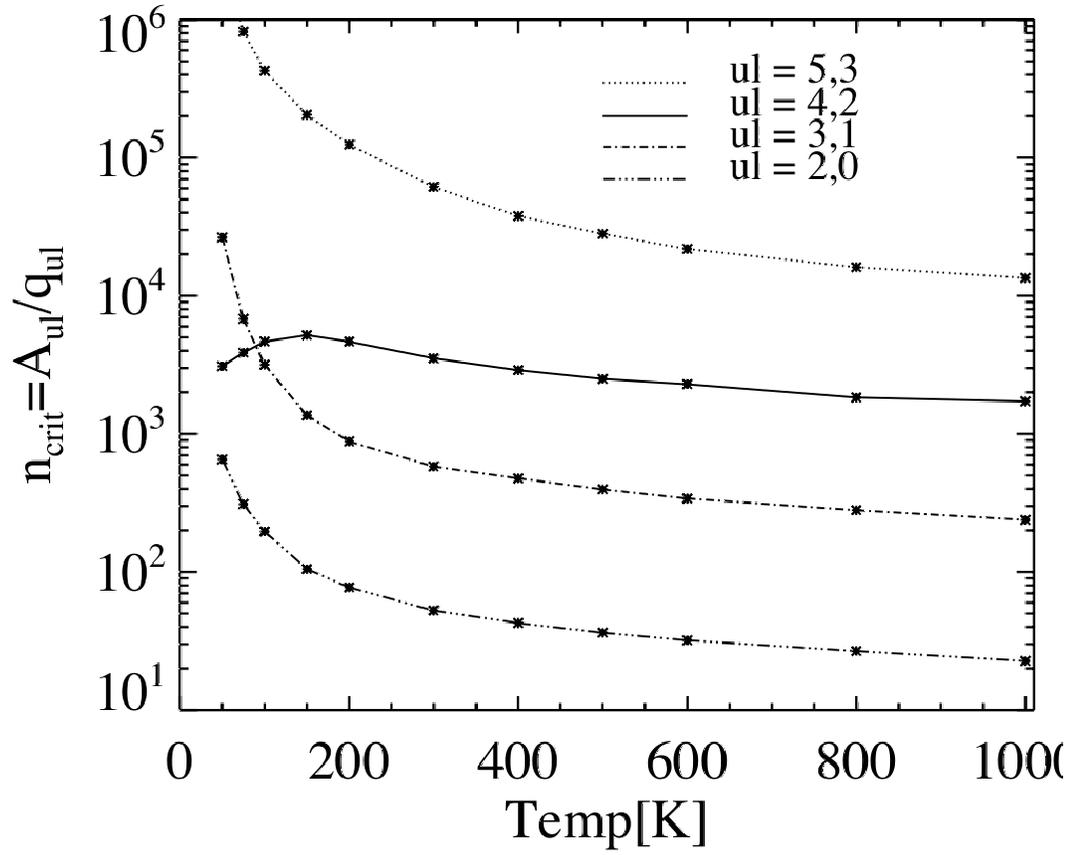}
\caption{Critical densities as a function of temperature for the \htwo\ rotational J states.  The subscripts u and l denote upper and lower respectively.
}
\label{fig:criticaldensity}
\end{figure}

Additionally, we can use the J = 2 rotational state to place an upper limit on the density if the excitation temperature, T$_{ex}^{02}$ is not equal to that of T$_{ex}^{01}$, indicating that the J = 2 state is not thermalized, and that the density is below n$_{crit}$, the critical density needed for thermalization.  We plot critical density as a function of temperature for the \htwo\ J states in Figure~\ref{fig:criticaldensity}.  In component 1 of \dla 0812$+$32, T$_{ex}^{02}$ = 132 K is greater than T$_{01}$ = 102 K, indicating that the density must be less than n$_{crit}$.  In this case we derive a upper limit on the density of component 1 of n $\lesssim$200 cm$^{-3}$, consistent with the derived densities for this component.  In component 3, T$_{ex}^{02}$ = 57 K is greater than T$_{01}$ = 47 K, also indicating an upper limit on the density of component 3 of n $\lesssim$700 cm$^{-3}$.  For component 2, T$_{02}$ is slightly less than T$_{01}$ -- indicating either that it is thermalized, or that this is within the errors.  In either case, the upper limit derived from the next higher state, T$_{03}$, would not be that restrictive as n$_{crit}$ $\sim$ 10$^3$cm$^{-3}$.  


\subsection{FJ0812$+$32, z$_{abs}$ =2.066780}

The \ciistr\ transition of this \dla\ falls in the \lya\ forest, and because the profile is quite different from that of the other low low ions (see Figure~\ref{fig:J0812_lowz_spec_lowions}), it is difficult to make a definitive estimate of the true \ciistr\ column density.  Instead, we attempt to estimate the star formation rate by assuming three models motivated by the bimodality of \dla\ cooling rates \citep{wolfe08}.  In the first case, case (a), 
we assume this is a 'low cool' \dla\ and use the average low-cool \lc\  = 10$^{-27.4}$\lcunit .  This assumption is likely the closest to the truth given that several physical traits of this \dla\ match those of the low-cool population of \dlas , i.e.  the small value of the low-ion velocity, $\Delta\ v$ = 26 \kms\ is more likely to be drawn from the low-cool sample, with median $\Delta\ v$ = 46 \kms , than from the high cool sample with median  $\Delta\ v$ = 104 \kms .  The \sii\ 1526 rest-frame equivalent width is also small at, $W_{1526}$ = 0.22 $\pm$ 0.01 \AA , compared with the low-cool median $W_{1526}$ = 0.26 $\pm$ 0.09 \AA , while the metallicity, [M/H] = $-$1.38 $\pm$ 0.01, is slightly higher than the median metallicity for the low-cool population [M/H] = $-$1.74 $\pm$ 0.19.  This discrepancy in metallicity may simply be the result of metallicity increasing with decreasing redshift, as the $z_{abs}$ = 2.06 of this \dla\ is lower than the median $z_{abs}$ = 2.85 of the \lc\ sample \citep{wolfe08}.    
Finally, the dust-to-gas ratio log$_{10}$$\kappa$=$-$2.74 is similar to the median low-cool dust-to-gas ratio, log$_{10}$$\kappa$=$-$2.57 $\pm$ 0.17.  In the second case, case (b), we will again make the low cool assumption, but instead of the standard $\alpha$-enhancement assumption of [Fe/Si] = $-$0.2, we will assume [Fe/Si] = 0.0.  Finally, in the last model, case (c), we will assume that the \dla\ is a `high cool' \dla\ and has an \lc\ equal to the median high cool \dla\ \lc\ = 10$^{-26.6}$\lcunit.  

We present the results for the three cases in Table~\ref{tab:cisolutions}.  
It is clear from the resultant densities that either cases (a) or (b) are more likely to be correct.  In case (c), there are no acceptable 1$\sigma$ solutions, and the range of densities at the 2$\sigma$ level (n(\hi ) $>$ 3800 cm$^{-3}$) seems unphysical.  Densities this high are unlikely to be observed because they imply a very small cloud size that would be unlikely to cover the background quasar, i.e. here, the cloud size is estimated as $\ell $ $\lesssim$ N(\hi )/n(\hi ) $\lesssim$ 10$^{21.5}$/3800 = 8.3 $\times$ 10$^{17}$cm $\lesssim$ 0.3 pc.  

Following our assumption that the minimal depletion model is more likely to be correct, we conclude that case (a) is the best physical approximation for this system.  If we now relax the constraint imposed by the \ciistr\ technique, and model a range of possible \jnutotci ,
we find that at the 2$\sigma$  level, we can restrict the density to n $\geq$ 32 cm cm$^{-3}$, T $\leq$ 1585 K and \jnutot\ = 0.41 $-$ 195 $\times$ 10$^{-19}$\junit , see Table~\ref{tab:cisolutions_fullgrid}.

\subsection{\dla\ 1331$+$17}~\label{sec:1331}


Here we discuss the analysis of component 1, at z$_{abs}$ = 1.77636, the only component with all three \ci\ fine structure lines detected.  The Haardt-Madau background at the redshift of this \dla\ is given by \jnubkd\  = 2.53$\times$10$^{-20}$ \junit .  
To estimate the local stellar contribution to the UV field we utilized the \ciistr\ technique.  While WPG03 report logN(\ciistr) = 14.05$\pm$0.05 cm$^{-2}$ for this object, the \ciistr\ transition is likely blended with a \lya\ forest line because its velocity profile is very different from that of the low ions, see Figure~\ref{fig:1331_spec_other_ions}.  Generally, the \ciistr\ velocity profile traces that of low ions such as SiII $\lambda$1808.  Therefore, we cannot exclude the possibility that the \ciistr\ transition is blended with a forest line, and we take the WPG03 value of logN(\ciistr) = 14.05$\pm$0.05 cm$^{-2}$ as an upper limit.  To obtain an estimate of the actual N(\ciistr), assuming that it is blended,
we (Carswell 2007; private communication) fixed the shape of the velocity profile to that of SiII $\lambda$1808 and normalized the \ciistr\ contribution by fitting the profile  simultaneously with a coincident \lya\ line.  
Summing over the \ciistr\ components results in an estimate of the true N(\ciistr ) of log N(\ciistr ) $\lesssim$ 13.56 cm$^{-2}$.  With this value of \ciistr\ absorption, the \ciistr\ technique results in a star formation rate that produces a radiation field of \jnulocciistr\  $\approx$ 3.1$\times$ 10$^{-19}$ \junit .  Therefore, the total input radiation field, that is, \jnubkd\  $+$ \jnulocciistr\ , is \jnutotciistr\  $\approx$ 3.3$\times$ 10$^{-19}$ \junit .  Finally, the \ci\ fine structure data constrained by ionization equilibrium, gives 2$\sigma$ results of 16 $\lesssim$ T $\lesssim$ 32 K, 91 $\lesssim$ n(\hi ) $\lesssim$ 363 cm$^{-3}$, and 3.36 \cmk $\lesssim$ log(P/k) $\lesssim$ 3.86 \cmk .  

We can compare our results to those of \cite{cui05}, who used CLOUDY to derive a best fit model to their \htwo\ data that resulted in a hydrogen number density n(\hi ) $\approx$ 0.2 cm$^{-3}$ and T $\approx$ 140 K for the \htwo\ bearing cloud at z$_{abs}$ = 1.776553.  This cloud is clearly not in pressure equilibrium with the \ci\ cloud of our analysis, component 1 at z$_{abs}$=1.77637, i.e. while the temperature is somewhat higher, the density is more than two orders of magnitude lower, resulting in pressures of P/k $\approx$ 28 cm$^{-3}$K for the \htwo\ -bearing cloud.  
The \ci\ -bearing cloud more closely resembles the \ci\ -bearing clouds in the local ISM as found by \cite{jenkins07} and \cite{jenkins01}. 
We discuss these similarities further in section ~\ref{sec:compareISM}. 

 \cite{cui05} determine a photoabsorption rate based on the population of \htwo\ in the J = 0 and J = 4 states, the latter of which is optically thin and therefore measures the intensity of radiation outside the cloud.  
Solving for this radiation field they derive \jnu ($\lambda$ = 1000$\AA$) $\approx$ 2.1 $\times 10^{-3}$ \jnu $_{, \odot}$ ($\lambda$ = 1000$\AA$), where \jnu $_{, \odot}$ ($\lambda$ = 1000$\AA$) is the UV intensity in the solar neighborhood at 1000 \AA\ and is, according to their paper, equal to \jnu $_{, \odot}$ ($\lambda$ = 1000$\AA$) $\approx$ 3.2$\times 10^{-20}$\junit .  They do recognize that the radiation field as determined by the \htwo\ J states is $\sim$3 orders of magnitude weaker than that determined by the \ciistr\ technique and comment that this is a discrepancy.  More importantly however, this radiation field is $\sim$3 orders of magnitude below the Haardt-Madau background, which sets a lower limit to the radiation field.  It is difficult to understand at first why the molecular hydrogen excitation is not consistent with this minimum radiation field.  A solution exists however, if the bulk of the \htwo\ gas resides in the narrow \ci\ velocity component, i.e. component 2, and hence, has been missed because of the effects of hidden saturation.  ~\cite{carswell09} have completed an analysis of this narrow component and include a detailed interpretation of the \htwo\ data that is consistent with both the Haardt-Madau background and the local stellar field derived by \ciistr .

When we relax the constraint on the radiation field as determined by the \ciistr\ technique and allow the field to vary, as already discussed in section~\ref{sec:ion_constrain}, we can constrain the radiation field to \jnutotci\ $\lesssim$ 1 $\times 10^{-19}$ \junit\ (1$\sigma$).  This limit on \jnutotci\ is actually quite strong considering that the Haardt-Madau background is only $\sim$1 order of magnitude lower than this value.  Note that the lower limit on the allowed \jnutotci\ is fixed by \jnubkd .  It is apparent from Figure~\ref{fig:select_jnus} that the density is limited to the range, 11 $\lesssim$  n(\hi ) $\lesssim$ 44 cm$^{-3}$ while the temperature is constrained to be 79 K $\lesssim$ T $\lesssim$ 794 K, and the pressure 3.5 $\lesssim$ log (P/k) $\lesssim$ 4.04 cm$^{-3}$ K.  We stress that these limits are derived independent of the results of the \ciistr\ technique, using \emph{only} the \ci\ fine structure data and the assumption of ionization equilibrium for a range of possible radiation fields.  The 2$\sigma$ results provide the less restrictive results \jnutot\ $\lesssim$ 8.6 $\times$10$^{-18}$ \junit , 
n(\hi ) $\gtrsim$ 10 cm$^{-3}$, and T $\lesssim$ 794 K.
 
\subsection{J2100$-$00}
While insufficiently good quality data hinder a full analysis of \dla\ 2100$-$00, we attempted an analysis of component 1, by treating the upper limit on N(\cistrstr ) as a detection.  The measured N(\ciistr ) produces a \jnulocciistr\ = 17 $\times$ 10$^{-19}$ \junit , also an upper limit.  The relatively large errors on the fine structure column densities do not produce very restrictive results.  However, the VPFIT derived Doppler parameter of this component, b = 0.2 $\pm$ 0.3 \kms , if real, 
places a strict upper limit on gas temperature of T$_{thermal}$ $\lesssim$ 29 K, which is consistent with the results from the \ci\ fine structure analysis.  Unfortunately, the weakness of the \ci\ line coupled with the low S/N of the data, makes a curve of growth analysis inconclusive.  For the purposes of this paper however, it is the \ci\ column densities that are important, and while they typically depend on the choice of Doppler parameter, we can obtain a sort of lower limit to the column density by fixing a relatively large Doppler parameter, i.e. b = 7 \kms .  We measure N(\ci ) = 12.12 cm$^{-2}$, which is well-within the errors quoted in Table~\ref{tab:cidata}, and see that in this case, the \ci\ column densities have little dependence on changes in b. 

\subsection{Q2231$-$00}
 
While there are two \ci\ components contained in the \dla\ at z$_{abs}$ $\sim$ 2.066, here we will focus only on component 2, which possess measurable \ci\ fine structure lines.  While not formally an upper limit, the detection of N(\cistrstr ) is relatively weak and has large error.  For the purpose of this analysis we will treat this value as a detection.   

We estimate the \jnulocciistr\ from the measured N(\ciistr ) with some confidence, even though it falls within the Lyman-$\alpha$ forest, because the velocity profile of \ciistr\ closely traces that of other low ions as expected.  Using this \jnulocciistr\ $\approx$ 24.7 $\times$10$^{-19}$ \junit , 
we find 199 $\cm{-3}$ $\lesssim$ n $\lesssim$ 1513 $\cm{-3}$, and 13 $\lesssim$ T  $\lesssim$ 25 K, see Table~\ref{tab:cisolutions}.  Relaxing the constraint of \jnulocciistr\ given by the \ciistr\ technique results in a lower limit on the density of n(\hi ) $\gtrsim$ 3 cm$^{-3}$ (2$\sigma$).  

\subsection{J2340$-$00}

\dla\ 2340$-$00 is a relatively complex system requiring nine \ci\ velocity components.  Given the strong blending of transitions and the resultant uncertainty on upper limits, we do not analyze those components for which N(\cistrstr ) is not detected, namely, components 1 and 7.  In addition, the column densities of component 6 have extremely large errors, making analysis pointless, as well as the unphysical condition of N(\cistr ) $>$ N(\ci ). 

This \dla\ contains a relatively high total column density of molecular hydrogen, logN(\htwo ) = 18.20 cm$^{-2}$, or f = 0.014.  
The neutral hydrogen column density, at log N(\hi ) = 20.35 $\pm$ 0.15 $\cm{-2}$, is close to the threshold defining a \dla\ (logN(\hi ) = 20.30 cm$^{-2}$), which supports the findings of ~\cite{noterdaeme08} that the probability of detecting \htwo\ does not strongly depend on N(\hi ).   
Furthermore, the low-ion velocity profile is large, with $\Delta$$v$ = 104 \kms , and the cooling rate, as determined by \ciistr\ over the entire profile, at log \lc\ =  $-$26.15 \lcunit , is among the highest of \dlas .   Because of the complex nature of the \ci , the \htwo , and the low-ion profiles, all requiring multiple components, and because of heavy blending and saturation in some components of the low-ions, we were not able to obtain unique fits to all low ion components using VPFIT.  Instead, we analyzed the \ci\ data using two different model assumptions. 

In the first case, we assume \jnutot\ just slightly above the background, \jnutot\ = 3.9 $\times$ 10$^{-20}$ \junit .  Because this is a lower limit to the radiation field, this analysis provides an upper limit on density (i.e. for all else being constant, if we increase the radiation field, the density required to collisionally excite the \ci\ fine structure levels is decreased).  The metallicity, dust-to-gas ratio and log$\frac{\cii }{\ci }$ are determined by AODM over the entire profile with measured results given in Table~\ref{tab:otherdata}. The densities and temperatures derived from the \ci\ fine structure data are given in Table~\ref{tab:cisolutions}. 

In an attempt to refine the model, in the second case, we used the AODM 
over three 'super-components' of the \ciistr , S II, Fe II, and other low-ion and resonance line transitions.  The choice of super-components was motivated by visual inspection of the spectra that reveal three large velocity components separated in velocity space (albeit with each component containing smaller substructure).  We arbitrarily chose $v = 0$ \kms at z$_{abs}$ = 2.054151, and defined the AODM super-components as follows: super-component (a) from $v$ = $-$30 to 15 \kms , super-component (b) from $v$ = 15 to 70 \kms , and super-component (c) from $v$ = 70 to 120 \kms , see Figure~\ref{fig:J2340_spec_otherions}.  Super-components a, b, and c coincide with \ci\ components (1, 2), (3, 4, 5, 6, 7) and (8, 9) respectively.  We then applied the metallicity, dust-to-gas ratio, and N(\ciistr ) measurement derived from the AODM 'super-component' to each associated \ci\ component. Table~\ref{tab:j2340_aodm} contains a summary of this analysis.
Looking at each super-component, we find the following:

$\bullet {Super-component \ (a):}$ We did not perform the \ciistr\ analysis on super-component (a) because we measure a super-solar [Fe/S] and [Ni/S].  
The absence of depletion detected for Fe and Ni implies that we cannot calculate the dust-to-gas ratio.  If real, this super-solar (or nearly solar) value of Fe II would require a different and special star formation history.  Given that we cannot calculate a dust-to-gas ratio or \jnulocciistr , and the fact that the \ci\ components 1 and 2 associated with super-component (a) contain the smallest amount \ci , we did not perform further analysis.

$\bullet {Super-component \ (b):}$ Super-component (b), which includes \ci\ components 3, 4, 5, 6 and 7, contains the bulk of the gas.  The \ciistr\ technique results in \jnutotciistr\ = 5.27$\times$10$^{-18}$ \junit.  
The combination of the large \jnutot\ and the low $\frac{C II}{C I}$ are not compatible with the measurements of \ci\ fine structure in components 3, 4, or 5 (6 and 7 are ruled out by upper limits).  
  
$\bullet {Super-component \ (c):}$  The \ciistr\ technique applied to super-component (c), which covers \ci\ components 8 and 9, results in a \jnutot\ = 1.1 $\times$10$^{-18}$\junit .  A reasonable range of results for components 8 and 9 are summarized in Table~\ref{tab:cisolutions}.

If we relax the constraint of \jnutotciistr\ as measured by the \ciistr\ technique, we can determine limits for each component for a range of possible radiation fields.  Results 
are given in Table~\ref{tab:cisolutions_fullgrid}.  In all cases we can put a lower limit on the density of n $>$ 7 cm$^{-3}$.  In all components except for component 2, the temperature can be constrained to be T $\lesssim$ 800 K and the \jnutotci\ $\leq$ 27 $\times$10$^{-19}$\junit , is 2 orders of magnitude larger than the Haardt-Madau background and consistent with the \jnutot\ derived by the \ciistr\ technique.

\subsubsection{On the possible ionization of \dla\ 2340$-$00}

\begin{figure}
\plotone{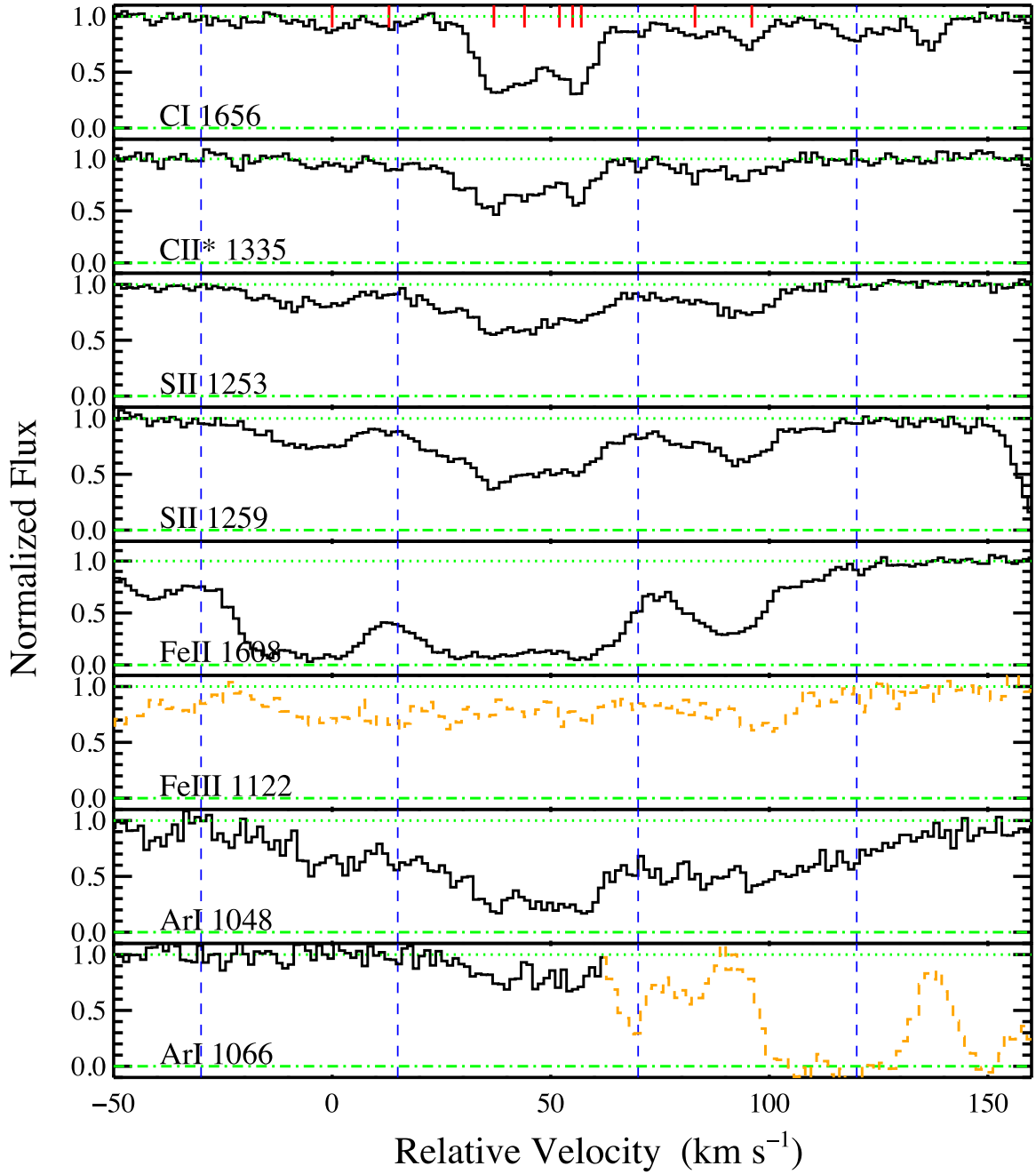}
\caption{\dla\ 2340$-$00 low ions, Fe III and Ar I.  Blends are indicated by orange dashed lines.  The velocity space defining the three super-components a, b, and c are separated by vertical blue dashed lines.  Relative to v = 0 at z$_{abs}$ = 2.054151, super-component (a) is from $-$30 - 15 \kms , super-component (b) from 15 - 70 \kms , and super-component (c) from 70 - 120 \kms .  For reference, the 9 \ci\ velocity components are indicated by red vertical tick marks.
}
\label{fig:J2340_spec_otherions}
\end{figure}

Given the potential presence of Fe III $\lambda$1122 and the fact that the column density of \dla\ 2340$-$00, at log N(\hi ) = 20.35 cm$^{-2}$, is just above the \dla\ threshold of 2 $\times$ 10$^{20}$ cm$^{-2}$, we considered the possibility of partial ionization of the gas.  We used the AODM technique to measure Fe III/Fe II in the same three 'super-components' discussed above, covering \ci\ components 1 and 2, components 3, 4, 5, 6 and 7 and components 8 and 9 respectively, see Figure~\ref{fig:J2340_spec_otherions}.  Because the Fe III profile does not trace that of the low ions (see Figure~\ref{fig:J2340_spec_otherions}), we cannot rule out the possibility of blending with the Lyman $\alpha$ forest and therefore, we treat our Fe III measurements as upper limits that ultimately give no information about possible ionization.

Instead, we can rule out a high ionization factor based on the [Ar/S] measurement.  Specifically, ~\cite{pro2002} state that photoionzation models in which [Ar I / S II] $>$ $-$0.2 dex require that x $<$ 0.1, in other words, require gas that is $>$ 90\% neutral.    
Using the Ar I $\lambda$$\lambda$ 1048, 1066 transitions, 
excluding blending in the super-component 3 of the $\lambda$1066 transition (see Figure~\ref{fig:J2340_spec_otherions}), we find  [Ar/S] $>$ $-$0.2 in all components (see Table~\ref{tab:j2340_aodm}), indicating that x $<$ 0.1 and that the gas is $>$90\% neutral.

\subsubsection{Molecular Hydrogen in \dla\ 2340$-$00}

\begin{figure}
\plotone{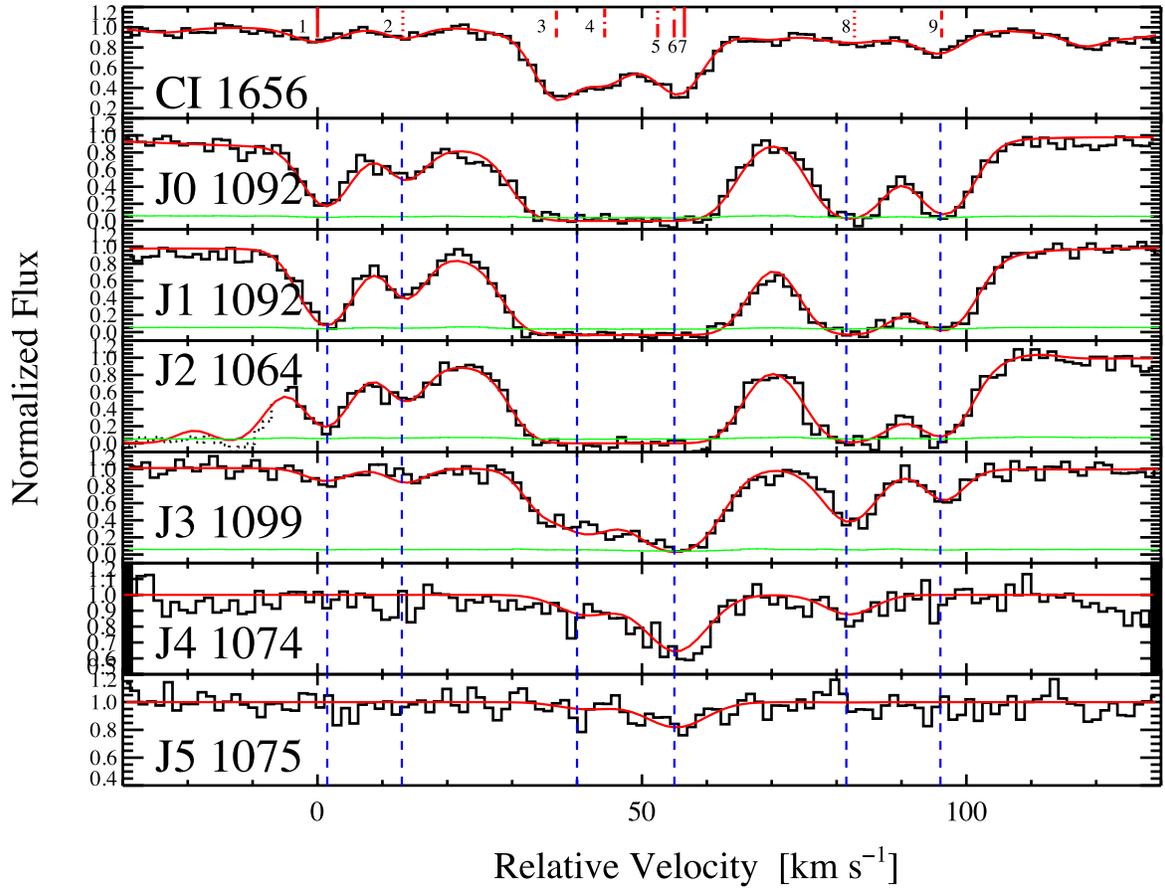}
\caption{Examples of the \htwo\ absorption in \dla\ 2340$-$00.  We show the relatively unblended sections here (blends are denoted by dotted lines) over several rotational J levels.  Blue vertical dashed lines mark the six \htwo\ components we call components 1, 2, 4, 6, 8 and 9.  
}
\label{fig:J2340_spec_h2}
\end{figure}

\begin{figure}
\plotone{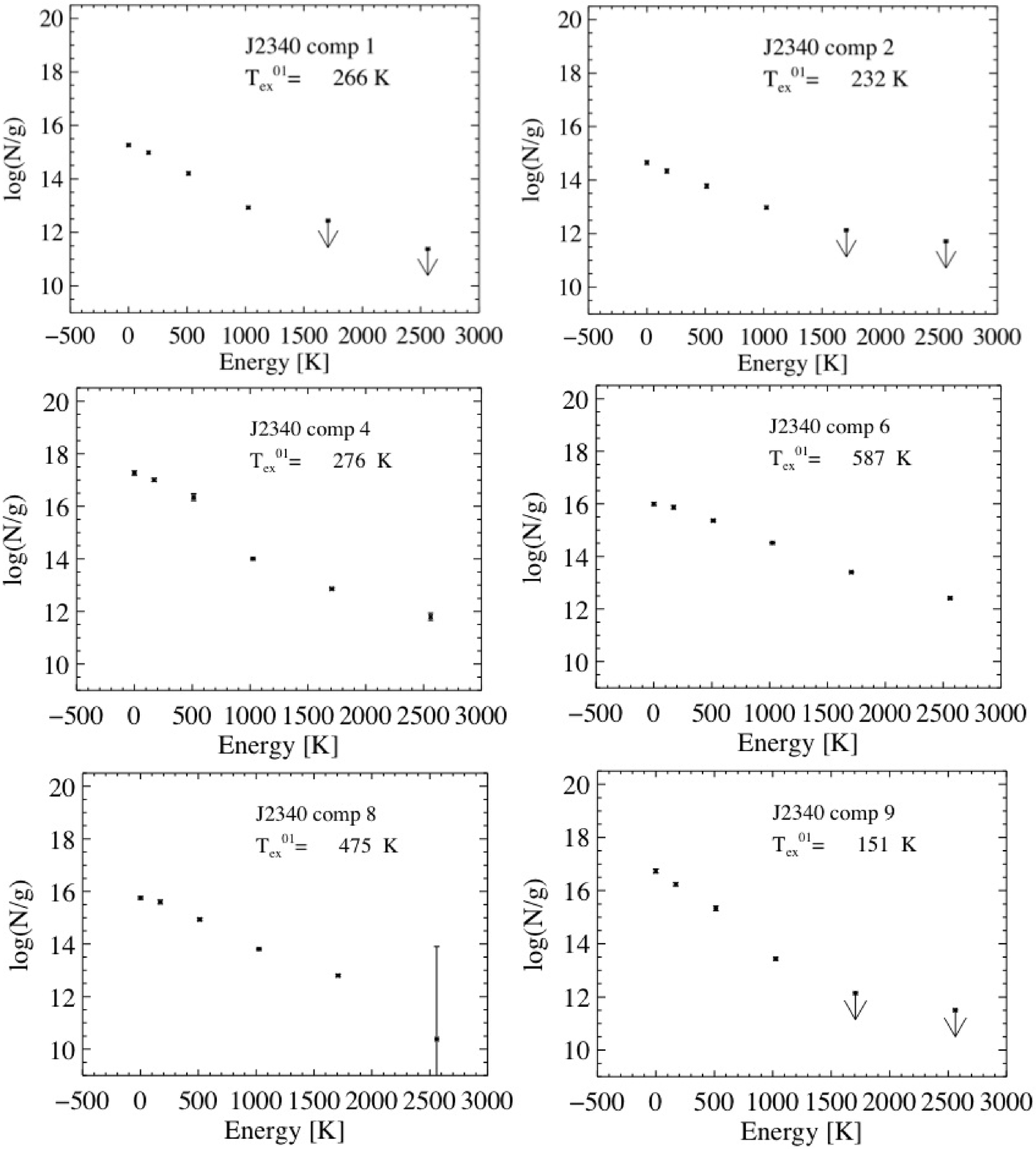}
\caption{The \htwo\ excitation diagrams for the 6 components of \dla\ 2340 $-$00 that contain \htwo .  
}
\label{fig:J2340_h2}
\end{figure}

The total column density of molecular hydrogen, logN(\htwo ) = 18.20 $\cm{-2}$, where f = 0.014, is large relative to most \htwo\ -bearing \dlas , where f is typically f $\sim$ 10$^{-5}$.  
 We have analyzed the \htwo\ using VPFIT.  To allow for the best fit we have let the \htwo\ component redshifts and b values vary independently of the \ci\ and low-ion components.
  We find that we require 6 \htwo\ components to achieve the best fit.  In redshift space, these components lie remarkably close to the \ci\ components 1, 2, 4, 6, 8 and 9, ($\Delta v$ =  $+$1.4, $+$0.6, $-$2.8, $+$0.3, $-$0.4, $+$0.4 \kms , respectively) and therefore we use this notation to refer to the \htwo\ components.  Details of the \htwo\ measurements are given in Table~\ref{tab:J2340_h2} and example spectra in Figure~\ref{fig:J2340_spec_h2}.  In Figure~\ref{fig:J2340_h2} we plot the excitation diagrams for each \htwo\ component and list the excitation temperature as determined by the J=0 and J=1 states.  Additionally, we use the population of the J=4 state to determine the radiation field as described in Appendix 3.  Details are given in Table~\ref{tab:j2340_h2_comp_analysis}.

 While the radiation fields derived from the \htwo\ data are in general consistent with the \ci\ constraints, it is interesting to note that the densities derived from the \htwo\ data alone, see Appendix 1 for details, while consistent with the \ci\ limits, tend to be significantly higher than that required for \ci .  
We summarize results for each component for which we could make comparisons between the different techniques:

$\bullet {Components\ 1 \& \ 2:}$ Excitation temperatures are in good agreement with constraints placed by the \ci\ data, however the \htwo\ analysis was not completed because of the super-solar [Fe/S] and [Ni/S]. 
 
$\bullet {Component\ 4:}$ The \htwo\ derived T = 276 K is consistent with the 2$\sigma$ \ci\ range of T = 40 $-$ 794 K.  The density derived from \htwo , n(\hi ) $\sim$1600 cm$^{-3}$ is not compatible with the \ci\ limits.  Finally, the \htwo\ derived \jnutothtwo\ = 2.25 $\times$ 10$^{-19}$ \junit\ is within range allowed by \ci\ and much lower 
than that predicted by \ciistr\ (\jnutotciistr\ = 52.7 $\times$ 10$^{-19}$ \junit\ ). 

$\bullet {Component\ 6:}$ Because we only obtained upper limits on the \cistrstr\ state of \ci\ component 6 we can not make direct comparisons in this case.  We can however, compare with \ci\ component 5.  In this case, the \htwo\ derived T = 587 K is consistent with the 2$\sigma$ \ci\ limit T $\leq$ 1259 K (or T $\leq$ 794 depending on \jnu\ ), and the \htwo\ derived n(\hi ) =  10, 509 cm$^{-3}$ is also consistent with the lower limits placed by \ci\ data.  The radiation field derived from the \htwo , \jnutothtwo\ = 18.8 $\times$ 10$^{-19}$ \junit , is consistent with the limits placed by \ci , however these are both much lower than the \jnutotciistr\ predicted by the \ciistr\ technique.  

$\bullet {Component\ 8:}$ The \htwo\ derived T = 475 K is inconsistent with the 2$\sigma$ \ci\ limit, T $\leq$ 158 K.  The \htwo\ derived density, n(\hi ) =  3595 cm$^{-3}$, is also incompatible with the range allowed by \ci\ (n(\hi ) =  12 - 209 cm$^{-3}$) .  However, the \htwo\ derived radiation field, \jnutothtwo\ $\leq$ 5.13 $\times$ 10$^{-19}$ \junit\ is compatible with the limits set by \ci , and similar to that predicted by the \ciistr\ technique, \jnutotciistr\ = 11  $\times$ 10$^{-19}$ \junit . 

$\bullet {Component\ 9:}$ The \htwo\ derived T = 151 K is consistent with the 2$\sigma$ \ci\ limit, T $\leq$ 398 K.  The \htwo\ derived density, n(\hi ) =  377 cm$^{-3}$, is also nearly compatible with the range allowed by \ci\ (n(\hi ) =  28 - 363 cm$^{-3}$).  However, the \htwo\ derived radiation field, \jnutothtwo\ $\leq$ 0.48 $\times$ 10$^{-19}$ \junit , while compatible with the limits set by \ci , is not compatible with that predicted by the \ciistr\ technique, \jnutotciistr\ = 11  $\times$ 10$^{-19}$ \junit . 

A striking difference is seen in the densities derived from the \ci\ and \htwo\ data.  A similar difference has been reported in sightlines towards Q0013$-$004 and Q1232$+$082 ~\citep{hira05} and HE0515$-$00 ~\citep{reimers03, quast02}.  They argue that if the density is this high, i.e. equal to the critical density,  then the states should be thermalized and there would be no difference in the excitation temperature as determined by the J = 1 and J = 2 states.  Given that the observed T$_{02}$ is always much greater than T$_{01}$, ~\cite{hira05} propose, following the suggestion of ~\cite{reimers03}, that a potential explanation may be that the \htwo\ formation rate (R$_{dust}$) may be larger than estimated, which would result from, for example, a smaller than estimated grain size.  
Therefore, if the formation rate is higher, you require a smaller density than that derived.   On the other hand, for each component of \dla\ 2340$-$00 studied here, the T$_{02}$ is either less than or approximately the same as T$_{01}$, implying that the critical density cannot be ruled out.  

\section{Discussion}~\label{sec:discussion}

We have used \ci\ fine structure absorption in high resolution, high signal-to-noise data to study the physical conditions in \dlas\ at high redshift.  Our work differs from previous studies of \ci\ fine structure absorption because we did not assume a gas temperature in order to derive the density.  Rather, we assume ionization equilibrium, which in conjunction with the \ci\ fine structure data, allows us to constrain \emph{both} the temperature and the density of the cloud.  In addition, we use the \ciistr\ technique to infer the local radiation field due to stars and include its contribution to the \ci\ fine structure excitation, thus providing a complete and fully self-consistent model of the gas.  In most cases, the \ci\ fine structure excitation is consistent with the \jnulocciistr , derived independently by the \ciistr\ technique. 

\subsection{Assessment of Systematic Errors}
To draw meaningful comparisons between the \ci\ results and those of the \ciistr\ technique we must first analyze our possible systematic errors.  The primary source of error in the \ci\ analysis stems from two sources --  1) measurement error of the fine structure column densities, including possible errors in the oscillator strengths (we report results for 2$\sigma$ errors, which should encompass these errors \footnote{For example, tests run using Jenkins 2006 f-values for \ci\ produced a Doppler parameter $\sim$ 15\% larger and column density $\sim$0.3 dex smaller than the Morton 2003 values in the case of the curve of growth analysis of the narrow, cold component in \dla\ 0812$+$32 \citep{jorg09}.}), and 2) the assumption that in solar units the carbon abundance is equal to that of Si II (or S II) $-$0.2 dex in case of minimal depletion.  The source of systematic errors in the \ciistr\ technique are 
more difficult to assess (see WGP03 for a complete discussion).  We explore the possible effects of these errors in the following section and demonstrate that, even considering these errors, the \ci\ data appear to be probing gas of higher densities and pressures -- likely small knots of gas within the larger \dla\ galaxies -- than that probed by the `global'  \ciistr\ technique.  

\subsubsection{Comparison with \ciistr\ technique model}\label{sec:model}
\begin{figure}
\plotone{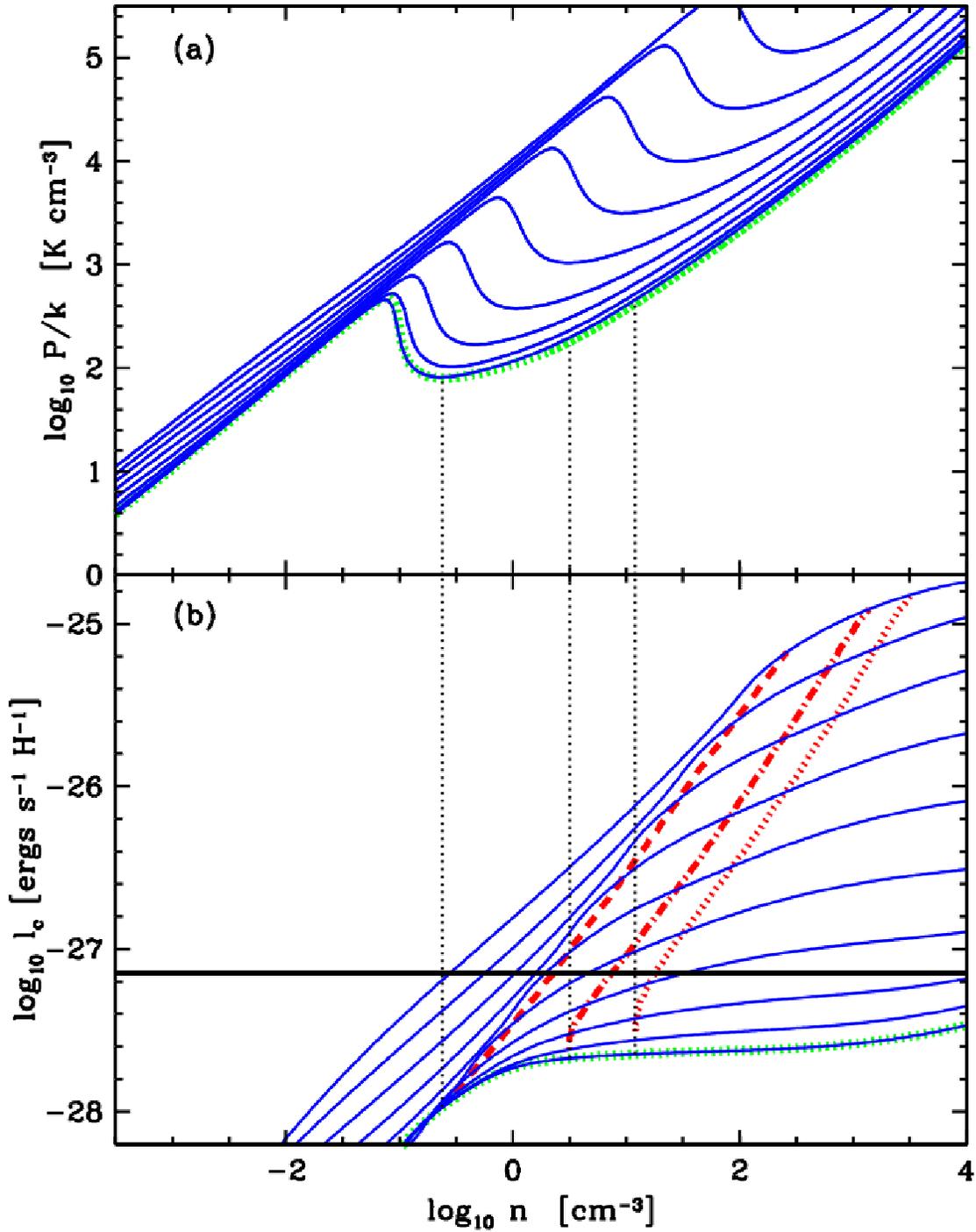}
\caption{Pressure curves and star formation rate solutions for \dla\ 1331$+$17.  The blue model curves for different star formation rates represent log$\Sigma _{SFR}$ = $-$$\infty$, $-$4, $-$3.5, ..., 0.0.  The CNM P$_{min}$, P$_{eq}$ and P$_{max}$ solutions are indicated by the red dashed lines.  Vertical, black dotted lines illustrate the location of the CNM stable points associated with, from left to right, P$_{min}$, P$_{geo}$ and P$_{max}$ for the case of background radiation only.
}
\label{fig:q1331_art}
\end{figure}

While it is difficult to assess the systematic errors involved in the \ciistr\ technique, some of the potential uncertainties have been removed since the work of WGP03.  For example, the SMC dust model is now assumed to be correct because of the non-detection of the 2175$\AA$ dust feature that would have indicated Galactic dust, while the reddening curve resembles SMC rather than Galactic or LMC ~\citep{vladilo08}.  Therefore, we focus on the two largest potential uncertainties:  1) the assumption of the equilibrium pressure, P$_{eq}$, and 2) the minimal versus maximal,
 depletion model (see discussion in section 2.3 of WGP03).  The standard \ciistr\ technique involves solving the equations of thermal and ionization equilibrium as described in WPG03 and ~\cite{wolfire95}.  A unique solution is determined by assuming that the equilibrium pressure is equal to the geometric mean between P$^{max}$ and P$^{min}$, P$_{eq}$ = P$_{geo}$ = (P$_{min}$P$_{max}$)$^{1/2}$, where P$_{min}$ and P$_{max}$ are the minimum and maximum pressures of the function P(n) where n is gas density.  This results in two stable solutions for a given star formation rate, one WNM and one CNM, and is the basis of the two-phase model.  However, a two-phase medium can achieve equilibrium with a pressure ranging from P$_{min}$ to P$_{max}$, and therefore, the assumption of P$_{eq}$ equal to the geometric mean, while reasonable (see Wolfire et al. 2003 and WPG03 discussion), is still an unproven assumption.  Following WGP03 we attempt to gain a sense of the possible systematic errors by allowing P$_{eq}$ to vary between P$_{min}$ and P$_{max}$.  In Figure~\ref{fig:q1331_art}, we show the standard two-phase diagram, plotting in (a), log(P/k) versus density for various star formation rates per unit area (which is proportional to \jnu ) which are constant along each P(n) curve, and which increase from bottom to top, and in (b) log(\lc ) versus density for \lc\ equilibrium solutions for those same star formation rates.  The green dashed line in each plot indicates the P(n) solution for heating by background radiation alone (i.e. log$\Sigma _{SFR}$ = $-$$\infty$).  The black horizontal line in (b) denotes the observed cooling rate of \dla\ 1331$+$17, log(\lc ) $\leq$ $-$27.14 \lcunit.  Three vertical, black, dotted lines illustrate the location of the CNM stable points associated with, from left to right, P$_{min}$, P$_{geo}$= (P$_{min}$P$_{max}$)$^{1/2}$, and P$_{max}$ for the case of background radiation alone.  The stable (\lc , n) pairs for the grid of star formation rates are denoted by the three red lines in (b), where the three different pressure assumptions have been made -- from left to right they are: dashed = P$_{min}$, dot-dashed = P$_{geo}$ and dotted = P$_{max}$.  Although not relevant for our current discussion, it is seen that P$_{min}$ requires a higher star formation rate and lower density to achieve an equilibrium solution with a cooling rate equal to that observed.  On the opposite extreme, P$_{max}$ requires a higher density and lower star formation rate.   

\begin{figure}
\plotone{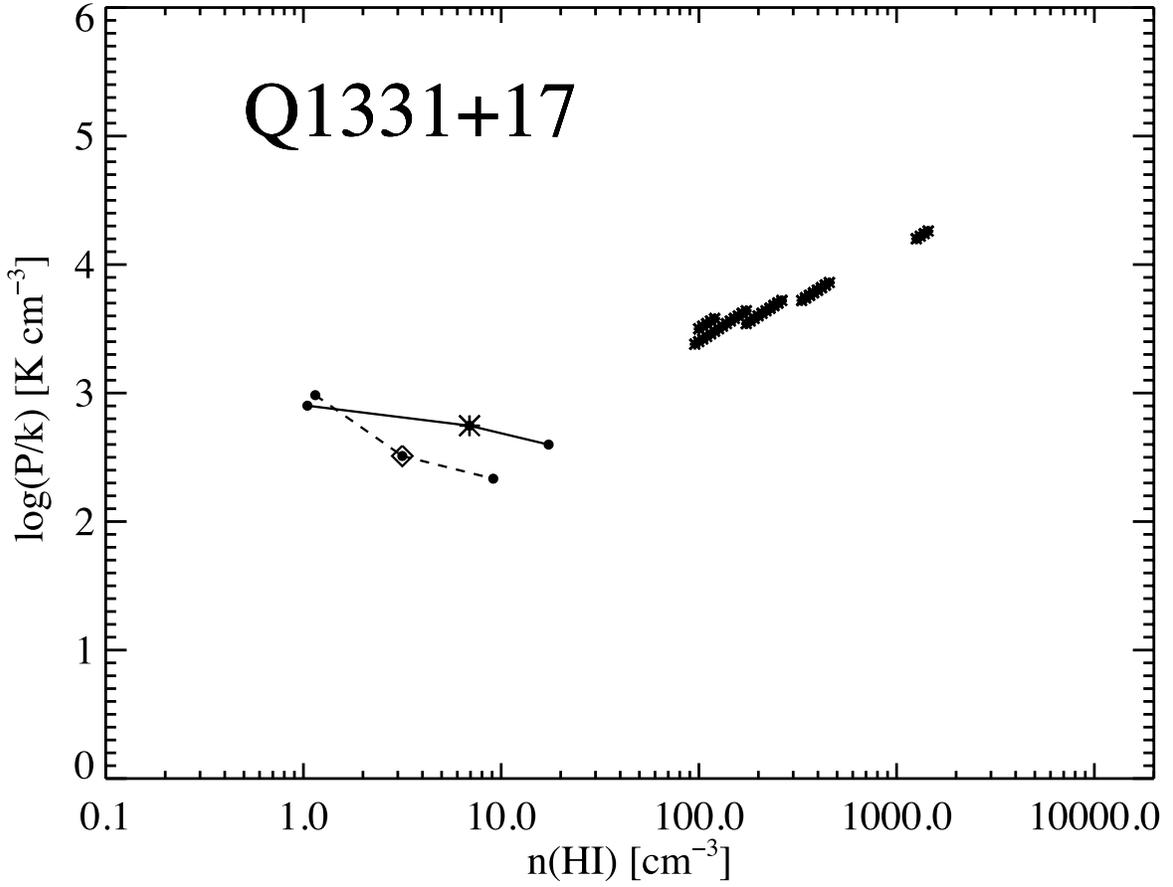}
\caption{\dla\ 1331$+$17 2$\sigma$ range of solutions for range of \ciistr\ technique solutions.  The minimal depletion \ciistr\ solution is indicated by the asterik, with P$_{min}$ and P$_{max}$ denoted by points at the ends of attached lines, while the maximal depletion model is denoted by a diamond, and the P$_{min}$ and P$_{max}$ at the ends of dashed lines.  For comparison, the full range of \ci\ solutions are given as a grouping of points in the upper right.  It is clear that these fall at higher pressures and densities than than the \ciistr\ results.
}
\label{fig:q1331_press}
\end{figure}

 For the purposes of this paper, we are interested in the range of densities and temperatures that result from these different model assumption inputs to the \ciistr\ technique.  In Figure~\ref{fig:q1331_press}, we plot the resultant log(P/k) versus density for our example case of \dla\ 1331$+$17.  For the minimal depletion model, the case of P$_{eq}$ = P$_{geo}$ = (P$_{min}$P$_{max}$)$^{1/2}$ is denoted by the asterisk.  Lines connect to the P$_{min}$ and P$_{max}$ solutions, providing a sense for the potential systematic "error" involved in the assumption of where the equilibrium pressure resides.  We have also plotted the results of the maximal depletion model, denoted by the diamond, with dashed lines connecting to the associated P$_{min}$ and P$_{max}$ solutions.  To correctly compare this range of solutions to that of the \ci\ data we must re-model the \ci\ theoretical curves in each case because the change in the assumed P$_{eq}$ results in a change of star formation rate, or \jnuloc , which is an input for the \ci\ theoretical curves.  We plot the 2$\sigma$ results of the \ci\ analysis for \jnuloc\ spanning that determined by P$_{min}$ and P$_{max}$.  We summarize the results of these two techniques for \dla\ 1331$+$17 in Table~\ref{tab:compare}.  
  
\begin{figure}
\plotone{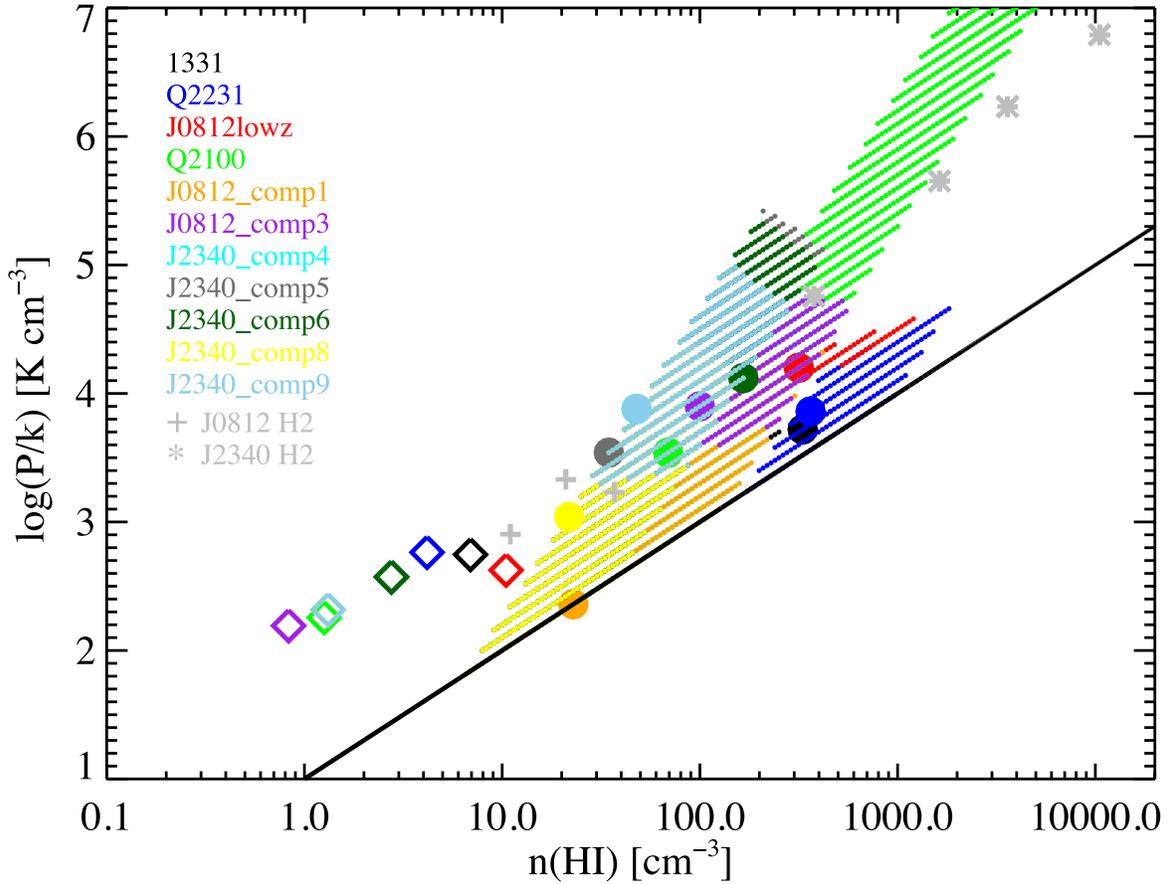}
\caption{Summary of all \dlas , compared with the results of the \ciistr\ technique (diamonds).  The 2$\sigma$ \ci\ results are the shaded regions, and the best-fit (minimized $\chi$$^2$) for each is marked by a filled circle.  It is clear that the \ci\ results are consistently higher in density and therefore pressure.  Note, some regions/solutions are overlapping.  The \htwo\ derived pressures are denoted by light grey crosses and asterisks for components in DLA 0812$+$32 and DLA 2340$-$00 respectively.  The solid black line denotes T = 10 K.   The resulting median values for the \ci\ sample are: $<$n(\hi )$>$ = 69 cm$^{-3}$, $<$T$>$ = 50 K, and $<$log(P/k)$>$ = 3.86 \cmk , with standard deviations,  $\sigma$$_{n(\hi )}$ = 134 cm$^{-3}$, $\sigma$$_T$ = 52 K, and $\sigma$$_{log(P/k)}$ = 3.68 \cmk.  This can be compared with the global \ciistr\ technique median values for the same \dlas : $<$n(\hi )$>$ = 2.8 cm$^{-3}$, $<$T$>$ = 139 K, and $<$log(P/k)$>$ = 2.57 \cmk , with standard deviations $\sigma$$_{n(\hi )}$ = 3.0 cm$^{-3}$, $\sigma$$_T$ = 43 K, and $\sigma$$_{log(P/k)}$ = 0.22 \cmk .   
}
\label{fig:all_press}
\end{figure}

In Figure~\ref{fig:all_press} we summarize the results of our \ci\ sample and compare them to the \ciistr -derived and \htwo -derived models of the same systems.  We plot the best-fit \ci\ solutions, as determined by $\chi$$^2$ minimization, as circles, while the shaded regions denote the 2$\sigma$ \ci\ solutions.  The \ciistr\ solutions are represented as diamonds.  
It is seen that the densities determined by the \ciistr\ technique are systematically lower than those determined by the \ci\ data.  As a result, the overall pressures are lower.  The temperatures vary depending on model assumptions, but are in general agreement or higher than the \ci\ results (see Table~\ref{tab:compare}).    We conclude that the \ci\ gas is tracing a denser region of the \dla\ than that traced by the global \ciistr\ technique: essentially, the \ci\ resides primarily in small, overdense knots.  
However, one problem with this picture is understanding how these two phases, the low pressure ambient medium and the higher pressure smaller 'clumps' of \ci\ -bearing gas, remain in pressure equilibrium.  We return to this issue in a following paper.  

These comparisons also reveal a discrepancy between the $\frac{\cii }{\ci }$ derived by the \ciistr\ technique analysis and the observed $\frac{\cii }{\ci }$.  In general, the observed $\frac{\cii }{\ci }$ ratio is approximately an order of magnitude smaller than that predicted by the \ciistr\ technique model (see Table~\ref{tab:compare}).  We avoid a detailed discussion here and refer the interested reader to WGP03, section 5.1, for a detailed discussion of the model inputs that affect $\frac{\cii }{\ci }$.  We only briefly mention that we have tested the results with cosmic rays turned on and off, and there is not a large effect on the results for the star formation rates we are considering.  However, this is not the case for large star formation rates, \jnutot\ $\sim$10$^{-18}$\ \junit , for which the effect of cosmic rays becomes more important.  Assuming that the \ciistr\ technique model is correct, we can understand this difference in terms of the new model in which the bulk of the measured \ci\ is localized in small, 
dense clumps, $\approx$ 1 $-$ 10 pc, relative to the larger \dla\ (for estimates of the cloud size, see last column in Table~\ref{tab:cisolutions}).  In this case, while the cloud is still optically thin and feels the same radiation field as the surrounding medium, the increased density, and hence, increased electron density, n$_e$, of the \ci\ cloud work to lower $\frac{\cii }{\ci }$ because $\frac{\cii }{\ci }$ is inversely proportional to density for a fixed radiation field.  In a sense, this conflict with the predictions of the 'global' \ciistr\ technique model, is an expected result of the overdense-\ci\ -region model. 

\subsubsection{Direct Estimate of Error on \jnuloc }\label{sec:direct}
While there are many assumptions and uncertainties in the \ciistr\ technique, we can make a direct estimate of the possible error on \jnuloc\ derived from the \ciistr\ technique by using the z$_{abs}$ =1.9 DLA 2206$-$19.  In ~\cite{wolfe04}, the FUV radiation field inferred from an image of a galaxy associated with this DLA  ~\citep{warren01} is compared with the radiation field inferred from the \ciistr\ technique and they are found to agree to within $\sim$ 50\%.  Therefore, we will assume that the error on \jnuloc\ is $\sim$ 50\% .  
 
\subsubsection{Distribution of \cii\ }\label{sec:ciiovci}
\begin{figure}
\plotone{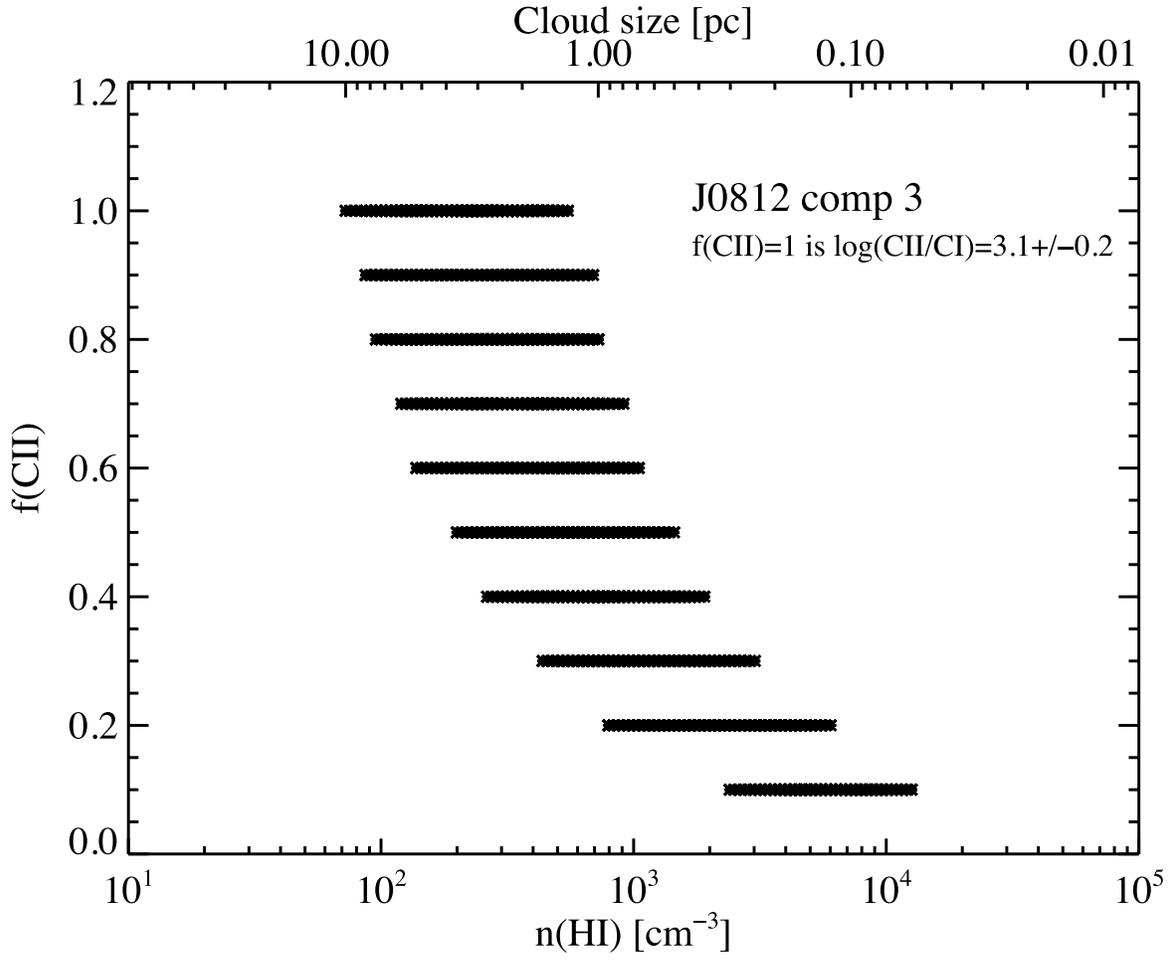}
\caption{The resultant cloud size in pc (top axis) as a function of the fraction of \cii , f(\cii ), associated with the \ci\ cloud. Here, f(\cii\ ) = 1 means that 100\% of the measured N(\cii ) is associated with the \ci\ cloud.  In this case, for DLA 0812$+$32, component 3, log($\frac{\cii }{\ci }$) = 3.1 $\pm$ 0.2.  As the fraction of associated \cii\ decreases, the n(\hi ) increases and hence, the resultant cloud size decreases.  
}
\label{fig:J0812_comp3_cloudsize}
\end{figure}

Throughout this work we assume that the measured N(\cii ) can be directly applied to each \ci\ cloud.  However, assuming that there is no hidden saturation -- which would actually increase the N(\cii ) -- this is actually an upper limit to the amount of N(\cii ) associated with each \ci\ cloud.  The distribution of \cii\ along the line of sight could be clumpy such that only a fraction of the measured N(\cii ) is associated with a given \ci\ cloud.  With respect to the analysis done in this paper, the effect of decreasing the amount of \cii\ associated with the \ci\ cloud causes an increase in the resultant volume density and a decrease in the temperature such that the pressure remains approximately constant.  The increase in volume density with decreasing fraction of \cii\ results in a decrease of the derived cloud size as shown in Figure~\ref{fig:J0812_comp3_cloudsize}.
 
\subsection{Relation to 'high-cool' \dla\ population}
With the exception of \dla\ 1331$+$17, all of the \ci\ -bearing objects not only contain \ciistr\ (as compared with $\sim$50\% of the general \dla\ population) but also have cooling rates, \lc , that place them firmly in the 'high-cool' range defined by ~\cite{wolfe08} (median 'high-cool' log\lc\ = $-$26.6 \lcunit ).  A literature search 
reveals that all previously published \dlas\ with positive detections of \ci\ fall into the 'high-cool' population as well.  While it is not clear how the bimodality discovered by ~\cite{wolfe08} is related to \ci , this correlation -- that almost all \ci\ -bearing \dlas\ are also high-cool \dlas\ -- could simply be a result of the higher metallicities and dust-to-gas ratios that are required to form and sustain, through dust shielding of UV radiation, measurable amounts of \htwo\  and  \ci .  
This trend, of higher metallicity \dlas\ being more likely to contain measurable \htwo , has been observed previously by ~\cite{petitjean06}.   Additionally, the 'high-cool' \dlas , shown by ~\cite{wolfe08} to consist of primarily CNM, simply might be an environment more conducive to the presence of \htwo\ and \ci .   


\subsection{Comparison with the local ISM}\label{sec:compareISM}
In this section we compare our \ci -bearing clouds to the ISM of the local Universe, namely, the Milky Way and the Small and Large Magellanic Clouds (the SMC and LMC respectively).  To draw these comparisons, we first determine the median n(\hi ), T, and P of our \ci\ sample.  Of course, this median is dependent upon which models we choose to include.  Here, we include the models for each \dla\ that are likely to be the most physically realistic, i.e. case (a), the low-\lc , minimal depletion model for \dla\ 0812$+$32, z$_{abs}$=2.06, and the 'global' models for both \dla\ 0812$+$32, z$_{abs}$ = 2.62 and \dla\ 2340$-$00.  We chose the 'global' models as the most likely to be correct because of the inherent uncertainty in determining the distribution of both the N(\hi ) and the metals amongst the velocity components necessary for the component by component analysis.  
A summary of these models is shown in Figure~\ref{fig:all_press}.  The resulting median values for this sample are:  $<$n(\hi )$>$ = 69 cm$^{-3}$, $<$T$>$ = 50 K, and $<$log(P/k)$>$ = 3.86 \cmk , with standard deviations,  $\sigma$$_{n(\hi )}$ = 134 cm$^{-3}$, $\sigma$$_T$ = 52 K, and $\sigma$$_{log(P/k)}$ = 3.68 \cmk.  

In the Milky Way, ~\cite{jenkins01} used very high resolution (R = 200,000) STIS data to analyze \ci\ fine structure populations and find a median log(P/k) = 3.35 \cmk .  This median pressure is similar to the pressures we derive in high redshift \ci\ -bearing \dlas , where the median pressures derived 
are typically log(P/k) = 3 $-$ 4 \cmk .  
While this would seem to indicate that physical conditions similar to the Milky Way exist in high-$z$ \dlas , we point out that in the case of the Milky Way, the high pressure is driven by the much higher heating rate (log\lc\ $\sim$$-$25 \lcunit\ ), a result of the higher dust-to-gas ratio (typically 30 times that of \dlas ).  In contrast, the heating rates in high redshift \dlas\ are generally 1$-$2 orders of magnitude smaller.  Hence, the pressures derived in the \ci\ -bearing clouds, while similar to those observed in the Milky Way, are not the result of the same physical conditions as those observed in the Milky Way.  ~\cite{jenkins01} also observe a small proportion of the gas in many sightlines to be at very high pressures, P/k $>$ 10$^5$ \cmk , which they speculate are caused by converging flows in a turbulent medium or in turbulent boundary layers: such pressures have not been detected in \dlas . 

It is perhaps more meaningful to compare \dlas\ with the Large and Small Magellanic Clouds that, like \dlas , are known to have sub-Milky Way dust-to-gas ratios and metallicities.~\cite{tumlinson2002} performed a FUSE survey of \htwo\ along 70 sightlines to the Small and Large Magellanic Clouds.   
 For all sightlines with logN(\htwo ) $\geq$ 16.5 cm$^{-2}$ they find $<T_{01}>$ = 82 $\pm$ 21 K, whereas for all sightlines (including those at lower columns), they find $<T_{01}>$ = 115 K.  This can be compared with the Galactic average, T = 77 $\pm$ 17 K ~\citep{savage77}. 
  The temperatures found by the \ci\ and \htwo\ data presented here are in broad agreement with these values.  
A noticeable exception to this agreement are the temperatures derived from the \htwo\ rotational states in \dla\  2340$-$00.  They are generally higher (T$_{ex}$$^{01}$ $\approx$ 150 - 600 K) than those found in the SMC/LMC and in better agreement with the mean kinetic temperature of the gas, T = 153 $\pm$ 78 K, found by ~\cite{srianand2005} in a sample of \htwo\ -bearing \dlas\ at high-$z$.

\section{Conclusions}~\label{sec:conclusion}
The goal of this paper is to present new detections of neutral carbon in high redshift \dlas\ and to present a new method for analyzing the \ci\ fine structure lines.  As done by several other authors, we utilize \ci\ fine structure lines to determine densities, however, instead of assuming a temperature, our work constrained the allowed density and temperature combinations using only the column density of \ci\ in the fine structure states and the assumption of ionization equilibrium.  In a second paper, we will incorporate these physical conditions into a general model for \ci -bearing \dlas .
  Our major conclusions are as follows:

\begin{enumerate}
\item The steady state analysis of \ci\ fine structure populations along with the assumption of ionization equilibrium provides realistic constraints on both the volume density \emph{and} temperature of high redshift \dlas .  The \ci\ data are in general consistent with the radiation fields, \jnulocciistr , derived from the \ciistr\ technique and provide further evidence of the presence of CNM in high redshift \dlas .  

\item The densities and pressures of the \ci\ - bearing gas are systematically higher than those of the 'global' \dla\ predicted by the \ciistr\ technique. 
We propose two physical scenarios that could be consistent with the data presented here.  First, the \ci\ could be tracing overdensities that are created by shocks, hence the \ci\ exists in the post-shock cool gas.  However, it seems likely that the post-shock gas would have a systemic velocity offset from the pre-shock gas.  This is not observed.  In fact, the general good agreement between the velocity centroids of the \ci\ and other resonance lines argues against the shock idea.  A second scenario is that the \ci\ exists in higher density, higher-pressure edge of a photodissociation region, i.e. the edge of a molecular cloud.  While the high optical depth through a classical molecular cloud would obscure a background quasar, the photodissociation region, or edge of the molecular, cloud could be optically thin enough to allow transmission of the background quasar light and would be consistent with the gas physics determined by the \ci\ fine structure lines. 

\item As noted by ~\cite{srianand2005}, all \ci\ -bearing \dlas\ also contain \ciistr\ absorption.  We find that, with only one exception, all \ci\ objects for which \ciistr\ coverage is available (5 presented in this paper, 1 yet to be published, and 7 from the literature) contain strong \ciistr\ absorption, placing them in the category of 'high-cool' \dlas .  This could be simply a consequence of the fact that \ci\ -bearing \dlas\ generally host larger fractions of \htwo\ 
whose formation is encouraged by the higher than average metallicities and dust-to-gas ratios, consistent with the 'high-cool' population of \dlas .

\item High resolution studies of neutral carbon lines reveal narrow, sub-1\kms , cold and unresolved components.  These components likely contain relatively large amounts of gas and are most likely cold, dense knots, perhaps photodissociation regions on the edges of star forming regions.  This would explain the presence of \ci , \htwo , and larger than average dust-to-gas ratios.  To date, two such components have been published; in \dla\ 0812$+$32 with a temperature of T $\leq$ 78 K ~\citep{jorg09} and in \dla\ 1331$+$17 with T $\leq$ 218 K ~\citep{carswell09} and we presented an additional candidate in this paper.  Such clouds may exist in all \ci\ systems.  Their non-detection in other \ci\ systems does not rule out their existence due to the difficulty in detecting such small equivalent widths that are likely blended with other velocity components.
It is possible that these narrow components are ubiquitous and contain significant amounts of gas that has been previously missed in lower resolution studies.  

\end{enumerate}

\acknowledgments
The authors wish to thank Bob Carswell for many useful discussions.  R.A.J. acknowledges support from the STFC-funded Galaxy Formation and Evolution programme at the Institute of Astronomy, University of Cambridge.  A.M.W., R.A.J., \& J.X.P. acknowledge partial support by NSF grant AST-07-09235.  The authors wish to recognize and acknowledge the very significant cultural role and reverence that the summit of Mauna Kea has always had within the indigenous Hawaiian community.  We are most fortunate to have the opportunity to conduct observations from this mountain.  

\section{Appendix 1: Molecular Hydrogen}

The purpose of this Appendix is to present an outline of the molecular hydrogen analysis performed in this work.  While the focus of this work was the analysis of neutral carbon, \dlas\ that contain neutral carbon are likely to also contain detectable \htwo .  
This is because neutral carbon and molecular hydrogen are photoionized and photodissociated respectively, by photons of the same energies and therefore, they are usually found together.   This is the case in several of the \dlas\ presented in this work, i.e. \dla\ 1331$+$17, \dla\ 0812$+$32, and \dla\ 2340$-$00.  For \dla\ 2231$-$00 and the low-$z$ \dla\ 0812$+$32, we did not have coverage of the \htwo\ region and therefore cannot determine anything about the presence of \htwo .  Here we present an outline of the \dla\ \htwo\ analysis frequently performed in the literature, most recently by works such as ~\cite{levshakov02}, ~\cite{hira05}, ~\cite{cui05}, ~\cite{noterdaeme07a}, and ~\cite{noterdaeme07b} and based upon work by ~\cite{spitzer74}, ~\cite{jura75a} and ~\cite{jura75b}.  Essentially, the measurement of \htwo\ in the different rotational J states allows for an independent estimation of the physical properties of the cloud such as density, temperature and the incident radiation field.  While the \htwo\ analysis arguably contains several uncertainties and assumptions, it is nonetheless interesting to compare the results from these two independent methods, the \htwo\ analysis we present below and the \ci\ fine structure analysis presented in this work.   

First, we can estimate the kinetic temperature of the cloud using the column densities of \htwo\ in the J=0 and higher J rotational states.  According to the Boltzmann distribution (see equation 8 in ~\cite{levshakov02}):

\begin{equation}
\frac{N(J)}{N(0)} = \frac{g(J)}{g(0)} e^ - {\frac{B_v J (J + 1)}{T_{ex}}}
\label{eqn:et}
\end{equation} 

\noindent where B$_v$ = 85.36 K for the vibrational ground state and g(J) is the degeneracy of level J, given by, for level J = x, g$_x$ = (2J$_x$ $+$ 1)(2I$_x$ $+$ 1) where I = 0 for even x and I = 1 for odd x (For states J = 0$-$5, g = [1, 9, 5, 21, 9, 33]).  The excitation diagram is typically plotted as log (N$_J$ / g$_J$) versus the relative energy between the level J and J = 0.  The excitation temperature, defined in equation ~\ref{eqn:et}, is inversely proportional to the negative slope of the line connecting the excitation diagram points,  i.e.

\begin{equation}
T_{ex}^{0J}  = -B_v J (J + 1) \frac{1}{\ln \frac{g_0 N_J}{g_J N_0}} 
\end{equation}

\noindent The typical assumption 
is that the kinetic temperature of the cloud can be estimated by the excitation temperature derived from the population of \htwo\ in the states J = 0 and J = 1, assuming that the J = 1 level is thermalized.  Because the critical density of the low J states is relatively small, the population of the low J levels is generally dominated by collisional excitation and therefore it reflects the kinetic temperature of the gas when in local thermodynamical equilibrium.  The higher J states (J $\geq$ 2) are typically characterized by a higher T$_{ex}$, or flatter slope, which is explained by population mechanisms other than collisions. 

The higher T$_{ex}$ derived from the population of \htwo\ in the higher rotational J states was originally unexpected ~\citep{spitzer74}.  After observations by Spitzer and Chochran in the 1970s in which they observed the high excitation temperatures derived from the high J states, ~\cite{spitzer74} proposed methods other than collisions that could populate the higher J states.  They proposed two methods other than collisions: 1) the cascade down from upper vibrational levels following the absorption and reemission of Lyman and Werner band photons -- i.e. the J = 0 molecule is excited to a higher vibrational state and then de-excites to a higher J state in the ground vibrational state (rather than staying in the original J=0  or J=1 state), or 2) the direct formation of \htwo\ in a higher state (i.e. the \htwo\ pops off the dust particle in an excited state and cascades to say J = 4).  Since these processes are believed to dominate collisions, the typical practice is to neglect collisions and to consider only these two processes, as we show in the following analysis.  

Assuming steady state and neglecting collisions, we use the two afore mentioned populating mechanisms and assume depopulation by spontaneous emission (true until the density is above 10$^4$), to write the following steady state equations for state J = 4 and J = 5 respectively (these are equations 2a and 2b from ~\cite{jura75b}),

\begin{equation}
p_{4,0} \beta _0 n(H_2, J=0) + 0.19 R n(H I) n(H)  = A_{4,2} n(H_2, J=4)
\label{eqn:level4}
\end{equation}

\noindent and

\begin{equation}
p_{5,0} \beta _1 n(H_2, J=1) + 0.44 R n(H I) n(H)  = A_{5,3} n(H_2, J=5)
\label{eqn:level5}
\end{equation}

\noindent  where p is the pumping coefficient or pumping efficiency (or by Jura, the redistribution probability) into the J=4 and J=5 levels from the J=0 and J=1 levels respectively, $\beta$ is the photoabsorption rate, R is the \htwo\ molecule formation rate (on dust grains, also known as R$_{dust}$), n(H I) is the neutral hydrogen number density, n(H) $\approx$n(H I) $+$ 2n(\htwo ), 
and A is the spontaneous transition probabilities.  In other words, the first term is the UV excitation and decay to higher J state, while term 2 represents the direct formation in the higher J state.  The values of the constants are as follows:  A$_{4,2}$ = 2.75 $\times$ 10$^{-9}$s$^{-1}$, A$_{5,3}$ = 9.9 $\times$ 10$^{-9}$s$^{-1}$ (Spitzer 1978), p$_{4,0}$ = 0.26, p$_{5.1}$ = 0.12 ~\citep{jura75b}.  Solving for $\beta$ will allow us to estimate the UV field incident on the cloud, while $\beta$ together with R will allow us to estimate the neutral hydrogen density.  

In order to solve for $\beta$ and to get rid of the N(\hi ) dependence in equation ~\ref{eqn:level4}, we take advantage of the assumption of equilibrium between the formation and the destruction of \htwo\ (which is reasonable because the time scales of \htwo\ formation and destruction are well below a dynamical time), as follows, 

\begin {equation}
R n(HI) n(H) = R_{diss} n(H_2)
\label{eqn:equ}
\end{equation}

\noindent  If we substitute equation ~\ref{eqn:equ} into equation ~\ref{eqn:level4}, and make the common assumption that 11\% of photoabsorption leads to photodissociation ~\citep{jura74b} , 

\begin{equation}
R_{diss} = 0.11 \beta\ 
\label{eqn:rdisstobeta}
\end{equation}

\noindent we obtain equilibrium equations that are independent of the neutral hydrogen column density \nhi , 

\begin{equation}
p_{4,0} \beta _0 \frac{N(H_2, J=0)}{N(H_{2})} + 0.021 \beta _0  = A_{42} \frac{N(H_2, J=4)}{N(H_{2})}
\label{eqn:level4_nonhi}
\end{equation}

\noindent and

\begin{equation}
p_{5,0} \beta _1 \frac{N(H_2, J=1)}{N(H_{2})} + 0.049 \beta _1  = A_{53} \frac{N(H_2, J=5)}{N(H_{2})}
\perd
\label{eqn:level5_nohi}
\end{equation}

\noindent Using the measured \htwo\ column densities we can then solve for $\beta$, the photoabsorption rate of \htwo\ in each component.  Note, that $\beta _0$ should equal $\beta _1$.  

Once we have solved for $\beta$ we can determine the incident radiation field by using the relation between the radiation field and the photodissociation rate that it induces on the molecular hydrogen.  Following ~\cite{abel97} and ~\cite{hira05},

\begin{equation}
R_{diss} = (4\pi ) 1.1 \times 10^{8} J_{\nu }^{LW} S_{shield} s^{-1}
\label{eqn:rdiss}
\end{equation}

\noindent where R$_{diss}$ is the photodissociation rate (= 0.11$\beta$ as above), \jnulw\ (LW stands for Lyman-Werner) is the UV intensity at h$\nu$ = 12.87 eV averaged over the solid angle (12.87 eV is the dominant energy at which the photodissociation happens).  
This can be compared with the \jnu\ that is typically calculated in the \ciistr\ technique at $\lambda$ = 1500\AA , or 8.27 eV).  S$_{shield}$ accounts for shielding due to two effects: 1) dust shielding and 2) self-sheilding.   In order to solve for \jnulw\  we must determine the effects of shielding in equation~\ref{eqn:rdiss}.  We estimate the shielding, following ~\cite{hira05}, as

\begin{equation}
S_{shield} = (\frac{N(H_{2})}{10^{14}cm^{-2}})^{-0.75} e^{-\sigma _{d} N_{d}}
\label{eqn:shield}
\end{equation}

\noindent where the first term expresses the self-shielding and the exponential term is the shielding due to dust.  N$_{d}$, the column density of dust, is related to the HI column density by: (4$/$3)$\pi$ a$^3$$\delta$N$_d$ = 1.4 m$_H$N$_H$D, where 1.4 is the correction for the helium content, and $\sigma _d$ is the cross-section of a grain, 
$\sigma _d$ = $\pi$ a$^2$. 
$\sigma _{d}$ N$_{d}$ = $\tau _{UV}$ is the optical depth in dust and is expressed by ~\cite{hira05} as,

\begin{equation}
\begin{array}{lr}
\tau_{UV} = \frac{4.2 N_H m_H D}{4 a \delta} = \\
0.879 (\frac{a}{0.1\micron})^{-1} (\frac{\delta}{2g\ cm^{-3}})^{-1} (\frac{D}{10^{-2}}) (\frac{N_H}{10^{21} cm^{-2}}) 
\end{array}
\label{eqn:tau}
\end{equation}

\noindent where a is the radius of a grain, $\delta$ is the grain material density, and D is the dust-to-gas mass ratio.  ~\cite{hira05} assume the Galactic (Milky Way) dust-to-gas mass ratio to be D$_\odot$ = 0.01.  They define the normalized dust-to-gas ratio $\kappa$ = D/D$_\odot$.  Assuming a=0.1 and $\delta$=2, equation~\ref{eqn:tau} can be written,

\begin{equation}
\tau_{UV} = 0.879  \kappa  (\frac{N_H}{10^{21} cm^{-2}}) 
\end{equation}

\noindent see ~\cite{cui05} equation 7.  

The first part of equation ~\ref{eqn:shield}, the self-sheidling of \htwo , is taken from an analytic approximation from ~\cite{draine96} and is valid for N(\htwo) $>$ 10$^{14}$ cm$^{-2}$.  We can therefore rewrite equation~\ref{eqn:shield} as,

\begin{equation}
S_{shield} = (\frac{N(H_{2})}{10^{14}cm^{-2}})^{-0.75} exp[- 0.879 \kappa (\frac{N_H}{10^{21} cm^{-2}})]
\label{eqn:shield2}
\end{equation}

\noindent Therefore, given $\beta$ and S$_{shield}$ we can use equation~\ref{eqn:rdiss} to solve for the ambient radiation field, \jnulw .  Note that this is the total radiation field, or \jnulw\ = \jnutot .

We can also use the measurements of N(\htwo ) to estimate the volume density of \hi .  We define the molecular fraction, f$_{H_2}$ as follows,

\begin{equation}
f_{H_2} = \frac{2n(H_2)}{n(H I) + 2n(H_2)} = \frac{2N(H_2)}{N(H I) + 2N(H_2)}
\label{eqn:h2frac}
\end{equation}

\noindent where we measure the N(H I) and the N(\htwo ) directly.  If we substitute ~\ref{eqn:h2frac} into ~\ref{eqn:equ} and remember that n(H) $\approx$n(H I) $+$ 2n(\htwo ), we can solve for the number density of hydrogen,

\begin{equation}
R  n(H I) = R_{diss} \frac{n(H_2)}{n(H I) + 2n(H_2)} = R_{diss} \frac{f_{H_2}}{2}
\end{equation}

or
\begin{equation}
n(H I) = \frac{R_{diss}}{R} \frac{f_{H_2}}{2}
\label{eqn:density}
\end{equation}
 
\noindent This is temperature dependent however, and we will use the detailed expression for R given by ~\cite{hira05} (note, they call it R$_{dust}$),

\begin{equation}
R = 4.1 \times 10^{-17} S_d (T) (\frac{a}{0.1\micron})^{-1} (\frac{D}{10^{-2}}) (\frac{T}{100 K})^{1/2} (\frac{\delta}{2g cm^{-3}})^{-1}
\end{equation}

\noindent where S$_d$(T) is the sticking coefficient of hydrogen atoms onto dust and everything else was defined previously.  The sticking coefficient is given by (~\cite{hollenbach79}; ~\cite{omukai00}

\begin{equation}
\begin{array}{lr}
S_d(T) = [1 + 0.04(T + T_d)^{0.5} + 2 \times 10^{-3}T + 8 \times 10^{-6} T^2]^{-1} \times \\
(1 + exp[7.5 \times 10^2(1/75 - 1/T_d)])^{-1}
\end{array}
\end{equation}

\noindent where T$_d$ is the dust temperature and is given by 

\begin{equation}
T_d = 12(\chi Q_{UV})^{1/6}  (\frac{A}{3.2 \times 10^{-3} cm})^{-1/6}(\frac{a}{0.1\micron})^{-1/6}K
\end{equation}

\noindent where A is a constant that depends on the optical properties of the dust grains.  For silicate grains, A = 1.34 $\times$ 10$^{-3}$ cm and for carbonaceous grains, A = 3.20 $\times$ 10$^{-3}$ cm. Following ~\cite{hira05} we assume Q$_{UV}$ = 1 (Q$_{UV}$ is the dimensionless absorption cross-section normalized by the geometrical cross-section) , A = 3.20 $\times$ 10$^{-3}$ cm and a = 0.1 $\micron$, while $\chi$, the normalized radiation field, was calculated previously from the \htwo\ levels ($\chi$ = \jnulw\ / \jnulw $_{\odot}$ where \jnulw $_{\odot}$ = 3.2 $\times$ 10$^{-20}$ \junit .
Therefore, we can determine R as a function of T.  In the present work, we assume that T is equal to the excitation temperature as derived from the J=0 and J=1 \htwo\ states.
Finally, we use equation~\ref{eqn:density} to estimate the neutral hydrogen density.  Note, this method of determining temperature, density and radiation field, is independent of the \ci\ fine structure data.

\clearpage

\clearpage

\clearpage

\clearpage

\begin{deluxetable}{ccccc}
\tablewidth{0pc}
\tablecaption{DETAILS OF OBSERVATIONS\label{tab:observations}}
\tabletypesize{\footnotesize}
\tablehead{\colhead{Quasar}&\colhead{Telescope/Instrument} &\colhead{Date}&\colhead{Resolution}&\colhead{Total Exp. Time}\\
& & & FWHM [\kms ] & [s] }
\startdata
FJ0812$+$32&Keck/HIRES&Mar. 16/17, 2005&8.33&14400\\
FJ0812$+$32&Keck/HIRES&Jan. 12, 2008&6.25& 7200\\
Q1331$+$17&Keck/HIRES&Jan. 10, 1994&6.25&3600\\
Q1331$+$17&VLT/UVES&Dec. 18, 2001&7.0&4500\\
J2100$-$06&Keck/HIRES&Sept. 18, 2007&6.25&10800\\
Q2231$-$00&Keck/HIRES&Jan. 11, 1995&8.33&5400\\
J2340$-$00&Keck/HIRES&Aug. 18/19, 2006&6.25&15000\\
\enddata
\end{deluxetable}

\begin{deluxetable}{lcccccccccc}
\tablewidth{0pc}
\tablecaption{\ci\ Data\label{tab:cidata}}
\tabletypesize{\scriptsize}
\tablehead{\colhead{Quasar } &\colhead{cmp}&\colhead{$z_{abs}$} &  
\colhead{FWHM$_{inst}$$^{a}$} & \colhead{b} &\colhead{logN(CI)}
&\colhead{logN(CI*)}&\colhead{logN(CI**)]}&\colhead{T$_{ex}^{01}$$^b$}&\colhead{(f1, f2)$^c$}\\
&&&[\kms]&[\kms]&[cm$^{-2}$]&[cm$^{-2}$]&[cm$^{-2}$]&[K]
}
\startdata
FJ0812$+$32&1&2.066780(7)&6.25/8.33&1.20 $\pm$ 0.14&12.96 $\pm$ 0.04&12.92 $\pm$ 0.02&12.35 $\pm$ 0.04&19.8&(0.43,0.11)\\
\cutinhead{}
FJ0812$+$32&1& 2.625808(2)&6.25/8.33&3.25 $\pm$ 1.00&12.13 $\pm$ 0.05&11.68 $\pm$ 0.16&11.37 $\pm$ 0.270&11.1&(0.23, 0.11)\\ 
FJ0812$+$32&2&2.6263247(8)&...&2.57 $\pm$ 0.56& 12.70 $\pm$ 0.02&12.32 $\pm$ 0.04&$<$12.39(1$\sigma$)&12.0&(0.22, 0.26)$^{d}$\\
FJ0812$+$32&3&2.626491(1)&...&0.33$^{e}$ $\pm$ 0.05&13.30 $\pm$ 0.23&13.02 $\pm$ 0.03&12.47 $\pm$ 0.05&13.5&(0.31,0.09)\\

\cutinhead{}
Q1331$+$17&1& 1.7763702(9) & 6.25/7.0&5.08 $\pm$ 0.24 & 13.08 $\pm$ 0.02&12.59 $\pm$ 0.02 &11.90 $\pm$ 0.14&10.6&(0.23, 0.05) \\ 
Q1331$+$17&2 &1.776524(1) & ...&0.55 $\pm$ 0.13&13.06 $\pm$ 0.13&12.05 $\pm$ 0.07&$<$11.7&6.9$^f$&-- \\ 
Q1331$+$17&3&1.77664(5)& ...& 24.45 $\pm$ 6.2&  12.51 $\pm$ 0.11&$<$12.2&$<$12.25&$<$13.0&--\\ 
\\

\cutinhead{}

Q2100$-$06&1&3.089776(7)&6.25& 0.20 $\pm$ 0.3& 12.28 $\pm$ 0.21&12.01 $\pm$ 0.13&11.56 $\pm$ 0.28&13.7&(0.31, 0.11)\\ 
Q2100$-$06&2&3.091463(4)&...& 3.89 $\pm$ 0.69& 12.57 $\pm$ 0.03& 12.31 $\pm$ 0.08&--&13.9&--\\
Q2100$-$06&3&3.09236(2)&...&  12.77 $\pm$ 1.84& 12.61 $\pm$ 0.05&12.16 $\pm$ 0.12&--&11.1&--\\

\\
\cutinhead{}
Q2231$-$00&1&2.06534(3)&8.33&19.08 $\pm$ 4.37 & 12.55 $\pm$ 0.08 &--&--&--\\ 
Q2231$-$00&2& 2.066122(5)&...& 5.36 $\pm$ 1.17 & 12.65 $\pm$ 0.05 & 12.41 $\pm$ 0.09& 11.60 $\pm$ 0.16&14.3&(0.35, 0.05)\\ 

\cutinhead{}

J2340$-$00 &1 & 2.054151(4) &6.25& 3.36 $\pm$ 0.80 & 12.32 $\pm$ 0.04  & 11.96 $\pm$ 0.12& --&12.2& --\\
J2340$-$00 &2 & 2.054285(3) & ...&0.24 $\pm$ 0.12 & 12.54 $\pm$ 0.17 & 12.28 $\pm$ 0.09 &11.84 $\pm$ 0.14&13.9&(0.32,0.11)\\
J2340$-$00 &3 & 2.054526(1) &  ...&2.55$\pm$ 0.16 & 13.33 $\pm$ 0.02  & 12.98 $\pm$ 0.02  & 11.52 $\pm$ 0.33&12.4&(0.31, 0.01)\\
J2340$-$00 &4 & 2.054602(3) & ...& 1.39 $\pm$ 0.70 & 12.97 $\pm$ 0.12 & 12.53 $\pm$ 0.12 & 12.12 $\pm$ 0.13&11.2&(0.24, 0.09) \\
J2340$-$00 &5 & 2.05469(3) & ...& 7.01 $\pm$ 3.21 & 13.03 $\pm$ 0.08 & 12.46 $\pm$ 0.31  & 12.02 $\pm$ 0.34&9.8&(0.20, 0.07) \\
J2340$-$00 &6 & 2.054711(9) &  ...&1.19 $\pm$ 2.31 & 12.54 $\pm$ 0.67  & 12.64 $\pm$ 0.36 & 11.96 $\pm$ 0.26 &27.2&(0.50, 0.11)\\
J2340$-$00 &7 & 2.054727(5) &  ...&0.61 $\pm$ 0.29 & 13.48 $\pm$ 0.10& 12.59 $\pm$ 0.37 & -- &7.5& --\\
J2340$-$00 &8 & 2.05499(1) &  ...&9.12 $\pm$ 2.11 & 12.56 $\pm$ 0.07& 11.96 $\pm$ 0.16 & 11.70 $\pm$ 0.29 &9.5& (0.18, 0.10)\\
J2340$-$00 &9 & 2.055131(4) &  ...&2.92 $\pm$ 0.90 & 12.47 $\pm$ 0.07& 12.15 $\pm$ 0.08 & 11.72 $\pm$ 0.16 &12.9&(0.29, 0.11)\\

\cutinhead{}

J2340$-$00 &1 & 1.36027 & 6.25 & 1.58 $\pm$ & 12.83 $\pm$ 0.07 & 12.28 $\pm$ 0.07  & 11.96 $\pm$ 0.14&10.0&(0.20, 0.10) \\
J2340$-$00 &2 & 1.36049 & ...& 1.03 $\pm$ & 12.68 $\pm$ 0.08 & 11.78 $\pm$ 0.29 & -- &7.4& -- \\
J2340$-$00 &3 & 1.36061 &  ...&9.75$\pm$  & 11.74 $\pm$ 0.27  & 12.42 $\pm$ 0.15  & 11.96 $\pm$ 0.26& ...$^g$ &(0.64, 0.22)\\
J2340$-$00 &4 & 1.36088 & ...&0.64 $\pm$  & 13.23 $\pm$ 0.28 & 12.77 $\pm$ 0.08 &11.81 $\pm$ 0.21&10.9& (0.25, 0.03)\\

\enddata
\tablenotetext{a}{Instrumental FWHM}
\tablenotetext{b}{T$_{ex}^{01}$ is the excitation temperature derived from the \ci\ and \cistr\ states.  This should be consistent with and generally is higher than T$_{CMB}$.}
\tablenotetext{c}{f1 and f2 as defined in the text}
\tablenotetext{d}{(f1, f2 ) derived using the upper limit on N(\cistrstr )}
\tablenotetext{e}{Doppler parameter determined from the curve of growth method in ~\cite{jorg09}}
\tablenotetext{f}{T$_{ex}$ $<$ T$_{CMB}$. This unphysical result is likely due to measurement error.  Indeed, accounting for the given error, T$_{ex}$ = 8 K consistent with T$_{CMB}$ = 7.6 K.}
\tablenotetext{g}{N(\cistr ) $>>$ N(\ci ) which implies T$_{ex}$ = $-$50 K.}
\end{deluxetable}

\clearpage
\begin{landscape}
\begin{deluxetable}{cccccccccc}
\tablewidth{0pc}
\tablecaption{Other Relevant DLA Data\label{tab:otherdata}}
\tabletypesize{\scriptsize}
\tablehead{\colhead{Quasar} & \colhead{$z_{abs}$} & \colhead{logN(H I)} & 
\colhead{logN(\ciistr )} &\colhead{$l_c$} & 
\colhead{[M/H]$^a$} &\colhead{[Fe/H]$^b$}  &\colhead{logN(CI$^{tot}$)}&\colhead{log ($\frac{\cii }{\ci }$)$^{c}$}&\colhead{$\Delta v$}\\
&&[cm$^{-2}$]&[cm$^{-2}$]&[ergs s$^{-1}$H$^{-1}$]&&&&[cm$^{-2}$]&[\kms ]\\
}
\startdata
FJ0812$+$32&2.066780&21.50 $\pm$ 0.20&13.62$^d$&$-$27.4$^d$&$-$1.83 $\pm$ 0.20&$-$2.06 $\pm$ 0.02&13.30&2.57 $\pm$ 0.20&26\\
...&...&...&14.42$^e$&$-$26.6$^e$&...&...&...&...&...\\
FJ0812$+$32&2.62633&21.35 $\pm$ 0.10&14.30 $\pm$ 0.01&$-$26.56 $\pm$ 0.10&$-$0.81$^{f}$ $\pm$ 0.10&$-$1.62$^{g}$ $\pm$ 0.10&13.63&3.10 $\pm$ 0.20&70\\
Q1331$+$17&1.77636&21.14 $\pm$ 0.08&$<$13.56$^{h}$&$<$$-$27.16&$-$1.37 $\pm$ 0.01&$-$1.97 $\pm$ 0.01&13.56&2.40 $\pm$ 0.20&72\\
Q2100$-$06&3.09237&21.05 $\pm$ 0.15 & 14.09 $\pm$ 0.01 & $-$26.48 $\pm$ 0.15 & $-0.73 \pm 0.15$ & $-$1.20 $\pm$ 0.02 & 13.17 & 3.39 $\pm$ 0.02&187\\
Q2231$-$00&2.06615&20.56 $\pm$ 0.10&13.71 $\pm$ 0.04&$-$26.37 $\pm$ 0.11&$-0.74 \pm 0.16$&$-1.40 \pm 0.07$&13.04&2.88 $\pm$ 0.20&122\\
J2340$-$00&2.05452&20.35 $\pm$ 0.15& 13.84 $\pm$ 0.04&$-$26.03 $\pm$ 0.15&$-0.74\pm0.16$&$-0.92\pm0.03$&14.09&1.86 $\pm$ 0.20&104\\
\enddata
\tablenotetext{a}{Logarithmic $\alpha$-metal abundance with respect to solar, determined using Si II unless otherwise labeled}
\tablenotetext{b}{Logarithmic Fe abundance with respect to solar}
\tablenotetext{c}{ N(CII) calculated assuming the "minimal depletion model" from \cite{wolfe04} where [C/H] = [Si/H] - 0.2.  See text for details.}
\tablenotetext{d}{Assume low cool \dla\ with average low cool \lc .  (The measured upper limit on log N(\ciistr ) = 14.06 cm$^{-2}$, which includes blending)}
\tablenotetext{e}{Assume high cool \dla\ with average high cool \lc\ }
\tablenotetext{f}{Global metallicity based on Zn II measurement.}
\tablenotetext{g}{Based on Cr II instead of Fe II}
\tablenotetext{h}{N(\ciistr ) as determined by fitting constrained by Si II components.  Blending means that this is still technically an upper limit. }
\end{deluxetable}

\clearpage
\end{landscape}

\begin{deluxetable}{cccc}
\tablewidth{0pc}
\tablecaption{CII* TECHNIQUE SOLUTIONS\label{tab:ciistarsolutions}}
\tablehead{
\colhead{Quasar } &\colhead{z$_{abs}$}&\colhead{\jnuloc\ $^a$ }&\colhead{\jnubkd $^{d}$}\\
&&/1$\times$10$^{-19}$&/1$\times$10$^{-20}$
} 
\startdata
FJ0812$+$32&2.06678&0.43$^b$&2.56\\
FJ0812$+$32&2.06678&3.46$^c$&2.56\\
FJ0812$+$32&2.06678&36.38$^d$&2.56\\
FJ0812$+$32$^{global}$&2.62633&7.17&2.45\\
FJ0812$+$32$^{div}_{comp1}$&2.62633&19.2&2.45\\
FJ0812$+$32$^{div}_{comp2}$&2.62633&6.3&2.45\\
FJ0812$+$32$^{div}_{comp3}$&2.62633&259.9&2.45\\
Q1331$+$17&1.77636&3.09&2.53\\
J2100$-$06&3.09237&17&2.33\\
Q2231$-$00&2.06615&24.7&2.56\\
J2340$-$00&2.05452&52.4$^e$,10.8$^f$&2.56\\
\enddata
\tablenotetext{a}{Evaluated at 1500$\AA$; Error is approximately $\pm$ 50\% }
\tablenotetext{b}{\ciistr\ in forest. Assumed low \lc\ =$-$27.4 and [Fe/Si]=0.0, the maximal depletion model}
\tablenotetext{c}{\ciistr\ in forest. Assumed low \lc\ =$-$27.4 and [Fe/Si]=$-$0.2, the minimal depletion model}
\tablenotetext{d}{\ciistr\ in forest. Assumed high \lc\ =$-$26.6 and [Fe/Si]=$-$0.2}
\tablenotetext{e}{\jnuloc\ determined for components 3, 4, 5}
\tablenotetext{f}{\jnuloc\ determined for components 6 \& 7}
\end{deluxetable}

\clearpage
\begin{landscape}
\tabletypesize{\scriptsize}
\begin{deluxetable}{cccccccccc}
\tablefontsize{\scriptsize}
\tablewidth{0pc}
\tablecaption{\ci\ TECHNIQUE SOLUTIONS \label{tab:cisolutions}}
\setlength{\tabcolsep}{0.02in} 
\tablehead{\colhead{DLA, comp. } &
	\colhead{\jnutot\ }        &
	\colhead{log($\frac{\cii\ }{\ci\ }$)}   &
	\multicolumn{3}{c}{ 1$\sigma $ (2$\sigma $) \ci\ constraints  }  &
	\multicolumn{3}{c}{best $\chi $$^2$ fit } &
	\colhead{$\ell$ $^{a}$ } \\                   
	\colhead{}                     &
	\colhead{$\times $10$^{-19}$} &%
	\colhead{}                     &
	\colhead{n(\hi\ )}           &
	\colhead{T}                   &
	\colhead{log(P/k)}       &
	\colhead{n(\hi\ )}           &
	\colhead{T}                   &
	\colhead{log(P/k)}       &\\
	\colhead{}                      &
	\colhead{}                      &
	\colhead{}                      &
	\colhead{[cm$^{-3}$]} &
	\colhead{[K]}                 &
	\colhead{[cm$^{-3}$K]}&
	\colhead{[cm$^{-3}$]}  &
	\colhead{[K]}                 &
	\colhead{[cm$^{-3}$K]} &
	\colhead{[pc]}  
}
\startdata
0812$+$32$_{z_{abs}=2.06}$,1(a)&3.72$^{b}$&2.57$\pm$0.2& 302 $-$ 1096(302 $-$ 1202) & 32 $-$ 50(32 $-$ 63) & 4.18 $-$ 4.54(4.10 $-$ 4.58) & 316 & 50 & 4.20&3.2\\

0812$+$32$_{z_{abs}=2.06}$,1(b)&0.69$^{c}$&2.77$\pm$0.2& 40 $-$ 83(35 $-$ 91)& 251 $-$ 1000(200 $-$ 1259) & 4.26 $-$ 4.66(4.08 $-$ 4.80) & 79 & 251 &  4.30&13.0\\

0812$+$32$_{z_{abs}=2.06}$,1(c)&36.64$^{d}$&2.57$\pm$0.2& -- (3802 $-$ 12589) &-- (20 $-$ 25) & -- (4.88 $-$ 5.40) & 12589 & 20 & 5.40 & 0.1 \\
\\


0812$+$32$_{z_{abs}=2.62}$$^{global}$,1&7.4&3.10$\pm$0.2&     72 $-$ 120(23 $-$ 417) & 32 $-$ 50(10 $-$ 126) & 3.48 $-$ 3.58(2.36 $-$ 4.34) & 76 & 40 & 3.49&9.5 \\
                                                                           &11.0$^e$&3.10$\pm$0.2&     105 $-$ 151(46 $-$ 602) & 25 $-$ 32(10 $-$ 79) & 3.52 $-$ 3.58(2.66 $-$ 4.38) &&&&\\
                                                                            &3.8$^f$&3.10$\pm$0.2&     50 $-$ 138(13 $-$ 275) & 32 $-$ 53(10 $-$ 200) & 3.50 $-$ 3.64(2.12 $-$ 4.42) &&&&\\

0812$+$32$_{z_{abs}=2.62}$$^{global}$,3&7.4&3.10$\pm$0.2& 87 $-$ 398(72 $-$ 549) & 32 $-$ 126(25 $-$ 251) & 3.74 $-$ 4.30(3.54 $-$ 4.80) & 100 & 79 &3.90&7.2 \\
                                                                           &11.0$^e$&3.10$\pm$0.2&     138 $-$ 602(105 $-$ 912) & 25 $-$ 79(20 $-$ 158) & 3.74 $-$ 4.38(3.52 $-$ 4.76) &&&&\\
                                                                            &3.8$^f$&3.10$\pm$0.2&     66 $-$ 275(55 $-$ 417) & 40 $-$ 158(32 $-$ 398) & 3.74 $-$ 4.28(3.54 $-$ 4.90) &&&&\\

0812$+$32$_{z_{abs}=2.62}$$^{div}$,1&19.4&3.71$\pm$0.2& 14 $-$ 57(5 $-$ 158)& 25 $-$ 79(10 $-$ 126)& 2.86 $-$ 3.32(1.66 $-$ 4.60) & 14 & 50 & 2.86&5.9\\
                                                                        &29.0$^e$&3.71$\pm$0.2& 8 $-$ 66(8 $-$ 209)& 10 $-$ 32(10 $-$ 251)& 1.92 $-$ 3.12(1.92 $-$ 4.40) & &&&\\
                                                                        &9.9$^f$&3.71$\pm$0.2& 14 $-$ 42(2 $-$ 105)& 63 $-$ 316(10 $-$ 1259)& 3.32 $-$ 3.80(1.22 $-$ 4.86) & &&&\\

0812$+$32$_{z_{abs}=2.62}$$^{div}$,3&260&3.06$\pm$0.2&1513$-$12020(1513$-$12587)&10 $-$ 20(10 $-$ 32)&4.18 $-$ 5.38(4.18 $-$ 5.60)& 3161 & 10 & 4.50&0.1 \\
                                                                        &390.1$^e$&3.06$\pm$0.2&2629$-$12587(2187$-$12587)&10 $-$ 20(10 $-$ 32)&4.42 $-$ 5.40(4.34 $-$ 5.60)& &&&\\
                                                                        &130.2$^f$&3.06$\pm$0.2&758$-$6024(758$-$8316)&10 $-$ 20(10 $-$ 40)&3.88 $-$ 5.08(3.88 $-$ 5.42)& &&&\\


\\
1331$+$17,1&3.3&2.40$\pm$0.2&-- (229 $-$ 457) & -- (16 $-$ 20) & -- (3.66 $-$ 3.86) & 331 & 16 & 3.72&1.4\\
                         &4.9$^e$&2.40$\pm$0.2&-- (417 $-$ 1738) & -- (13 $-$ 16) & -- (3.82 $-$ 4.34) & &&&\\
                         &1.8$^f$&2.40$\pm$0.2&-- (100 $-$ 398) & -- (16 $-$ 32) & -- (3.40 $-$ 3.80) & &&&\\

\\


2100$-$06,1&17.2& 3.39$\pm$0.20& 22 $-$ 525(22 $-$ 12587) & 10 $-$ 251(10 $-$ 10000) & 2.34 $-$ 4.78(2.34 $-$ 8.10) & 69 & 50 & 3.54&5.3 \\
                       &25.7$^e$& 3.39$\pm$0.20& 44 $-$ 871(44 $-$ 12587) & 10 $-$ 158(10 $-$ 10000) & 2.64 $-$ 4.74(2.64 $-$ 8.10) &&&&\\
                       &8.7$^f$& 3.39$\pm$0.20& 24 $-$ 347(10 $-$ 7584) & 20 $-$ 501(10 $-$ 10000) & 2.88 $-$ 4.94(2.02 $-$ 7.88) &&&&\\

\\


2231$-$00,2&25.0&2.88$\pm$0.2 & 199 $-$ 1513(174 $-$ 1819) & 13 $-$ 25(10 $-$ 32) & 3.40 $-$ 4.48(3.24 $-$ 4.66) & 363 & 20 & 3.86&0.3 \\
                       &37.3$^e$&2.88$\pm$0.2 & 347 $-$ 2630(347 $-$ 3019) & 10 $-$ 20(10 $-$ 25) & 3.54 $-$ 4.62(3.54 $-$ 4.78) &&&&\\
                       &12.6$^f$&2.88$\pm$0.2 & 151 $-$ 758(79 $-$ 832) & 16 $-$ 40(10 $-$ 50) & 3.54 $-$ 4.28(2.90 $-$ 4.42) & &&&\\
\\
2340$-$00,2&0.39$^g$&1.86$\pm$0.2& 48 $-$ 240(30 $-$ 525) & 63 $-$ 501(40 $-$ 1585) & 3.84 $-$ 4.80(3.38 $-$ 5.68) & 66 & 200 & 4.12&1.1\\
2340$-$00,3&"&"&--&--&--&--&--&--&--\\
2340$-$00,4&"&"&48 $-$ 120(33 $-$ 316) & 63 $-$ 158(40 $-$ 794) & 3.78 $-$ 3.98(3.42 $-$ 5.12) & 48 & 158 & 3.88&1.5 \\
2340$-$00,5&"&"&22 $-$ 120(7 $-$ 417) & 25 $-$ 158(10 $-$ 1259) & 3.04 $-$ 4.00(1.82 $-$ 5.42) & 35 & 100 & 3.54&2.1 \\
2340$-$00,8&"&"& 35 $-$ 91(7 $-$ 166) & 40 $-$ 100(10 $-$ 316) & 3.44 $-$ 3.62(1.82 $-$ 4.40) & 35 & 79 & 3.44&2.1 \\
2340$-$00,9&"&"& 48 $-$ 138(29 $-$ 275) & 63 $-$ 200(32 $-$ 631) & 3.82 $-$ 4.10(3.30 $-$ 4.98) & 48 & 158 & 3.88&1.5 \\
&\\
2340$-$00,2&--$^h$&2.24$\pm$0.2&--&--&--&--&--&--&--\\
2340$-$00,3&52.7&1.74$\pm$0.2&--&--&--&--&--&--&--\\
2340$-$00,4&52.7&1.74$\pm$0.2&--&--&--&--&--&--&--\\
2340$-$00,5&52.7&1.74$\pm$0.2&--&--&--&--&--&--&--\\
2340$-$00,8&11.0&2.34$\pm$0.2&  -- (661 $-$ 5011)& -- (10 $-$ 20) & -- (3.82 $-$ 5.0) & 661 & 10 & 3.82&0.03\\

                       &14.5$^e$&2.34$\pm$0.2& -- (1318 $-$ 7943)& -- (10 $-$ 20) & -- (4.12 $-$ 5.10) & &&&\\
                       &5.7$^f$&2.34$\pm$0.2&  -- (219 $-$ 1819)& -- (10 $-$ 32) & -- (3.34 $-$ 4.56) & &&&\\

2340$-$00,9&11&2.34$\pm$0.2&  -- (912 $-$ 6606)& -- (13 $-$ 40) & -- (4.12 $-$ 5.32) & 912 & 16 & 4.16 &0.02\\

                       &14.5$^e$&2.34$\pm$0.2& -- (1513 $-$ 12588)& -- (13 $-$ 40) & -- (4.28 $-$ 5.60) & &&&\\
                       &5.7$^f$&2.34$\pm$0.2& -- (363 $-$ 2884)& -- (16 $-$ 63) & -- (3.80 $-$ 5.06) & &&&\\
\enddata
\tablenotetext{a}{This estimated cloud size, $\ell $ = N(\hi\ )/n(\hi\ ), is technically an upper limit.}
\tablenotetext{b}{The most likely model: Minimal depletion model, low \lc\ object with [Fe/Si] = $-$0.2}
\tablenotetext{c}{Maximal depletion model, assuming the low cool \lc\ with [Fe/Si] = 0.0}
\tablenotetext{d}{Minimal depletion model, assuming the high cool \lc\ with [Fe/Si] = $-$0.2}
\tablenotetext{e}{Includes $+$50\% error on \jnuloc }
\tablenotetext{f}{Includes $-$50\% error on \jnuloc }
\tablenotetext{g}{Minimum background assumption \jnu\ $\sim$ Haardt-Madau for this object}
\tablenotetext{h}{Solar (or super-solar) Fe/H precludes calculation. }
\end{deluxetable}

\clearpage
\end{landscape}

\clearpage
\begin{landscape}
\begin{deluxetable}{cccccccc}
\tablewidth{0pc}
\tablecaption{\ci\ Technique solutions: full \jnutot\ grid\label{tab:cisolutions_fullgrid}}
\tabletypesize{\scriptsize}
\tablehead{\colhead{\dla\ } &\colhead{comp.}&\colhead{log($\frac{\cii }{\ci }$)} &\colhead{n(\hi )(2$\sigma$)}&\colhead{T(2$\sigma$)}&\colhead{log(P/k)(2$\sigma$)}&\colhead{\jnutot\ /1$\times 10^{-19}$(2$\sigma$)} &\colhead{Consistent?$^b$}\\
&&&[cm$^{-3}$]&[K]&[cm$^{-3}$K]&$a$&}
\startdata
0812$+$32$_{z_{abs}=2.06}$ &1&2.57$\pm$0.2& $\geq$36 ($\geq$32) & 20$-$1258 (13$-$1585) &4.08$-$5.42 (4.0$-$5.50)& 0.41$-$195 (0.41$-$275) & yes  \\
\\
0812$+$32$_{z_{abs}=2.62}$$^{global}$&1&3.10$\pm$0.2& $\geq$6 ($\geq$0.1) & $\leq$4000(-) & 3.04$-$5.10 (0.08$-$ 5.70) &  $\leq$773 ($\leq$773) & yes  \\
0812$+$32$_{z_{abs}=2.62}$$^{global}$&3&3.10$\pm$0.2& $\geq$9 ($\geq$7) & $\leq$6300(-) & 3.30$-$5.40 (3.10$-$ 5.60) &  $\leq$773 ($\leq$773)  & yes \\
\\
0812$+$32$_{z_{abs}=2.62}$$^{div}$&1&3.71$\pm$0.2& 2-57(0.002-4166) & - (-) & 1.72$-$4.68 ($\leq$5.34) &  $\leq$34 ($\leq$1082) & yes  \\ 
0812$+$32$_{z_{abs}=2.62}$$^{div}$&3&3.06$\pm$0.2 & $\geq$7 ($\geq$6) & $\leq$7943(-) & 3.40$-$5.40 (3.18$-$ 5.60) &  $\leq$923 ($\leq$923)  & yes \\
\\
1331$+$17&1&$2.40\pm0.2$&11$-$44 (10$-$12021) & 79$-$794 ($\leq$794) & 3.50$-$4.04 (3.20$-$5.08) & 0.34$-$0.94 (0.34$-$86) & yes \\
\\
2100$-$06&1&3.39$\pm$0.20& $\geq$2 ($\geq$0.03) & -- & 2.14$-$5.60 (--) & -- \\
\\
2231$-$00&2&2.88$\pm$0.2& $\geq$6 ($\geq$3) & $\leq$6300 ($\leq$7950) & 3.40$-$5.30 (2.54$-$5.40) & $\leq$751 ($\leq$751) & yes \\
\\
2340$-$00&2&1.86$\pm$0.2& $\geq$48 ($\geq$30) & 20$-$500 (16$-$1585) & 3.82$-$5.50 (3.38$-$5.90) & $\leq$19 ($\leq$21) & -- \\  
2340$-$00&3&"& $\geq$132 ($\geq$110) & 13$-$32 (13$-$32) & 3.62$-$5.20 (3.54$-$5.20) & 0.92$-$67 (0.44$-$24) & no \\ 
2340$-$00&4&"& 48$-$190 ($\geq$33) & 40$-$158 (16$-$794) & 3.78$-$3.98 (3.42$-$5.68) & 0.39$-$1.1 (0.39$-$21) & no \\
2340$-$00&5&"& $\geq$22 ($\geq$7) & 13$-$158 ($\leq$1259) & 3.04$-$5.20 (1.82$-$5.80) & 0.39$-$23.8 (0.39$-$26.7) & no \\ 
2340$-$00&8&"& 35$-$120 ($\geq$7) & 32$-$100 ($\leq$316) & 3.44$-$3.62 (1.82$-$5.30) & 0.39$-$0.92 (0.39$-$26.7) & yes \\ 
2340$-$00&9&"& 47$-$316 ($\geq$29) & 32$-$200 (13$-$631) & 3.82$-$4.10 (3.30$-$5.60) & 0.39$-$1.4 (0.39$-$23.8) & yes \\ 
\\
2340$-$00&2&2.24$\pm$0.2&$\geq$24 ($\geq$13) & 20$-$1995 (13$-$6310) & 3.82$-$5.50 (3.38$-$5.90) & 0.39$-$53(0.39$-$66.7)  \\
2340$-$00&3&1.74$\pm$0.2&$\geq$132 ($\geq$110) & 13$-$32 (13$-$32) & 3.62$-$5.20 (3.54$-$5.20) & 0.92$-$67 (0.39$-$24) \\
2340$-$00&4&1.74$\pm$0.2& 60$-$190 ($\geq$44) & 40$-$126 (16$-$501) & 3.78$-$3.94 (3.44$-$5.68) & $\leq$1.1 ($\leq$21) \\
2340$-$00&5&1.74$\pm$0.2&$\geq$32 ($\geq$12) & 13$-$126 ($\leq$794) & 3.10$-$5.20 (2.06$-$5.80) & $\leq$23.8 ($\leq$26.7)\\
2340$-$00&8&2.34$\pm$0.2& 34$-$120 ($\geq$7) & 32$-$100 ($\leq$316) & 3.44$-$3.62 (1.82$-$5.30)& $\leq$9.2 ($\leq$26.7)\\
2340$-$00&9&2.34$\pm$0.2& 48$-$316 ($\geq$29) &32$-$200 (13$-$631) & 3.82$-$4.10 (3.30$-$5.60) & $\leq$1.4 ($\leq$23.8)\\
\enddata
\tablenotetext{a}{[\junit ]}
\tablenotetext{b}{Is the radiation field derived via the \ciistr\ technique consistent with the range allowed by the \ci\ data?}
\end{deluxetable}

\clearpage
\end{landscape}

 \begin{deluxetable}{ccccc}
\tablewidth{0pc}
\tablecaption{Dust to Gas Ratio Component Analysis of \dla\ 0812$+$32
\label{tab:J0812_dust2gas_comps}}
\tabletypesize{\scriptsize}
\tablehead{ &\colhead{comp. 1} &\colhead{comp. 2} &\colhead{comp. 3} &\colhead{comp. 4}\\
&\colhead{z$_{abs}$=2.625890} &\colhead{z$_{abs}$=2.626310} &\colhead{z$_{abs}$=2.626491}&\colhead{z$_{abs}$=2.626447}\\
&\colhead{b = 18.99$\pm$1.91 [\kms ]} &\colhead{b = 6.23$\pm$0.35 [\kms ]} &\colhead{b = 0.33$\pm$0.05$^a$ [\kms ]} &\colhead{b = 5.6$\pm$3.3 [\kms ]} 
    } 
\startdata
N(Cr II) [cm$^{-2}$]&12.91 $\pm$ 0.04 & 13.15 $\pm$ 0.02 & $\leq$11.71 (1$\sigma$) & 12.52 $\pm$ 0.07\\
N(Zn II) [cm$^{-2}$]&12.44 $\pm$ 0.04 & 12.96 $\pm$ 0.02 & 13.00$^b$ & 12.32 $\pm$ 0.24 \\
N(\ciistr ) [cm$^{-2}$]& 13.56$^{c}$ $\pm$ 0.2 & 13.69 $\pm$ 0.01 & 15.14 $\pm$ 0.17 & 13.08 $\pm$ 0.05\\
N(\hi )$^{d}$ [cm$^{-2}$]& 20.41 & 20.93 & 20.97 & 20.29 \\
$[Zn/H]$&  $-$0.58 $\pm$ 0.04 & $-$0.58 $\pm$ 0.02 & $-$0.58  & $-$0.58 $\pm$ 0.24 \\
$[Cr/H]$&  $-$1.13 $\pm$ 0.04 & $-$1.41 $\pm$ 0.02 & $\leq$$-$2.89 & $-$1.40 $\pm$ 0.07 \\
$[Cr/Zn]$& $-$0.55 $\pm$ 0.06 & $-$0.83 $\pm$ 0.03 & $\leq$$-$2.31 & $-$0.82 $\pm$ 0.12\\
$\kappa$$^{e}$& 0.09 & 0.13 & 0.16 & 0.13 \\
log$_{10}$$\kappa$ & $-$1.04 & $-$0.90 & $-$0.78 & $-$0.90\\
$\frac{CII}{CI}$& 3.71 $\pm$ 0.2 & 3.56 $\pm$ 0.2 & 3.06 $\pm$ 0.2 & --\\
$l_c$& $-$26.37 & $-$26.76 & $-$25.35 & $-$26.73\\
\jnuloc\  $/$10$^{-19}$$^f$ &19.2 & 6.3 & 259.9 & 2.5\\
f$_{H_2}$& 2.34$\times$10$^{-5}$& 9.0$\times$10$^{-3}$ & 0.14 & ... \\
$\beta _0$ [s$^{-1}$]&3.53$\times$10$^{-11}$&6.74$\times$10$^{-14}$&1.53$\times$10$^{-14}$ & ... \\
$\beta _1$ [s$^{-1}$]& 5.97$\times$10$^{-11}$&8.96$\times$10$^{-14}$& ... & ... \\
S$_{self}$&0.08 & 3.6$\times$10$^{-4}$ & 3.9$\times$10$^{-5}$ & ... \\
S$_{dust}$&0.98 & 0.91 & 0.88 & ...  \\
S$_{total}$&0.08 & 3.3$\times$10$^{-4}$ & 3.41$\times$10$^{-5}$ & ... \\
T$_{ex}^{01}$ [K]& 102 & 64 & 47 & ... \\
$J$$_{\nu }^{LW}$$/$10$^{-19}$$^c$&0.37&0.17&0.36 & ... \\
n(\hi )$^{g}$ [cm$^{-3}$]& 21 & 11 & 37  & ... \\
\enddata
\tablenotetext{a}{Doppler parameter fixed to match that of \ci\ as determined by ~\cite{jorg09} }
\tablenotetext{b}{N(Zn II) fixed by upper limit on N(O I) assuming solar relative abundances. See text for details. }
\tablenotetext{c}{Summed over the two components required by VPFIT, as explained in the text.}
\tablenotetext{d}{N(\hi ) scaled to trace N(Zn II)}
\tablenotetext{e}{Dust to gas ratio relative to Milky Way, as defined in the text.}
\tablenotetext{f}{[\junit ]}
\tablenotetext{g}{Density derived from the \htwo\ as explained in the text.}
\end{deluxetable}

 \begin{deluxetable}{ccccccccc}
\tablewidth{0pc}
\tablecaption{\dla\ 0812$+$32 Molecular Hydrogen \label{tab:J0812}}
\tabletypesize{\scriptsize}
\tablehead{\colhead{comp}&\colhead{ion} & \colhead{z$_{abs}$} &\colhead{$\sigma$$_{z_{abs}}$} & \colhead{b} &\colhead{$\sigma$$_{b}$} & \colhead{log N} &\colhead{$\sigma$$_{log N}$}  &\colhead{T$_{0-J}$$^a$}  \\
&&&&[\kms ] &&[cm$^{-2}$]& & [K]
  } 
\startdata
comp 1&&&&&&&\\
&H2J0 & 2.625812 & 0.000011 & 1.03 & 0.11 & 15.03 & 0.15 & --\\
&H2J1 &  "  & " &  "  & " & 15.26 & 0.15 & 102 $^{+1}_{-1}$ \\
&H2J2 &  "  & " &  "  & " & 14.04 & 0.06 & 132 $^{+7}_{-6}$\\
&H2J3 &  "  & " &  "  & " & 13.39 & 0.10 & 150 $^{+2}_{-2}$\\
&H2J4 & "   &" &   "  & " & $<$12.64& -- & 222 $^{+10}_{-9}$ \\
&H2J5 &  "  & " &  "  & " & 12.22 & 0.61 & 257 $^{+31}_{-25}$\\
&total N(H$_2$) = &  3.01 $\times$10$^{15}$\\
&log (total N(H$_2$)) = &    15.48\\
&f$^b$ $\geq$ & 2.69$\times$10$^{-6}$\\
comp 2&&&&&&&\\
&H2J0 & 2.626326 & 0.000001 & 2.67 & 0.09 & $<$18.45$^c$ & -- & --\\
&H2J1 & " & " &  "  & " & 18.25 & 0.04 & 64 $^{+24}_{-14}$ \\
&H2J2 & " & " &  "  & " & 15.49 & 0.06 & 61 $^{+4}_{-4}$ \\
&H2J3 & " & " &  "  & " & 14.43 & 0.01 & 83 $^{+5}_{-4}$  \\
&H2J4 & " & " &  "  & " & 13.18 & 0.10 & 119 $^{+4}_{-4}$ \\
&H2J5 &    " & "&      "  &   " &$<$12.60 & -- & 151 $^{+6}_{-6}$ \\ 
&total N(H$_2$) $<$ & 4.61$\times$10$^{18}$\\
&log (total N(H$_2$)) $<$ &      18.66\\
&f$^b$ $\geq$ & 4.1$\times$10$^{-3}$\\
comp 3, model 1$^d$&&&&&&&\\
&H2J0 & 2.626491& 0.000001 & 0.81$^d$ & -- & 19.79 & 0.03 & -- \\
&H2J1 & "  & "   &  " & " & 19.15 & 0.03 & 47 $^{+1}_{-1}$  \\
&H2J2 & "  & "  &  " & " & 16.60 & 0.03 & 57 $^{+1}_{-1}$ \\
&H2J3 & "  & "  &  "& " & 15.11 & 0.05 & 74 $^{+1}_{-1}$ \\
&H2J4 & "  & "  &  " & " & 13.99 & 0.04 & 110 $^{+1}_{-1}$ \\
&total N(H$_2$) = &  7.58$\times$10$^{19}$ \\
&log (total N(H$_2$)) =&      19.88\\
&f$^b$ $\geq$ &  0.06 \\
comp 3, model 2$^e$&&&&&&&\\
&H2J0 & 2.626494 & 0.000001 & 1.19$^e$ & -- & 19.81 & 0.03 & -- \\
&H2J1 & "  & "   &  " & " & 19.13 & 0.03 & 45 $^{+1}_{-1}$  \\
&H2J2 & "  & "  &  " & " & 16.21 & 0.11 & 52 $^{+1}_{-2}$ \\
&H2J3 & "  & "  &  "& " & 14.88 & 0.07 & 71 $^{+1}_{-1}$ \\
&H2J4 & "  & "  &  " & " & 13.97 & 0.04 & 109 $^{+1}_{-1}$ \\
&total N(H$_2$) = &  7.81$\times$10$^{19}$ \\
&log (total N(H$_2$)) =&      19.89\\
&f$^b$ $\geq$ &  0.07 \\
\enddata
\tablenotetext{a}{Excitation temperature between rotational level J and J = 0.}
\tablenotetext{b}{Molecular fraction f calculated using N(\hi )$^{total}$ = 21.35 cm$^{-2}$, so that these values are technically lower limits. }
\tablenotetext{c}{We report the 2$\sigma$ upper limit because blending with the stronger J=0 line of component 3 make the formal errors large.}
\tablenotetext{d}{Model 1: Doppler parameter tied to that of \ci .  Note that the J=0 and J=1 transitions are heavily saturated and therefore the resultant log N is determined by the damping wings and is insensitive to the choice of b.}
\tablenotetext{e}{Model 2: Doppler parameter determined by the J=3 state.}
\end{deluxetable}

\begin{deluxetable}{cccc}
\tablewidth{0pc}
\tablecaption{AODM Component Analysis of \dla\ 2340$-$00
\label{tab:j2340_aodm}}
\tabletypesize{\scriptsize}
\tablehead{\colhead{Super-Component}&\colhead{a}&\colhead{b}&\colhead{c}\\
\colhead{ $\Delta$$v$$_{90}$}&$-$30$-$15 \kms&15$-$70 \kms&70$-$120 \kms \\
 \colhead{\ci\ Component}&\colhead{(1, 2)} &\colhead{(3,4,5,6,7)} &\colhead{(8,9)} \\
    } 
\startdata
N(Fe II) [cm$^{-2}$]&14.55 $\pm$ 0.05&14.63 $\pm$ 0.04&14.06 $\pm$ 0.01\\
N(S II) [cm$^{-2}$]&14.17 $\pm$ 0.01&14.73 $\pm$ 0.01&14.31 $\pm$ 0.01\\
N(\ciistr ) [cm$^{-2}$]&12.53 $\pm$ 0.09& 13.60 $\pm$  0.01 & 12.99 $\pm$ 0.04  \\
N(\hi )$^{a}$ [cm$^{-2}$]&19.58&20.13&19.72 \\
$[Fe/H]$&$-$0.47 $\pm$ 0.05&$-$0.95 $\pm$ 0.04&  $-$1.10 $\pm$ 0.01\\
$[S/H]$& $-$0.56 $\pm$ 0.01& $-$0.56 $\pm$ 0.01& $-$0.56 $\pm$ 0.01\\
$[Fe/Met]$&0.09 $\pm$ 0.05&  $-$0.39 $\pm$ 0.04&$-$0.54 $\pm$ 0.01\\
$\kappa$$^{b}$&$-$0.06& 0.06&  0.09\\
log$_{10}$$\kappa$ &--& $-$1.21&$-$1.03\\
N(\ci )$^{total}$& 12.96 & 14.01 & 13.00 \\
log$\frac{CII}{CI}$&2.24&1.74&2.34\\
$l_c$&  $-$26.56& $-$26.05& $-$26.24\\
\jnuloc\ $/$10$^{-19}$$^c$ & --$^d$ &52.4&10.8\\
N(Fe III)[cm$^{-2}$]&$<$13.87$^{e}$  &$<$ 13.94$^{e}$  &$<$13.89$^{e}$ \\ 
Fe III/Fe II$^{f}$&$<$0.21&$<$0.21& $<$0.66\\
N(Ar I)[cm$^{-2}$]& 13.31&13.83&13.65\\
$[Ar/S]^g$&0.12 & 0.08 &  0.32\\
&\\
N(Ni II) [cm$^{-2}$]&13.21 $\pm$ 0.02 & 13.50 $\pm$ 0.01 & 13.14 $\pm$ 0.02 \\
$[Ni/H]$&$-$0.55 $\pm$ 0.02 & $-$0.82 $\pm$ 0.01 & $-$0.79 $\pm$ 0.02 \\
$[N/Met]$ & 0.01 $\pm$ 0.02 & $-$ 0.26 $\pm$ 0.01 & $-$ 0.23 $\pm$ 0.03\\
$\kappa$&$-$0.01 &   0.02 &  0.01 \\
log$\kappa$& --&  $-$1.67 & $-$ 1.97 \\
\enddata
\tablenotetext{a}{N(\hi ) scaled to trace N(S II)}
\tablenotetext{b}{Dust to gas ratio relative to Milky Way, defined in the text.  Here we have used S II instead of Si II. }
\tablenotetext{c}{[\junit ]}
\tablenotetext{d}{We did not determine \jnuloc\ for this component because the super-solar Fe II measurement.  }
\tablenotetext{e}{AODM measurements taken as upper limits because of possible blending with the forest.}
\tablenotetext{f}{Fe III/Fe II $\sim$0.3 means partially ionized, HI/H = 0.5}
\tablenotetext{g}{If [Ar/S] $>$ $-$0.2 then x $<$0.1  (but low Ar/S does not require x$>>$0).}
\end{deluxetable}

 \begin{deluxetable}{lcccccccc}
\tablewidth{0pc}
\tablecaption{\dla\ 2340$-$00 Molecular Hydrogen \label{tab:J2340_h2}}
\tabletypesize{\scriptsize}
\tablehead{\colhead{component$^{a}$ } &\colhead{ion} & \colhead{z$_{abs}$} &\colhead{$\sigma$$_{z_{abs}}$} & \colhead{b} &\colhead{$\sigma$$_{b}$} & \colhead{log N} &\colhead{$\sigma$$_{log N}$}  &\colhead{T$_{0-J}$$^b$}   \\
&&&&[\kms ] &&[cm$^{-2}$]& & [K]
 } 
\startdata
comp 1 &  &  &  &  &  &  &  \\
& H2J0 & 2.054165 & 0.000001 & 2.31 & 0.07 & 15.262 & 0.040 & -- \\
& H2J1 & 2.054165 & ... & 2.31 & 0.00 & 15.94 & 0.049 & 266  $^{+10}_{-9}$\\
& H2J2 & 2.054165 & ... & 2.31 & 0.00 & 14.904 & 0.065 & 211 $^{+5}_{-5}$ \\
& H2J3 & 2.054165 & ... & 2.31 & 0.00 & 14.251 & 0.049 & 191 $^{+1}_{-1}$ \\
& H2J4 & 2.054165 & ... & 2.31 & 0.00 & $<$13.39 & --  & 262 $^{+4}_{-4}$\\
& H2J5 & 2.054165 & ... & 2.31 & 0.00 & $<$12.90 & -- & 287 $^{+3}_{-3}$\\
&log (total N(H$_2$)) = &16.06&\\
&f$^{c}$ =& 1.03$\times$10$^{-4}$\\
  \\
comp 2 &  &  &  &  \\
& H2J0 & 2.054291 & 0.000001 & 1.25 & 0.09 & 14.657 & 0.073 & --  \\
& H2J1 & 2.054291 & ... & 1.25 & 0.00 & 15.292 & 0.079 & 232 $^{+4}_{-4}$ \\
& H2J2 & 2.054291 & ... & 1.25 & 0.00 & 14.478 & 0.079 & 253  $^{+2}_{-2}$\\
& H2J3 & 2.054291 & ... & 1.25 & 0.00 & 14.299 & 0.059 & 265  $^{+2}_{-2}$\\
& H2J4 & 2.054291& ... & 1.25 & 0.00 & $<$13.08 & -- & 293 $^{+9}_{-8}$ \\
 & H2J5 & 2.054291& ... & 1.25 & 0.00 & $<$13.23 & -- & 378  $^{+10}_{-9}$\\
&log (total N(H$_2$)) = &15.47&\\
&f$^{c}$ =& 2.63$\times$10$^{-5}$\\
 &  &  &  &  &  &  &  \\
comp 4 &  &  &  &  &  &  &  \\
& H2J0 & 2.054573 & 0.000002 & 4.62 & 0.16 & 17.269 & 0.082  & -- \\
& H2J1 & 2.054573& ... & 4.62 & 0.00& 17.955 & 0.054 & 276  $^{+32}_{-26}$\\
& H2J2 & 2.054573 & ... & 4.62 & 0.00 & 17.045 & 0.129 & 241  $^{+13}_{-12}$ \\
& H2J3 & 2.054573& ... & 4.62 & 0.00 & 15.326 & 0.028 & 136 $^{+2}_{-2}$ \\
& H2J4 & 2.054573 & ... & 4.62 & 0.00 & 13.812 & 0.041 & 168   $^{+2}_{-2}$\\
& H2J5 & 2.054573 & ... & 4.62 & 0.00 & 13.317 & 0.139 & 203  $^{+2}_{-2}$\\
&log (total N(H$_2$)) = &18.08&\\
&f$^{c}$ =& 1.06$\times$10$^{-2}$\\
  \\
 &  &  &  &  &  &  &  \\
comp 6 &  &  &  &  &  &  &  \\
& H2J0 & 2.054714 & 0.000001 & 5.06 & 0.11 & 15.997 & 0.047 & --  \\
& H2J1 & 2.054714 & ... & 5.06 & 0.00 & 16.825 & 0.065 & 587 $^{+99}_{-74}$ \\
& H2J2 & 2.054714 & ... & 5.06 & 0.00 & 16.059 & 0.039 & 349  $^{+5}_{-5}$\\
& H2J3 & 2.054714 & ... & 5.06 & 0.00 & 15.836 & 0.019 & 300  $^{+6}_{-6}$\\
& H2J4 & 2.054714 & ... & 5.06 & 0.00 & 14.355 & 0.015 & 286  $^{+3}_{-4}$\\
& H2J5 & 2.054714& ... & 5.06 & 0.00 & 13.932 & 0.040  & 310 $^{+1}_{-1}$\\
&log (total N(H$_2$) = &16.98&\\
&f$^{c}$ =& 8.51$\times$10$^{-4}$\\
 \\
comp 8 &  &  &  &  &  &  &  \\
& H2J0 & 2.054986 & 0.000001 & 3.64 & 0.10 & 15.76 & 0.042  & --\\
& H2J1 & 2.054986 & ... & 3.64 & 0.00 & 16.55 & 0.067 & 475 $^{+93}_{-67}$\\
& H2J2 & 2.054986 & ... & 3.64 & 0.00 & 15.63 & 0.047 & 271  $^{+2}_{-2}$\\
& H2J3 & 2.054986 & ... & 3.64 & 0.00 & 15.13 & 0.027 & 229 $^{+2}_{-2}$ \\
& H2J4 & 2.054986 & ... & 3.64& 0.00 & 13.75 & 0.046  & 251 $^{+1}_{-1}$\\
& H2J5 & 2.054986 & ... & 3.64 & 0.00 & 11.90 & 3.520 & 207 $^{+380}_{-81}$ \\
&log (total N(H$_2$) = &16.67&\\
&f$^{c}$ =& 4.21$\times$10$^{-4}$\\
 &  &  &  &  &  &  &  \\
comp 9 &  &  &  &  &  &  &  \\
& H2J0 & 2.055135 & 0.000001 & 1.80 & 0.07 & 16.735 & 0.074  & --\\
& H2J1 & 2.055135 & ... & 1.80 & 0.00 & 17.200 & 0.057 &151 $^{+6}_{-5}$ \\
& H2J2 & 2.055135 & ... & 1.80 & 0.00 & 16.035 & 0.088 & 159 $^{+2}_{-2}$\\
& H2J3 & 2.055135 & ... & 1.80 & 0.00 & 14.753 & 0.049 & 135 $^{+1}_{-1}$\\
 & H2J4 & 2.055135& ... & 1.80 & 0.00 & $<$13.10 & -- & 162 $^{+3}_{-3}$ \\
  & H2J5 & 2.055135 & ... & 1.80 & 0.00 &$<$13.02& -- & 212 $^{+3}_{-3}$\\
&log (total N(H$_2$) = &17.35&\\
&f$^{c}$ =& 2.00$\times$10$^{-3}$\\
\enddata
\tablenotetext{a}{Excitation temperature between rotational level J and J = 0.}
\tablenotetext{b}{Components are numbered by the closest z \ci\ component }
\tablenotetext{c}{f calculated assuming logN(HI) = 20.35}
\end{deluxetable}

\begin{deluxetable}{ccccccc}
\tablewidth{0pc}
\tablecaption{\htwo\ Component Analysis of \dla\ 2340$-$00 $-$ using AODM values
\label{tab:j2340_h2_comp_analysis}}
\tabletypesize{\scriptsize}
\tablehead{\colhead{Super-Component}&\colhead{a}&\colhead{a}&\colhead{b}&\colhead{b}&\colhead{c}&\colhead{c}\\ 
\colhead{component}&\colhead{1$^a$} &\colhead{2$^a$} &\colhead{4} &\colhead{6}&\colhead{8}&\colhead{9}
    } 
\startdata
f$_{H_2}$$^b$&6.04$\times$10$^{-4}$& 1.55$\times$10$^{-4}$&  1.74$\times$10$^{-2}$& 1.41$\times$10$^{-3}$& 1.79$\times$10$^{-3}$& 8.47$\times$10$^{-3}$\\

T$_{ex}^{01}$ [K]& 266& 232&276&587&  475& 151\\

$\beta _0$ [s$^{-1}$]& 9.44$\times$10$^{-11}$&1.84$\times$10$^{-10}$&  2.43$\times$10$^{-12}$& 1.36$\times$10$^{-10}$&6.31$\times$10$^{-11}$&1.84$\times$10$^{-12}$\\

$\beta _1$ [s$^{-1}$]& 1.00$\times$10$^{-10}$& 8.42$\times$10$^{-10}$&  2.52$\times$10$^{-12}$&1.44$\times$10$^{-10}$& 2.62$\times$10$^{-12}$& 5.89$\times$10$^{-12}$\\

S$_{self}$&2.85$\times$10$^{-2}$&7.92$\times$10$^{-2}$ &8.72$\times$10$^{-4}$& 5.83$\times$10$^{-3}$& 9.88$\times$10$^{-3}$& 3.07$\times$10$^{-3}$\\

S$_{dust}$& --$^a$&  --$^a$&0.993&0.993&  0.996&0.996\\

S$_{total}$& --&--&  8.66$\times$10$^{-4}$&5.78$\times$10$^{-3}$& 9.84$\times$10$^{-3}$ & 3.06$\times$10$^{-3}$\\

$J$$_{\nu }^{LW}$$/$10$^{-19}$$^c$& --& --&      2.25&   18.84& 5.13&$<$ 0.48\\

n(\hi )$^d$ [cm$^{-3}$]& --& --&   1629 &  10509 &    3595 &   377 \\

\enddata
\tablenotetext{a}{Super-solar Fe II measurement, therefore $\kappa$ was not sensible}
\tablenotetext{b}{f is calculated using the N(\hi ) of the Super-Component, i.e. N(\hi ) = 19.58 for (1, 2), N(\hi ) = 20.13 for (3, 4, 5, 6, 7) and N(\hi ) = 19.72 for (8, 9). }
\tablenotetext{c}{[\junit ]}
\tablenotetext{d}{Density derived from the \htwo\ as explained in Appendix 3.}
\end{deluxetable}

\begin{deluxetable}{cccccc}
\tablewidth{0pc}
\tablecaption{Comparison of \ci\ $^a$ and \ciistr\ Technique Models for \dla\ 1331$+$17 \label{tab:compare}}
\tabletypesize{\scriptsize}
\tablehead{
\colhead{model } &\colhead{\jnuloc\ /10$^{-19}$}&\colhead{T } &\colhead{n} &\colhead{log(P/k)}&\colhead{log($\frac{\cii }{\ci }$)}\\
&[\junit ]&[K]&[cm$^{-3}$]&[\cmk ] \\
}
\startdata
\ciistr , P$_{eq}^b$, min$^c$&    3.09&80.6& 6.9&2.75&3.70\\
\ci , P$_{eq}$, min&    3.09& 16$-$20 & 229$-$457 & 3.66$-$3.86 & 2.40\\
\\
\ciistr , P$_{max}$, min&1.74 &22.9&17.4&2.6&3.23\\
\ci , P$_{max}$, min&1.74 & 20$-$32 & 95$-$263 & 3.38$-$3.72 & 2.40\\
\\
\ciistr , P$_{min}$, min&4.90 &  761.5&  1.0  &2.88&4.51\\
\ci , P$_{min}$, min&4.90 & 13$-$16 & 347$-$1445 & 3.74$-$4.26 & 2.40\\
\\
\ciistr , P$_{eq}$, max$^d$&1.64&  102.4&   3.2    &2.52& 3.69   \\
\ci , P$_{eq}$, max&1.64 & 25$-$63 & 42$-$158 & 3.22$-$3.60 & 2.60\\
\\
\ciistr , P$_{max}$, max&  0.65& 23.6&   9.1    &2.33&3.04    \\
\ci , P$_{max}$, max&  0.65 & 25$-$63 & 46$-$158 & 3.26$-$3.60 & 2.60\\
\\
\ciistr , P$_{min}$, max& 7.34 &  838.8&1.1   &2.97& 4.62  \\
\ci , P$_{min}$, max& 7.34 & 13$-$16 & 263$-$1096 & 3.62$-$4.14  & 2.60\\
\enddata
\tablenotetext{a}{\ci\ results are all 2 $\sigma$}
\tablenotetext{b}{P$_{eq}$ = ( P$_{min}$ P$_{max}$ )$^{1/2}$}
\tablenotetext{c}{Minimal Depletion Model}
\tablenotetext{d}{Maximal Depletion Model}
\end{deluxetable}

\clearpage





 \end{document}